\documentclass{PoS}

\title{Quantum fluctuations around low-dimensional topological defects}

\ShortTitle{Quantum topological defects}

\author{\speaker{Juan Mateos Guilarte}\thanks{We are grateful to the chairman and co-chairman
of the School, Professors Carlos Pinheiro and Gentil O. Pires, for
inviting us to lecture at such a stimulating event. We also
appreciate the work of the local organizers Professors Alberto
Arruda and Harold Blas, who have succeeded in achieving, together
with the main organizers and other collaborators, a remarkable
School of Physics. Concerning the material presented in these
Lectures we are grateful above all to P. van Nieuwenhuizen for
solid, illuminating, and constructive criticism of our work in this
field, communicated to us by electronic mail. We thank also e-mail
correspondences and/or oral conversations with A. Rebhan, R. Wimmer,
M. Bordag, D. Vassilevich and H. Gies on this subject. Finally, we
acknowledge the Spanish Ministerio de Educacion y Ciencia and Junta
de Castilla y Leon for partial financial support under grants
FIS2006-09417, GR224, and SA034A08. J. M. G. thanks the ESF Research
Network CASIMIR for providing excellent opportunities for discussion
on the Casimir effect and related topics like topological defect
fluctuations.
}\\
        Departamento de Fisica Fundamental and IUFFyM, Universidad de Salamanca, SPAIN\\
        E-mail: \email{guilarte@usal.es}}

\author{A. Alonso Izquierdo\\
        Departamento de Matematica Aplicada and IUFFyM, Universidad de Salamanca, SPAIN\\
        E-mail: \email{alonsoiz@usal.es}}

\author{W. Garcia Fuertes\\
        Departamento de Fisica, Universidad de Oviedo, SPAIN\\
        E-mail: \email{wifredo@uniovi.es}}

\author{M. de la Torre Mayado\\
        Departamento de Fisica Fundamental and IUFFyM, Universidad de Salamanca, SPAIN\\
        E-mail: \email{marina@usal.es}}

\author{M. J. Senosiain\\
        Departamento de Matematicas, Universidad de Salamanca, SPAIN\\
        E-mail: \email{idiazabal@usal.es}}

\abstract{In these Lectures a method is described to analyze the
effect of quantum fluctuations on topological defect backgrounds up
to the one-loop level. The method is based on the spectral heat
kernel/zeta function regularization procedure, and it is first
applied to various types of kinks arising in several deformed linear
and non-linear sigma models with different numbers of scalar fields.
In the second part, the same conceptual framework is constructed for
the topological solitons of the planar semilocal Abelian Higgs
model, built from a doublet of complex scalar fields and one $U(1)$
gauge field.}

\FullConference{5th International School on Field Theory and Gravitation,\\
         April 20 - 24 2009\\
         Cuiabà ¡ city, Brazil}

\begin{document}

\section{Introduction}

The concept of solitary waves was brought to light in 1834 by the
Scottish civil engineer Scott Rusell, chasing a single \lq\lq wave
of translation" on a horse along a channel in Edinburgh. Unlike
ordinary dispersive waves, these non-linear waves do not fade away
and they preserve their shape, size and speed even they undergo weak
perturbations. A major step forward in the conceptual understanding
of this counterintuitive phenomenon was the discovery of the
Korteweg-de Vries equation in 1895 which admits solutions exhibiting
the features of solitary waves. Moreover, there are stronger
relatives of solitary waves among the solutions of the KdV equation
-the solitons- which even survive collisions amongst themshelves.
During the celebrated sixties of the past century, a powerful
technique was invented, the inverse scattering method, which allowed
the solution of many \lq\lq integrable" non-linear partial
differential equations in the class of the KdV equations. For a
recent review on the concept of soliton associated with all these
non-linear equations, see, e.g., Reference \cite{TAO}.

Our theme in these Lectures, however, is the analysis of the quantum
descendants of these classical non-linear solitary waves/topological
defects in one and two spatial dimensions. In 1961, Skyrme
\cite{SKYR} discovered that a certain extension of the non-linear
sigma model, the so called Skyrme model, has both 3D dispersive and
solitary waves among their solutions. Because the model attempted to
describe the low-energy hadron phenomenology, and because solitary
waves are formed from a heavy classical lump of energy, the idea is
natural: upon quantization, dispersive waves become light mesons
-pions- and solitary waves give rise to heavy baryons -protons,
neutrons-. This bold idea prompted the task of investigating
solitary waves in the quantum domain, mainly performed in
\cite{DHN}, \cite{FK}, and \cite{CCG} as far as our Lectures are
concerned. The main examples were reported and the conceptual
framework was extraordinarily well clarified in \cite{SFC1}. We
insist that there are many more authors who contributed to
developing this research topic. Some of them took another approach,
and there are many very good reviews in the Proceedings of several
Schools, and even important old and modern books. Most of the
pertinent bibliography is collected in Reference \cite{AMAJJW}.
Here, we only cite the papers and reviews with an approach close to
ours: the $\hbar$/weak coupling/semiclassical expansion.

The plan of these Lectures is to describe a method for computing
one-loop fluctuations in one-dimensional and two-dimensional
topological defects based on the heat kernel/zeta function
regularization of ultraviolet divergences. The method began in
Reference \cite{AMAJW} with the calculation of the one-loop mass
shift to the masses of the paradigmatic sine-Gordon and
$\lambda\phi^4_2$ kink as a test. In the same paper, some of us
calculated the one-loop mass corrections to other two-scalar field
theoretical models with only one field but with insufficient
information about the kink fluctuation spectrum to apply the
conventional Dashen-Hasslacher-Neveu approach. In References
\cite{AMAJW1} and \cite{AMAJW2}, the same group generalized the
method to provide one-loop mass corrections for two-component
topological kinks: i.e., the models addressed have two scalar fields
and the kinks considered are such that the two components of the
scalar field are not zero for the kink solution. The kink
fluctuations of such type of kinks are determined by non-diagonal
Schr$\ddot{\rm o}$dinger operators and the only possibility for
managing the spectral information needed to compute one-loop mass
shifts is the heat kernel expansion.

Prior to our work, at the end of the last century, there had been a
renaissance in interest about the problem of the quantization of
classical lumps. The new impetus came from the subtleties arising in
the quantization of supersymmetric kinks. Several groups at Stony
Brook/Wien, \cite{RvN}, \cite{NSRvN}, Minnesota \cite{SVV}, and MIT
\cite{GJ} addressed mixed issues in the problem by studying the
impact made by using different types of boundary conditions - PBC,
Dirichlet, Robin-, regularization methods -energy cutoff, mode
number cutoff, high-derivatives-, and/or performed phase shift
analyses, in connection with possible modifications due to the
quantum effects of the central charge of the SUSY algebra. There are
several valuable reviews, of different character and scope, in the
literature on these developments: e.g., \cite{RW}, \cite{GRvNW}, and
\cite{RvNW}.

The stimulus to our work on the quantization of solitons came,
however, from the discovery of kinks in theories with two scalar
fields living on a infinite line. In this type of model, there are
often solitary wave solutions such that the two scalar fields are
space-dependent for the kink profiles, see e.g. \cite{SV},
\cite{BNRT}, and \cite{AMAJ}. The knowledge of the spectrum of these
two-component kink fluctuations is insufficient to profit from any
type of Dashen-Hasslacher-Neveu approach. The only possibility is to
use the spectral zeta function obtained from the heat kernel
asymptotic (high-temperature) expansion. This framework was
precisely chosen in the SUSY kink problem in Reference \cite{BGvNV}.

Kinks are one-dimensional topological defects, but extremely
interesting two-dimensional extended structures were discovered by
Abrikosov in Type II superconductors \cite{AAA}. The
phenomenological Ginzburg-Landau theory allowed the existence of
magnetic flux lines when applied to this type of superconducting
materials. Relativistic cousins of Abrikosov strings exist in the
Abelian Higgs model and were proposed by Nielsen and Olesen in 1973,
see \cite{NO}, as plausible candidates as the basic objects in the
early string theory approach to hadron physics. More recently
Achucarro and Vachaspati have discovered even more complex
two-dimensional topological defects in the so called semilocal
Abelian Higgs model, the bosonic sector of electro-weak theory when
the weak (Weinberg, mixing) angle is $\frac{\pi}{2}$ \cite{AV}. This
model enjoys a symmetry group that is the direct product of two
groups: ${\mathbb S}{\mathbb U}(2)\otimes{\mathbb U}(1)$. The
non-Abelian group ${\mathbb S}{\mathbb U}(2)$ engenders a global
symmetry whereas the other symmetry generated by the ${\mathbb
U}(1)$ factor is local (or gauge). Henceforth, Achucarro and
Vachaspati christened the topological solitons of this system as
semilocal strings. Given the important r$\hat{\rm o}$le that these
models play in our present understanding of the Standard model, it
is convenient to address the problem of studying the quantum
behavior of these two-dimensional solitons, sometimes referred to as
ANO vortices or semilocal vortices because the vector (gauge) field
of these solutions is purely vorticial (rotational). This task was
successfully accomplished in the ${\cal N}=2$ SUSY Higgs model
independently by Vassilevich, \cite{DV}, and Rebhan, van
Nieuwenhuizen, and Wimmer, \cite{RvNW1}.

In the bosonic setting, however, without fermions to cancel a good
deal of the bosonic  fluctuations, the problem is much more
difficult. A good step forward towards this goal was given in
Reference \cite{BD} in which the authors calculate the energy of the
fermionic ANO vortex fluctuations. We profited from our experience
with multi-component kinks to compute the one-loop mass shifts of
ANO vortices with a quantum of magnetic flux in a purely bosonic
setting in \cite{AMJW}. To this end we used the heat kernel/zeta
function regularization method, jumping painfully from one to two
dimensions. In \cite{AMJW1} our calculations were extended to
superposed vortices up to four quanta of magnetic flux and we
attacked the problem of computing one-loop mass shifts to the
topological solitons of the generalized Abelian Higgs model in
Reference \cite{AMJW3}. We summarized all this material in the
Proceedings of QFEXT05 and QFEXT07 published in \cite{AMJW2} and
\cite{AMJW4}.

One might think that this is a very narrow and highly focused
subject. This way of thinking is not completely true, for two
reasons. First, knowledge of quantum field theories with topological
sectors other than vacuum sectors is not fully settled down, at
least at the level of perturbation theory around the ground state.
Second, study of the quantum fluctuations around topological defects
is a problem in the kinship of very important physical phenomena,
such as the cosmological constant problem and the Casimir effect.
Vacuum fluctuations (loop graphs) give rise to a non-zero constant
term in the Lagrangian of the Standard Model. Coupling of this
Lagrangian to gravity means that the constant term is a cosmological
constant induced by the quantum fluctuations of the particle fields
of an order of magnitude greater than the experimental value of
$\sim 10^{60}$. The Casimir effect is an even closer physical
phenomenon. Vacuum fluctuations also play a central r$\hat{\rm
o}$le. Here, the idea is to sum the effect of the vacuum
fluctuations in the presence of some set up -parallel plates,
cylinders, spheres- measured with respect to the vacuum. The outcome
is the appearance of physical forces on the plates emerging from the
vacuum.

Our goal in this report is to present an analysis of the quantum
corrections to the mass of topological defects developed in
different one-dimensional and two-dimensional systems in the set of
References cited above in a manner as unified as possible. The
contribution is divided into two separate parts. In the first, we
deal with one-dimensional relativistic scalar fields. We explain the
problem, the method of solution chosen, and the derivation of a
compact formula for one-loop kink mass shifts in a multi-parametric
family of deformed linear ${\mathbb O}(N)$-sigma models. The
$\lambda\phi^4$ model is a member of this family for only one scalar
field: $N=1$. Our approach is tested in this prototypical case, and
detailed computations are offered. We also describe the results
achieved in another member of this family with $N=2$ scalar fields
having degenerate families of topological kinks. The first part ends
with an analysis of the kink one-loop mass shifts of the topological
kinks in the massive non-linear ${\mathbb S}^2$-sigma model studied
in Reference \cite{AMAJ2}. Again the model is embedded in the family
of linear ${\mathbb O}(N)$-sigma models, taking the formal
$\lambda\to\infty$ in the case of $N=3$ scalar fields, a process
that ends with a non-linear field theory. Following \cite{AMAJ3}, we
compute the mass shift using the Cahill-Comtet-Glauber formula, see
\cite{CCG}.

Part two is devoted to understanding the quantum fluctuations of
semilocal strings and Nielsen-Olesen vortices. The action of the
semilocal Abelian Higgs model is considered when the space-time is
the $(2+1)$-dimensional Minkowski ${\mathbb R}^{2,1}$. The mix of
local and global symmetries, the Higgs mechanism in the 't Hooft
renormalizable gauge, and the Feynman rules are discussed. We then
go on  to study the very rich moduli space of topological soliton
solutions, all of them having vorticial vector fields. Numerical
solutions of the first-order field equations arising at the critical
point between Type I and Type II superconductivity are calculated in
the case of circular symmetry. These planar solitons become strings
seen from $(3+1)$ dimensions. The next task is an analysis of the
semilocal self-dual vortex fluctuations and the subsequent vortex
Casimir and mass renormalization energies. These ultraviolet
divergent quantities are regularized via the spectral zeta function
of the second-order fluctuation operators as in kink cases. Unlike
kink cases, the pertinent differentials are $6\times 6$- Matrix
second-order PD operators. In the background gauge there are
fluctuations corresponding to the Higgs field, a real Higgs ghost
field, a complex massless scalar field, and the two polarizations of
the massive vector field. There are also fluctuations of the
Faddeev-Popov ghost field that restore unitarity by compensation of
the real Higgs ghost field fluctuations. The heat kernel expansion
allows us the calculation of the one-loop vortex mass shift in terms
of Seeley coefficients and incomplete Euler Gamma functions.
Finally, we provide numerical results that suggest the breakdown of
the classical degeneracy of the semi-local vortices in favour of the
embedded Nielsen-Olesen vortices.

\section{Scalar field models in a line}

In this first part we report the procedure of computing the one-loop
mass shifts for one-dimensional topological defects developed in
\cite{AMAJW}, \cite{AMAJW1}, and \cite{AMAJW2}. Ultraviolet
divergences are regularized using heat kernel/zeta function methods
- comprehensive reviews of these techniques are \cite{EORBZ},
\cite{KK}, \cite{DVV}- and the models considered belong to
(1+1)-dimensional scalar field theory. We shall always deal with
topological kinks, but in a particular model of $N=2$ scalar fields
we must struggle with the problem of studying fluctuations of a
degenerate-in-energy continuous family of kinks.

\subsection{Deformed linear ${\mathbb O}(N)$-sigma models}
We shall focus on a multi-parametric family of deformed linear
${\mathbb O}(N)$-sigma models. The target (isospin, internal,
$\cdots$) space is ${\mathbb R}^N$. Let $\chi_a, a=1,2, \cdots ,N$,
denote the coordinates of a point in ${\mathbb R}^N$. The
(multi-component) scalar fields are maps from the
$(1+1)$-dimensional Minkowski space to the target space:
$\chi_a(y^\mu)\in {\rm Maps}({\mathbb R}^{1,1},{\mathbb R}^N)$. Here
$y_\mu$, \, $\mu=0,1$, denote the coordinates of a point in the
Minkowski space-time ${\mathbb R}^{1,1}$. We shall use the following
conventions for the metric and volume element:
\begin{eqnarray*}
y^\mu y_\mu=y_0^2-y_1^2=g^{\mu\nu}y_\mu y_\nu \qquad  &,& \qquad
\qquad g^{\mu\nu}=g_{\mu\nu}={\rm diag}[1,-1]
\\
dy^2=dy_0dy_1 \hspace{1.0cm} &,& \hspace{1.5cm}
\frac{\partial\chi_a}{\partial
y^\mu}\cdot\frac{\partial\chi_a}{\partial
y_\mu}=\frac{\partial\chi_a}{\partial
y_0}\frac{\partial\chi_a}{\partial
y_0}-\frac{\partial\chi_a}{\partial
y_1}\frac{\partial\chi_a}{\partial y_1} \qquad .
\end{eqnarray*}
The action governing the dynamics of the deformed linear ${\mathbb
O}(N)$-sigma model is:
\begin{eqnarray*}
S[\chi_1, \chi_2, \cdots , \chi_N]&=& \frac{1}{2}\int \, dy^2 \,
\left\{\sum_{a=1}^N \, \frac{\partial\chi_a}{\partial
y^\mu}\cdot\frac{\partial\chi_a}{\partial
y_\mu}-\frac{\lambda}{2}\left(\sum_{a=1}^N \,
\chi_a\chi_a-\frac{m^2}{\lambda}\right)^2 \right. \\ &-& \left.
\begin{array}{c} \\ \sum_a\sum_b \\
a\leq b\end{array}\lambda_{ab}\chi_a^2\chi_b^2-\sum_{a=1}^N \, m_a^2
\chi_a\chi_a\right\} \qquad .
\end{eqnarray*}
We choose a system of units where the speed of light is $c=1$, but
we keep $\hbar$ explicit because we shall work in the framework of
$\hbar$-expansion. In this system of units the dimension of $\hbar$
is mass$\times$length, $[\hbar]=ML$, whereas the dimension of the
fields and parameters are:
\[
[\chi_a]=M^{\frac{1}{2}}L^{\frac{1}{2}} \qquad , \qquad
[\lambda]=[\lambda_{ab}]=M^{-1}L^{-3} \qquad , \qquad
[m]=[m_a]=L^{-1} \qquad .
\]
Defining non-dimensional coordinates, fields, and coupling constants
\[
x^\mu=\frac{m}{\sqrt{2}}y^\mu \qquad , \qquad
\phi_a(x^\mu)=\frac{\sqrt{\lambda}}{m}\chi_a(y^\mu) \qquad , \qquad
 \sigma_{ab}=\frac{\lambda_{ab}}{\lambda} \qquad , \qquad
\sigma_a^2=\frac{m_a^2}{m^2}
\]
the action reads:
\begin{eqnarray} S[\phi_1, \phi_2, \cdots , \phi_N]&=&
\frac{m^2}{\lambda}\int \, dx^2 \, \left\{\frac{1}{2}\sum_{a=1}^N \,
\frac{\partial\phi_a}{\partial
x^\mu}\cdot\frac{\partial\phi_a}{\partial
x_\mu}-V(\phi_a;\sigma_{ab},\sigma_a^2)\right\} \label{acton}\\
V(\phi_1(x^\mu), \phi_2(x^\mu), \cdots , \phi_N(x^\mu))&=&
\frac{1}{2}\left(\sum_{a=1}^N \, \phi_a\phi_a -1\right)^2
+\begin{array}{c} \\ \sum_a\sum_b \\
a\leq b\end{array}\sigma_{ab}\phi_a^2\phi_b^2+\sum_{a=1}^N \,
\sigma_a^2 \phi_a\phi_a \nonumber \qquad .
\end{eqnarray}

Besides the usual $m$ parameter, which sets the length scale of the
system, and the $\lambda$, which sets the strength of the isotropic
quartic couplings, there are $\frac{N(N+1)}{2}$ non-isotropic
quartic couplings $\sigma_{ab}$ and $N$ $\sigma_a^2$ parameters
giving possible quadratic anisotropies {\footnote{A warning: despite
the notation, the possibility that some $\sigma_a^2$ might be
negative will not be ruled out.}}. The rationale behind the choice
of this family of models is as follows: the set of parameters
$\sigma_{ab}=\sigma_a^2=0, \forall a,b$ corresponds to the linear
${\mathbb O}(N)$-sigma model. In this case, there are $N-1$
Goldstone bosons, owing to the spontaneous symmetry breaking of the
global ${\mathbb O}(N)$ symmetry. Goldstone bosons do not exist in
$(1+1)$ dimensions, see \cite{SFC}. Even if we were to start from
the linear ${\mathbb O}(N)$-sigma model, the $(1+1)$-dimensional
infrared asymptotics of these massless fields would generate
anisotropic quartic and/or quadratic terms, such that no global
${\mathbb O}(N)$ symmetry (or any of its continuous subgroups) would
remain. Note that no cubic or linear terms in the fields are allowed
because the discrete subgroup ${\mathbb Z}_2^N$ of ${\mathbb O}(N)$
generated by the internal reflections $\phi_a \, \, \to \, \,
-\phi_a, \forall a$, would be explicitly broken.

\subsection{Vacuum fluctuations}

In the parameter range
\[
1 > \sigma_a^2 > -\infty, \quad \sigma_{ab}
> {\rm max}\left(\frac{1-\sigma_a^2}{1-\sigma_b^2}(1+2\sigma_{bb}),
\frac{1-\sigma_b^2}{1-\sigma_a^2}(1+2\sigma_{aa})\right)
> 0 ,  a\neq b, \quad 1+2\sigma_{aa}>0
\]
the constant configurations
\[
(\phi_1^{V^{(a)}}=0, \cdots ,
\phi_a^{V^{(a)}}=\pm\sqrt{\frac{1-\sigma_a^2}{1+2\sigma_{aa}}},
\cdots, \phi_N^{V^{(a)}}=0) \, \, , \, \, \quad a=1,2, \cdots , N
\]
are stable solutions of the time-independent and static field
equations
\[
\frac{\partial V}{\partial\phi_a}=2\phi_a\left[\sum_{b=1}^N \,
\phi_b\phi_b+\begin{array}{c}\\ \sum \\ a\leq b
\end{array}\sigma_{ab}\phi_b^2+\sigma_a^2-1\right]=0 \qquad .
\]
These non-zero constant solutions are thus the $2N$ classical minima
of the system. In the quantum domain only the absolute minima
\[
V(\phi_c^{V^{(c)}})=\frac{\sigma_c^2(2-\sigma_c^2)+2\sigma_{cc}}{2(1+2\sigma_{cc})}
\quad , \quad S^{(0)}[\phi_a^{V^{(c)}}]=\frac{m^2}{\lambda}\int \,
d^2x \, V(0, \cdots , \pm\sqrt{\frac{1-\sigma_c^2}{1+2\sigma_{cc}}},
\cdots , 0)
\]
are the true vacua. Tunnel effects triggered by bounces turn the
relative minima $V(\phi_a^{V^{(a)}})>V(\phi_c^{V^{(c)}})$ into false
vacua \cite{BK}. The ground states are built from at least one pair
of minima (more than a pair of absolute minima could exist) and the
${\mathbb Z}_2^N$ symmetry is broken spontaneously (at least) to
${\mathbb Z}_2^{N-1}$.

Let us denote space-time coordinates in the form $x_0=t \, , \,
x_1=x$ and let us consider small fluctuations $\phi_a(t,x)=
\phi_a^{V^{(c)}}+\eta_a(t,x)$ around a true vacuum. The action at
the quadratic order is
\begin{equation}
S^{(2)}[\phi_a^{V^{(c)}}; \eta_1, \cdots , \eta_N]
=\frac{m^2}{2\lambda} \int \, dx^2 \, \sum_{a=1}^N \,
\left[\frac{\partial\eta_a}{\partial
t}\frac{\partial\eta_a}{\partial
t}-\eta_a\left(-\frac{\partial^2}{\partial
x^2}+\mu_a^2\right)\eta_a\right]+{\cal O}(\eta^3)\label{qacton}
\end{equation}
where
\[
\mu_a^2=\left.\frac{\partial^2
V}{\partial\phi_a^2}\right|_{V^{(c)}}=
2(\frac{1-\sigma_c^2}{1+2\sigma_{cc}}(1+\sigma_{ac})-(1-\sigma_a^2))\,\,
, \, \, a\neq c \quad , \quad \mu_c^2=\left.\frac{\partial^2
V}{\partial\phi_c^2}\right|_{V^{(c)}}=4(1-\sigma_c^2)
\]
are the particle masses.

The normal modes of these system with a infinite number of degrees
of freedom are determined in terms of the eigenfunctions of the
differential operator:
\[ K_0 =
  \left( \begin{array}{cccc}
 -\frac{d^2}{dx^2}+\mu_1^2 & \dots & \dots
& 0\\
 0 & -\frac{d^2}{dx^2}+\mu_2^2 & \dots
& 0  \\
\vdots&\vdots&\ddots&\vdots \\
0 & \dots & \dots & -\frac{d^2}{dx^2}+\mu_N^2 \end{array} \right)
\qquad , \qquad f_n^a(x)= \frac{1}{\sqrt{l}}e^{ik_nx} \qquad .
\]
To avoid problems with the continuous spectrum, we choose a 1D
\lq\lq box" of very large but finite length $l=\frac{mL}{\sqrt{2}}$,
and we impose periodic boundary conditions: $f^a(x+l)=f^a(x)$. $K_0$
therefore acts  on the Hilbert space $L^2=\bigoplus_{a=1}^N \,
L^2_a({\mathbb S}^1)$ and the eigenvalues
\[
K_0 f^a_n(x)=\omega_a^2(k_n)f^a_n(x) \qquad , \qquad
\omega_a^2(k_n)=k_n^2+\mu_a^2 \qquad , \qquad k_n=\frac{2\pi}{l}n
\qquad , \qquad n\in {\mathbb Z}
\]
are obtained from wave numbers labeled by the integers.

Classically the system is tantamount to a infinite numerable set of
uncoupled oscillators with frequencies given by the eigenvalues of
$K_0$ that become quantum oscillators upon canonical quantization.
The free quantum Hamilton
\begin{equation}
\hat{H}^{(2)}=\frac{\hbar m}{\sqrt{2}}\sum_{a=1}^N \,
\sum_{n\in{\mathbb Z}} \,
\omega_a(k_n)\left(\hat{b}_a^\dagger(k_n)\hat{b}_a(k_n)+\frac{1}{2}\right)\label{olve}
\end{equation}
is given in terms of the creation and annihilation operators,
$[\hat{b}_a^\dagger(k_n),\hat{b}_c(k_m)]=\delta_{ac}\delta_{mn}$:
the quantum disguise of the Fourier coefficients. Note that
$\hat{H}^{(2)}$ is proportional to $\hbar$. In general, the operator
$\hat{H}^{(2j)}$ coming from the $2-j$th-order fluctuation term in
the expansion of the classical action is proportional to
$\hbar^{j}$. Therefore, result (\ref{olve}) is obtained in the
first-order (one-loop) of the $\hbar$-expansion (loop expansion).

The vacuum state, annihilated by all the destruction operators, is a
coherent state, and is an eigen-state of the field operator:
\[
\hat{b}_a(k_n)|V; 0\rangle=0 \, \, \, , \, \, \, \forall a \, , \,
\forall k_n \qquad , \qquad \hat{\phi}_a |V; 0\rangle=0, a\neq c, \,
\, \, \, \, \, \, \hat{\phi}_c |V; 0\rangle=
\sqrt{\frac{1-\sigma_c^2}{1+2\sigma_{cc}}}\, \,|V; 0\rangle \qquad .
\]
It is clearly a ground state of the quantum system with energy:
\[
\langle 0 ; V|\hat{H}^{(2)}|V; 0\rangle =\frac{\hbar
m}{2\sqrt{2}}{\rm Tr}_{L^2}K_0^{\frac{1}{2}} \qquad .
\]

\subsubsection{Spectral zeta function regularization,
the $K_0$-heat equation kernel, and the $K_0$-heat trace}

We usually measure the energy of any state in QFT with respect to
the vacuum or ground state, or, equivalently, we set
$\hat{H}^{(2)}|V; 0\rangle =\frac{\hbar m}{2\sqrt{2}}{\rm
Tr}_{L^2}K_0^{\frac{1}{2}}|V; 0\rangle$ as the zero-energy level.
Important physical phenomena such as the Casimir effect or the
cosmological constant problem have taught us that we must be
cautious about using this calibration. In particular, in our system
there are other topological sectors and it is convenient not to set
the energy of the ground state in the vacuum  sector to zero a
priori in order to allow a comparison with the energy of the ground
state in the kink sector.

The problem is that $\frac{\hbar m}{2\sqrt{2}}{\rm
Tr}_{L^2}K_0^{\frac{1}{2}}$ is a divergent quantity that, one way or
another must be regularized. Several regularization methods have
been proposed in the literature: cut-offs in the energy or the
number of modes, high-derivatives, etcetera, see, e. g., \cite{DHN},
\cite{FK}, \cite{GJ}, \cite{RW}, \cite{RvNW}. As in Reference
\cite{BGvNV}, however, we shall regularize the vacuum energy using
the zeta regularization method. Instead of computing the $L^2$-trace
of the square root of the $K_0$ operator, we calculate the spectral
zeta function -the $-s$-complex power of the $K_0$ operator:
\begin{equation}
\zeta_{K_0}(s)={\rm
Tr}_{L^2}K_0^{-s}=\sum_{a=1}^N\sum_{n=-\infty}^\infty \,
\frac{1}{(\frac{4\pi^2}{l^2}n^2+\mu_a^2)^s}=\sum_{a=1}^N \,
E(s,\mu_a^2|\frac{4\pi^2}{l^2}) \, \,  , \qquad s\in{\mathbb C}
\qquad . \label{zeta}
\end{equation}
The series in (\ref{zeta}) are convergent only if ${\rm Re}\,
s> \frac{1}{2}$ although conventionally they are analytically
continued to the whole $s-{\rm complex}$ plane to find the
Epstein zeta functions, all of which are meromorphic functions of
$s$, see \cite{KK}. The central idea of the zeta function
regularization method is to assign to the divergent vacuum energy
the finite value
\[
\langle 0;V|\left(\hat{H}^{(2)}\right)^{-s}|V;0\rangle
=2^{s-1}\hbar\mu\left(\frac{\mu^2}{m^2}\right)^s\zeta_{K_0}(s)
\]
at a regular point $s\in{\mathbb C}$ of $\zeta_{K_0}(s)$. $\mu$ is a
parameter of dimension $L^{-1}$ necessary to keep track of the right
dimensions.

The analysis of the associated $K_0$-heat equation kernel
\[
\displaystyle
\left(\frac{\partial}{\partial\beta}+K_0\right)K_{K_0}(x,y;\beta)=0
\qquad , \qquad K_{K_0}(x,y;0)=\delta(x-y) \qquad
\]
will help us to unveil the structure of $\zeta_{K_0}(s)$ as a
meromorphic function. Here $\beta=\frac{\hbar m}{k_B T}$ is a
non-dimensional inverse temperature because the dimension of the
Boltzmann constant in our system of units is $[k_B]=ML$. In terms of
the eigenfunctions and eigenvalues of $K_0$, one can express the
heat kernel in the form:
\[
{\rm tr}K_{K_0}(x,y;\beta)=\frac{1}{l}\sum_{a=1}^N
e^{-\beta\mu_a^2}\sum_{n\in{\mathbb Z}}e^{\frac{2\pi}{l}i
n(x-y)}e^{-\beta\frac{4\pi^2}{l^2}n^2}=\frac{1}{l}
\Theta\left[\begin{array}{c} 0 \\
0\end{array}\right](\frac{x-y}{l}|i\frac{4\pi}{l^2}\beta)\sum_{a=1}^N
e^{-\beta\mu_a^2} \qquad .
\]
The notation used for the Jacobi Theta function, see e.g. \cite{MS},
is:
\[
\Theta\left[\begin{array}{c} a \\
b\end{array}\right](z|\tau)=\sum_{n\in{\mathbb Z}}e^{2\pi
i(n+a)(z+b)+\frac{(n+a)^2}{2}\tau)} \quad ; \quad a,b=0,\frac{1}{2}
\, \, \, \, \, , \, \, \, \, \, z\in{\mathbb C} \quad , \quad
\tau\in{\mathbb C} \, \, , \, \, {\rm Im}\tau>0 \qquad .
\]
The \lq\lq modular" transformation $\tau=i\frac{4\pi}{l^2}\beta\to
-\frac{1}{\tau}=i\frac{l^2}{4\pi\beta}$ allows us to write the heat
kernel in the form:
\[
{\rm
tr}K_{K_0}(x,y;\beta)=\frac{e^{-\frac{(x-y)^2}{\beta}}}{\sqrt{4\pi\beta}}\cdot
\Theta\left[\begin{array}{c} 0 \\
0\end{array}\right](-i\frac{l(x-y)}{\beta}|i\frac{l^2}{4\pi\beta})\cdot\sum_{a=1}^N
e^{-\beta\mu_a^2}
\]
because the Jacobi Theta function is a modular form of weight
$\frac{1}{2}$ (alternatively, this equivalence could be derived from
the Poisson summation formula). There are thus two ways of writing
the $K_0$-heat trace (related by the modular transformation):

\[
 {\rm Tr}_{L^2}e^{-\beta K_0}=\int_{-\frac{l}{2}}^{\frac{l}{2}}\,
dx \, {\rm
tr}K_{K_0}(x,x;\beta)=\Theta\left[\begin{array}{c} 0 \\
0\end{array}\right](0|i\frac{4\pi}{l^2}\beta)\cdot\sum_{a=1}^N
e^{-\beta\mu_a^2}=\sum_{a=1}^N e^{-\beta\mu_a^2}\,
e^{-\beta\frac{4\pi^2}{l^2}n^2}\quad , \quad
\]
\[
{\rm Tr}_{L^2}e^{-\beta K_0}=\frac{l}{\sqrt{4\pi\beta}}\cdot\Theta\left[\begin{array}{c} 0 \\
0\end{array}\right](0|i\frac{l^2}{4\pi\beta})\cdot\sum_{a=1}^N
e^{-\beta\mu_a^2}=\frac{l}{\sqrt{4\pi\beta}}\sum_{a=1}^N
\sum_{n=-\infty}^\infty \, e^{-\mu_a^2\beta}\,
e^{-\frac{l^2}{4\beta}n^2} \qquad .
\]
The Mellin transform of the first form gives the sum of Epstein
functions:
\[
\displaystyle\zeta_{K_0}(s)=\frac{1}{\Gamma(s)}\int_0^\infty \,
d\beta \, \beta^{s-1}\ {\rm Tr}_{L^2}e^{-\beta K_0}\sum_{a=1}^N \,
E(s,\mu_a^2|\frac{4\pi^2}{l^2})=\sum_{a=1}^N \,
E(s,\mu_a^2|\frac{4\pi^2}{l^2}) \qquad .
\]
However, Mellin's transform of the Poisson inverted version
\begin{eqnarray}
\zeta_{K_0}(s)&=&\frac{1}{\Gamma(s)}\cdot \int_0^\infty \, d\beta \,
\beta^{s-1}\, \sum_{a=1}^N \, e^{-\mu_a^2\beta}\,
\left({l\over\sqrt{4 \pi}}\beta^{-{1\over
2}}\sum_{n=-\infty}^\infty\, e^{-{l^2\over
\beta}n^2}\right)\nonumber\\&=&\frac{l}{\sqrt{4\pi}\Gamma(s)}\cdot
\sum_{a=1}^N \,\left(\frac{\Gamma(s-\frac{1}{2})}{\mu_a^{2s-1}} +\,
2\, \sum_{n\in{\mathbb Z}/\{0\}} \, \left(\frac{n
l}{\mu_a}\right)^{s-1/2}\, K_{1/2-s}(2 \mu_a n l)\right)\quad .
\label{dzfk}
\end{eqnarray}
identifies the spectral zeta function as a series of modified Bessel
functions of the second type, see \cite{KK}. Moreover, formula
(\ref{dzfk}) shows that there are poles of $\zeta_{K_0}(s)$ only at
the poles of the Euler Gamma function $\Gamma(s-\frac{1}{2})$,
\[
s=\frac{1}{2}, -\frac{1}{2}, -\frac{3}{2}, -\frac{5}{2},
-\frac{7}{2}, \cdots, -\frac{2j+1}{2}, \cdots, \qquad j\in{\mathbb
Z}^+ \qquad ,
\]
because $K_{1/2-s}(2 \mu_a l n)$ are transcendental entire
functions, i.e., holomorphic functions of $s$ in ${\mathbb
C}/{\infty}$ with an essential singularity at $s=\infty$. The
behavior of the heat trace at high-temperature is determined by the
asymptotic formula up to exponentially small terms:
\[
\Theta\left[\begin{array}{c} 0 \\
0\end{array}\right](0|i\frac{l^2}{4\pi\beta})=\sum_{n\in{\mathbb
Z}}\, e^{-\frac{l^2}{4\beta}n^2}\cong_{\beta\to 0}\, 1+{\cal
O}(e^{-\frac{c}{\beta}}) \qquad ,
\]
which also characterizes the behavior of ${\rm Tr}_{L^2}\, e^{-\beta
K_0}$ for very large $l$.

\subsection{ Kink fluctuations}
\subsubsection{Topological kinks}
Because of the ground- state structure of the deformed linear
${\mathbb O}(N)$-sigma model, there are other static solutions that
are not homogeneous. These classical lumps are \lq\lq one-component
topological kinks" (TK1) We shall refer to this type of kinks in
this way because: 1) They are topological. Their profiles connect
one ground state with another when plotted from $x=-\infty$ to
$x=\infty$. 2) They have only one component of the iso-vector field
different from zero.

In order to find their profile, one simply looks for solutions of
the field equations such that all the field components except the
one accommodating the absolute minima are zero: $\phi_a^{K^{(c)}}=0
\,, \, \forall a\neq c$. Under this assumption the classical energy
can be written in the Bogomolny form:
\[
E=\frac{m^3}{\sqrt{2}\lambda}\left[\int \, dx \,
\frac{1}{2}\left(\frac{d\phi_c}{dx}\mp
\sqrt{2\left(V(\phi_c)-V(\phi_c^{V^{(c)}})\right)}\right)^2\pm
\int_{-\phi_c^{V^{(c)}}}^{\phi_c^{V^{(c)}}}\, d\phi_c
\,\sqrt{2\left(V(\phi_c)-V(\phi_c^{V^{(c)}})\right)}\right] \quad .
\]
Therefore, the solutions
\[
\phi_c^{K^{(c)}}(x)=\pm \phi_c^{V^{(c)}}{\rm
tanh}[\frac{\mu_c}{2}(x-x_0)]
\]
of the first-order equations
\[
\frac{d\phi_c}{dx}=\pm\sqrt{2\left(V(\phi_c)-V(\phi_c^{V^{(c)}})\right)}
\]
are absolute minima of the energy that solve the static second-order
field equations. TK1 kinks are thus space-dependent solutions that
interpolate between the two ground states, reached by the kink in
different components of the boundary of the spatial line at
infinity. They have finite energy,
\[
E[\phi_c^{K^{(c)}}]=2\frac{m^3}{\sqrt{2}\lambda}\int_{-\infty}^\infty
\, dx \, \left[V(\phi_c^{K^{(c)}})-V(\phi_c^{V^{(c)}})\right]=
\frac{4}{3}\frac{1}{\sqrt{1-\sigma_c^2}}\frac{m^3}{\sqrt{2}\lambda}
\qquad ,
\]
and their energy density is spatially distributed. Despite these
features these classical lumps are stable because of topological
reasons: they belong to topological sectors disconnected from the
vacuum sectors in the configuration space.
\subsubsection{Small kink fluctuations}
Our goal is to study these sectors in the quantum domain. We start
from the fact that the quantum descendants of TK1 kinks are the
ground states in the topological sectors.  Small fluctuations around
TK1 kinks,
\[
\phi_a(t,x)= \phi_a^{K^{(c)}}(x)+\eta_a(t,x)
\]
are governed by the quadratic action
\[
S^{(2)}[\phi_a^{K^{(c)}}(x); \eta_1, \cdots , \eta_N]
=\frac{m^2}{2\lambda} \int \, dx^2 \, \sum_{a=1}^N \,
\left[\frac{\partial\eta_a}{\partial
t}\frac{\partial\eta_a}{\partial t}-\eta_a K \eta_a\right]+{\cal
O}(\eta^3) \qquad ,
\]
where the second-order fluctuation operator is a diagonal matrix of
P$\ddot{\rm o}$sch-Teller Schr$\ddot{\rm o}$dinger operators :
{\small \[ K= \left(
\begin{array}{ccccc}
 -\frac{d^2}{dx^2}+\mu_1^2-c_{1c}
 \cdot{\rm sech}^2(\frac{\mu_c}{2}x)& \dots & \dots
 & \dots
& 0\\ \vdots&\ddots&\vdots & \vdots&\vdots \\
 0 & \dots &-\frac{d^2}{dx^2}+\mu_c^2 -\frac{3}{2}\mu_c^2 \cdot {\rm sech}^2[\frac{\mu_c}{2}x]& \dots
& 0  \\
\vdots &\vdots &\vdots &\ddots & \vdots \\
0 & \dots & \dots &\dots & -\frac{d^2}{dx^2}+\mu_N^2-c_{Nc}\cdot{\rm
sech}^2[\frac{\mu_c}{2}x] \end{array} \right)\, .
\]}
The bottom of the wells with respect to the thresholds $\mu_a^2$ are
respectively: $-\frac{3}{2}\mu_c^2$ and
$-c_{ac}=-2(1+\sigma_{ac})\left(\phi_c^{V^{(c)}}\right)^2$.

Like the vacuum fluctuations, the normal modes of kink fluctuations
are obtained from the eigenfunctions of the $K$ operator: $K_{aa}
f^a_n(x)=\varepsilon_a^2(n)f^a_n(x)$. We impose periodic boundary
conditions $f^a(x+l)=f^a(x)$ to escape the problems of the
continuous spectrum, and $K$ also acts on: $L^2=\bigoplus_{a=1}^N \,
L^2_a({\mathbb S}^1)$.

The spectrum is slightly different for the operator acting on
fluctuations along the TK1 kink orbit:
\[
K_{cc}= -\frac{d^2}{dx^2}+\mu_c^2 -\frac{3}{2}\mu_c^2 \cdot {\rm
sech}^2[\frac{\mu_c}{2}x] \qquad ,
\]
which is summarized in the next Table.

\begin{table}[h]
\hspace{1.5cm}\begin{tabular}{l|c} \hline
{\bf Eigenvalues} & {\bf Eigenfunctions} \\[0.4cm]  \hline \rule{0.cm}{0.8cm} $\varepsilon_c^2(0)=0$ &
$f_0^c(x)={\rm sech}^2[\frac{\mu_c}{2}x]$\\[0.4cm] $\varepsilon_c^2(3)=\frac{3}{4}\mu_c^2$ &
$f_3^c(x)={\rm sinh}[\frac{\mu_c}{2}x]{\rm sech}^2[\frac{\mu_c}{2}x]$\\[0.4cm]
 $\varepsilon_c^2(k)=k^2+\mu_c^2$ & {\large  $f_k^c(x)=e^{ikx}(3{\rm
tanh}^2[\frac{\mu_c}{2}x]-1-\frac{6}{\mu_c}ik{\rm tanh}[\frac{\mu_c}{2}x]-\frac{4}{\mu_c^2}k^2)$}\\[0.4cm] \hline
\end{tabular}\caption{Spectrum for the Hessian operator acting  on fluctuations along the TK1 kink orbit}
\end{table}

There are two bound states, one of zero eigenvalue due to the
breaking of spatial translation invariance by the kink, and
scattering states with thresholds at $\varepsilon_c^2(0)=\mu_c^2$.
From the scattering eigenfunctions one reads the phase shifts:
\[
f_k^c(x) \stackrel{x\to\pm\infty}{\cong}e^{i(k
x\pm\frac{1}{2}\delta_c(k))}\qquad \Rightarrow \qquad
\delta_c(k)=-2{\rm arctan}\frac{3\mu_c k}{\mu_c^2-2 k^2}\qquad .
\]
The periodic boundary conditions select the even ground states  and
force the momenta to satisfy a transcendent equation:
\[
k l+\delta_c(k)=2\pi n \, \, , \, \, \qquad , \qquad k_n -\frac{2\pi
}{l}n=\frac{2}{l}\cdot {\rm arctan}\frac{3\mu_c k_n}{\mu_c^2-2
k_n^2} \qquad , \qquad n\in {\mathbb Z} \qquad .
\]
Although the solutions form an infinite discrete set, they can only
be identified graphically. The physicist's loophole is to work with
very large $l$ such that the information about the continuous
spectrum is codified in the spectral density:
\[
\rho_{K_{cc}}(k)=\frac{1}{2\pi}(l+\frac{d\delta_c(k)}{dk})=-2\mu_c\left(\frac{1}{k^2+\mu_c^2}+\frac{2}{4
k^2+\mu_c^2}\right) \qquad .
\]
The operators acting in the orthogonal directions to the orbit,
$a\neq c$, are of the type:
\[
K_{aa}= -\frac{d^2}{dx^2}+\mu_a^2 -c_{ac} \cdot {\rm
sech}^2[\frac{\mu_c}{2}x] \qquad , \qquad
c_{ac}=2(1-\sigma_c^2)\frac{1+\sigma_{ac}}{1+2\sigma_{cc}} \qquad ,
\]
again of P$\ddot{\rm o}$sch-Teller type, but a with non-null
reflection coefficient in general. To describe the spectrum we
define the parameter
$A=\sqrt{\frac{1+\sigma_{ac}}{1+2\sigma_{cc}}+\frac{1}{4}}$.

\vspace{0.2cm}

\begin{table}[h]
{\small
\begin{tabular}{l|c} \hline
{\bf Eigenvalues} & {\bf Eigenfunctions} \\[0.4cm]  \hline \rule{0.cm}{0.8cm}
$\varepsilon_a^2(j)=\left(\mu_a^2-\frac{\mu_c^2
\left(A-(j+\frac{1}{2})\right)^2}{4}\right)$ & $f_j^a(x)=[{\rm
sech}(\frac{\mu_c}{2}x)]^{j+\frac{1}{2}-A}{}_2F_1[-j+2a,-j,\frac{1}{2}-j+A;\frac{1}{2}
(1+{\rm tanh}(\frac{\mu_c}{2}x))]$ \\[0.4cm]
 $\varepsilon_a^2(k)=k^2+\mu_a^2$ & $f_k^a(x)=[{\rm
 sech}(\frac{\mu_c}{2}x)]^{ik}
 {}_2F_1[\frac{1}{2}-ik+A,\frac{1}{2}-ik+A,1-ik;\frac{1}{2}(1+{\rm tanh}(\frac{\mu_c}{2}x))]$\\[0.4cm] \hline
\end{tabular}} \caption{Spectrum for the Hessian operator acting  on orthogonal fluctuations to the TK1 kink orbit}
\end{table}

The integers $j=0,1,2, \cdots ; I[A-\frac{1}{2}]$ label the
eigenvalues and eigenfunctions of the integer part of the
$A+\frac{1}{2}$ bound states. There are also scattering
eigenfunctions. Both bound states and scattering eigenfunctions are
Gauss hypergeometric functions \cite{ABRAM} times some power of the
hyperbolic secant.

From this information, we obtain the reflection and transmission
coefficients, as well as the phase shifts, the transcendent equation
for momenta complying with periodic boundary conditions (PBC), and
the spectral densities:
\[
T_a(k)=\frac{\Gamma(\frac{1}{2}-ik+A)\Gamma(\frac{1}{2}-ik-A)}{\Gamma(1-ik)\Gamma(-ik)}
\qquad , \qquad R_a(k)=\frac{T_a(k)\Gamma(1-ik)\Gamma(ik)}
{\Gamma(\frac{1}{2}-A)\Gamma(\frac{1}{2}+A)} \qquad ,
\]
\[ \delta_a(k)=\delta_a^+(k)+\delta_a^-(k) \qquad ; \qquad
\delta_a^\pm(k)={1\over 4}{\rm arctan}\left(\frac{{\rm Im}(T_a(k)\pm
R_a(k))}{{\rm Re}(T_a(k)\pm R_a(k)}\right) \qquad ,
\]
\[
k_n l+\delta_a(k_n)=2\pi n \qquad , \qquad
\rho_{K_{aa}}(k)=\frac{1}{2\pi}(l+\frac{d\delta_a}{dk}(k)) \qquad .
\]

\subsubsection{ Spectral kink zeta function regularization}

Let us assume, temporarily, that all the eigenvalues orthogonal to
the TK1 kink orbit are positive: $\varepsilon_a(j)>0, \forall a$.
This means that the TK1 kink is isolated in the configuration space
and stable. We expand the small kink fluctuations in a basis in
$L^2$ formed by the eigenfunctions of $K$:
\begin{eqnarray*}
\eta_a(t,x)&=&\sqrt{\frac{\lambda\hbar}{m
l}}\left\{\sum_{j=0}^{N_b(a)}\, \frac{1}{\sqrt{2
\varepsilon_a(j)}}\,\left(B_a(j)^*\,
e^{-i\varepsilon_a(j)t}+B_a(j)\,
e^{i\varepsilon_a(j)t}\right)f^a_j(x)\right.\\
&+& \left.\sum_{k_n}\,\frac{1}{\sqrt{2
\varepsilon_a(k_n)}}\,\left(B_a(k_n)^*\, e^{-i\varepsilon_a(k_n)t}
f^{*a}_{k_n}(x)+B_a(k_n)\,
e^{i\varepsilon_a(k_n^2)t}f^{a}_{k_n}(x)\right)\right\} \qquad .
\end{eqnarray*}
We denote here the number of even bound states in the direction of
component $a$ as $N_b(a)$ and remark that the zero-mode
eigenfunction in the direction of the kink orbit $c$ does not enter
 this formula because zero modes only contribute at higher orders in
the loop expansion.

Promoting the coefficients of the expansion to the quantum operators
\[
[\hat{B}_a^\dagger(k_n),\hat{B}_c(k_m)]=\delta_{ac}\delta_{mn} \quad
, \quad [\hat{B}_a^\dagger(j),\hat{B}_c(l)]=\delta_{ac}\delta_{jl}
\]
one obtains the free Hamiltonian in the kink sector, assuming
ortho-normality and completeness of the chosen basis:
\[
\hat{H}^{(2)}=\frac{\hbar m}{\sqrt{2}}\sum_{a=1}^N
\,\left[\sum_{j=1}^{N_{b_+}(a)}\,
\varepsilon_a(j)\left(\hat{B}^\dagger_a(j)\hat{B}_a(j)+\frac{1}{2}\right)+
\sum_{k_n} \,
\varepsilon_a(k_n)\left(\hat{B}_a^\dagger(k_n)\hat{B}_a(k_n)+\frac{1}{2}\right)\right]
\qquad .
\]
The ground state in this sector (no kink fluctuation at all) is also
a coherent state
\[
\hat{B}_a(k_n)|K ; 0\rangle=0,  \, \, \forall a \, , \, \forall k_n
\, \, ; \quad \hat{\phi}_a|K ; 0\rangle=0, \, a\neq c \, , \quad
\hat{\phi}_c|K ; 0\rangle=\pm \phi_c^{V^{(c)}}{\rm
tanh}[\frac{\mu_c}{2}(x-x_0)]K ; 0\rangle \quad .
\]
The kink ground-state energy is divergent
\[
\langle 0; K|\hat{H}^{(2)}|K ; 0\rangle =\frac{\hbar
m}{2\sqrt{2}}{\rm Tr}_{L^2}K^{\frac{1}{2}}
\]
but we shall regularize it by means of the zeta function
prescription: let us take the value of
\[
\langle 0;K|\left(\hat{H}^{(2)}\right)^{-s}| K ;0 \rangle
=2^{s-1}\hbar\mu\left(\frac{\mu^2}{m^2}\right)^s{\rm Tr}_{L^2}\,
K^{-s}= 2^{s-1}\hbar\mu\left(\frac{\mu^2}{m^2}\right)^s\zeta_K(s)
\]
at a regular point $s\in{\mathbb C}$ of $\zeta_K(s)$. The problem is
that, since the wave numbers $k_n$ are determined by a transcendent
equation, there is no way of writing the ${\rm Tr}$ as any
manageable series. We shall rely on the less rigorous (physicists)
formula:
\[ \zeta_K(s)={\rm
Tr}_{L^2} K^{-s}=\sum_{a=1}^N \left(\sum_{j=1}^{N_b(a)}\,
\frac{1}{\varepsilon_a^s(j)}+\int_{-\infty}^\infty \, dk \,
\rho_{K_{aa}}(k)\frac{1}{(k^2+\mu_a^2)^{\frac{s}{2}}}\right) \qquad
,
\]
where in the $l=\infty$ limit all the bound states
$N_b(a)=N_{b_+}(a)+N_{b_-}(a)$ must be accounted for and the \lq\lq
sum" over the continuous spectra must be weighted with the
appropriate spectral density.

As in the vacuum sector, there is the $K$-heat equation kernel
\[ \left(\frac{\partial}{\partial\beta}+K\right)K_K(x,y;\beta)=0
\qquad , \qquad K_K(x,y;0)=\delta(x-y)
\]
solved for very large $l$ by:\[ {\rm
tr}K_K(x,y;\beta)=\sum_{a=1}^{N}\left[\sum_{j=1}^{N_b(a)}\left(f^a_j(x)\right)^*f^a_j(y)
e^{-\beta\varepsilon_a(j)}+\frac{1}{2\pi}\int_{-\infty}^\infty \,
\frac{dk}{N_a(k)} \, \left(f^a_k(x)\right)^*f^a_k(y)e^{-\beta
(k^2+\mu_a^2)}\right] \qquad ,
\]
where $N_a(k)$ is a normalization factor that depends on the wave
number. The heat trace reads:
\[
{\rm Tr}_{L^2}e^{-\beta K}=\int_{-\frac{l}{2}}^{\frac{l}{2}}\, dx
\,{\rm
tr}K_K(x,x;\beta)=\sum_{a=1}^{N}\left[\sum_{j=1}^{N_b(a)}e^{-\beta\varepsilon_a(j)}+
\int_{\infty}^\infty \, dk \, \rho_{K_{aa}}(k)e^{-\beta
(k^2+\mu_a^2)}\right] \qquad ,
\]
which can be used to determine the kink spectral zeta function via
the Mellin transform:

\[ \zeta_K(s)=\frac{1}{\Gamma(s)}\int_{0}^\infty \, d\beta \, {\rm
Tr}_{L^2}e^{-\beta K} \qquad .
\]

\subsubsection{ High-temperature expansion of the $K$-heat kernel and truncated zeta
functions}

We have met two levels of information concerning the spectra of
small fluctuations. The spectrum of vacuum fluctuations is fully
known and it is possible to analytically determine the zeta function
regularization of the sum over eigenfrequencies in terms of (famous)
meromorphic functions. This possibility fails dealing with
fluctuations around stable and isolated TK1 kinks, but knowledge of
the scattering data, bound state eigenvalues, phase shifts and
spectral densities, allows us to obtain formulas for the spectral
zeta function in the form of integrals (rather than series).

The situation is less favourable in the following three cases:
\begin{enumerate}
\item There exist one or several zero modes
but no negative eigenvalues in orthogonal directions to the TK1
orbit: $\varepsilon_a(0)=0, \quad a\neq c$. The TK1 kink is
\underline{stable} but degenerated in energy with a $k$-parametric
family of $k$-component topological kinks (TKk), also stable. $k$ is
the number of non-translational zero modes.

\item There exist one or several negative modes
but no zero eigenvalues in directions orthogonal to the TK1 orbit:
$\varepsilon_a(0)<0 , \quad a\neq c$. The TK1 kink is
\underline{unstable} and decay to some \underline{stable} and
isolated two-component topological kink (TK2).

\item There exist one or several zero modes
and one or several negative eigenvalues in  directions orthogonal to
the TK1 orbit. Isolated or degenerated families of TKk stable kinks
arise.

\end{enumerate}
The outcome is the same in these three situations when
multi-component stable kinks arise: $K$ is a \underline{non-diagonal
matrix Schr$\ddot{\rm o}$dinger operator} and the spectrum is
\underline{unknown}.

To cope with the problem of multi-component topological kinks we
write the $K$-heat equation in the form
\[
\left(\frac{\partial}{\partial\beta}+K_0-U(x)\right)K_K(x,y;\beta)=0
\]
and look for a solution based on the $K_0$-heat kernel:
\[
K_K(x,y;\beta)=C_K(x,y;\beta)K_{K_0}(x,y;\beta)\qquad ,
\]
where the density $C_K(x,y;\beta)$ satisfies the infinite
temperature condition $C_K(x,y;0)={\mathbb I}_{N\times N}$ and the
transfer equation:
\begin{equation}
\left(\frac{\partial}{\partial\beta} +\frac{x-y}{\beta}
\frac{\partial}{\partial x}-\frac{\partial^2}{\partial
x^2}\right)C_{K_{ab}}(x,y;\beta)=\sum_{c=1}^N\,
U_{ac}(x)C_{K_{cb}}(x,y;\beta)+(\mu_b^2-\mu_a^2)C_{K_{ab}}(x,y;\beta)
\label{trans} \quad .
\end{equation}
Then, we seek a power series solution
$C_{K_{ab}}(x,y;\beta)=\sum_{n=0}^\infty \, c_n^{ab}(x,y)\beta^n$ of
(\ref{trans}), which is tantamount to the recurrence relations
\begin{equation}
n c_n^{ab}(x,y)+(x-y)\frac{\partial c_n^{ab}}{\partial
x}(x,y)=\frac{\partial^2 c_{n-1}^{ab}}{\partial
x^2}(x,y)+\sum_{c=1}^N \,
U_{ac}(x)c_{n-1}^{cb}(x,y)+(\mu_b^2-\mu_a^2)c_{n-1}^{ab}(x,y)
\label{recr} \quad .
\end{equation}
In fact, only the densities at coincident points $x=y$ on the line
are needed. We introduce the notation
${}^{(k)}C_n^{ab}(x)=\lim_{x\to y}\frac{\partial^k
c_n^{ab}}{\partial x^k}(x,y)$ to write the recurrence relations
between these densities and their derivatives in the abbreviated
form
\[
{}^{(k)}C_n^{ab}(x)=\frac{1}{n}\left\{{}^{(k+2)}C_{n-1}^{ab}(x)+\sum_{c=1}^N
\,
U_{ac}(x){}^{(k)}C_{n-1}^{cb}(x)+(\mu_b^2-\mu_a^2){}^{(k)}C_{n-1}^{ab}(x)\right\}
,\] to be solved starting from:
${}^{(k)}C_0^{ab}(x)=\delta^{k0}\delta_{ab}$. The $n$-th density is
determined in terms of the $(n-1)$-order coefficients, their second
derivatives, and the potential matrix elements $U_{ab}(x)$.

The solution of these recurrence relations is achieved by using a
symbolic program run in Mathematica. We list the lower densities up
to the third order:
\begin{eqnarray*}
c_1^{ab}(x)&=&U_{ab}(x) \qquad \qquad , \qquad \qquad U_{ab}^{(k)}(x)=\frac{d^kU_{ab}}{dx^k}(x)\\
c_2^{ab}(x)&=&\frac{1}{6}U^{(2)}_{ab}(x)+\frac{1}{2}U^2_{ab}(x)+\frac{1}{2}(\mu_b^2-\mu_a^2)U_{ab}(x)\\
c_3^{ab}(x)&=&\frac{1}{60}U_{ab}^{(4)}(x)+\frac{1}{12}\left(\left(U^{(2)}U+UU^{(2)}\right)_{ab}(x)
+\left(U^{(1)}U^{(1)}\right)_{ab}(x)\right)+\frac{1}{6}U^3_{ab}(x)+\\&+&
\frac{1}{12}(\mu_b^2-\mu_a^2)\left(U^{(2)}_{ab}(x)+2U^2_{ab}(x)\right)+\frac{1}{6}\left(\sum_{c=1}^N\,
(\mu_b^2-\mu_c^2)U_{ac}(x)U_{cb}(x)+(\mu_b^2-\mu_a^2)U_{ab}(x)\right)
\, \, .
\end{eqnarray*}
There is an interesting point about these densities: the diagonal
components are the infinite conserved charges of some  matrix KdV
equation, see \cite{IGA}. Consider the family of differential
operators
\[
K(\beta)=-\frac{\partial^2}{\partial x^2}+{\rm diag}(\mu_1^2,
\cdots, \mu_N^2)-U(x,\beta),
\]
where the family of potentials solve the matrix KdV equation:
\[
\frac{\partial U}{\partial\beta}+3\left(U\frac{\partial U}{\partial
x}+\frac{\partial U}{\partial x}U\right)+\frac{\partial^3U}{\partial
x^3}=0 \qquad .
\]
The $\beta$-evolution of $K(\beta)$ can be expressed in the Lax pair
form
\[
\frac{\partial K}{\partial\beta}+[K,M]=0 \qquad , \qquad
M(\beta)=4\frac{\partial^3}{\partial
x^3}-3\left(U\frac{\partial}{\partial x}+\frac{\partial}{\partial
x}U\right)+B(\beta)
\]
such that it is iso-spectral. The diagonal densities codify the
spectrum of $K$. Ergo, $c_n^{aa}(x,x)\neq f(\beta)$.

Integration of the densities gives the asymptotic expansion of the
$K$-heat trace:
\[
{\rm Tr}_{L^2}e^{-\beta K}={\rm tr}\int_{-\infty}^\infty \, dx \,
K_K(x,x;\beta)=\sum_{a=1}^N \,
\frac{e^{-\beta\mu_a^2}}{\sqrt{4\pi}}\sum_{n=0}^\infty \,
c_n^{aa}[K]\, \beta^{n-\frac{1}{2}} \quad , \quad
c_n^{aa}[K]=\lim_{l\to\infty}\int_{-\frac{l}{2}}^\frac{l}{2}\, dx \,
c_n^{aa}(x) \qquad .
\]

\subsubsection{Truncated zeta function}
Finally, we split the Mellin transform
\begin{eqnarray*}
\zeta_K^*(s)&=&\frac{1}{\Gamma(s)}\int_0^\infty \, d\beta \,
\beta^{s-1}\left({\rm Tr}_{L^2}e^{-\beta
K}-n_0\right)=\frac{1}{\Gamma(s)}\int_0^b \, d\beta \,
\beta^{s-1}\left({\rm Tr}_{L^2}e^{-\beta
K}-n_0\right)+B_K(s;b)\\&=&\frac{1}{\Gamma(s)}\int_0^b \, d\beta \,
\beta^{s-1}\left(\sum_{a=1}^N\,
\frac{e^{-\beta\mu_a^2}}{\sqrt{4\pi}}\sum_{n=0}^\infty\,
c_n^{aa}[K]\beta^{n-\frac{1}{2}}-n_0\right)+\frac{1}{\Gamma(s)}\int_b^\infty
\, d\beta \, \beta^{s-1}\left({\rm Tr}_{L^2}e^{-\beta K}-n_0\right)
\end{eqnarray*}
into meromorphic and entire parts by choosing an upper limit in the
$\beta$-integration to be optimized in each particular problem. The
number of zero modes, $n_0$,  must be subtracted. Truncation of the
number of terms kept in the meromorphic part
\[
\zeta_K^*(s;b,N_0)=\frac{1}{\sqrt{4\pi}\Gamma(s)}\left(\sum_{a=1}^N\sum_{n=0}^{N_0}\,
c_n^{aa}[K]\frac{\gamma[n+s-1/2;b\mu_a^2]}{(b\mu_a^2)^{s+n-1/2}}-\frac{n_0}{b^ss}\right)
\]
provides us with a practical formula for the spectral zeta function
to be used in our computations in terms of the incomplete Euler
Gamma functions: $\gamma[z;c]$ . The two free parameters $b$ and
$N_0$ are correlated: the larger $b$, the greater the $N_0$ must be
chosen to achieve good approximations.

The subtraction of zero modes that we have performed is a very
tricky affair. We split the improper integral into two parts:
\[
I[n_0]=I_1[n_0]+I_2[n_0]=\lim_{\varepsilon\to
0}\frac{n_0}{\Gamma(s)}\left[\int_0^b\, d\beta \, \beta^{s-1}\,
e^{-\varepsilon\beta}+\int_b^\infty\, d\beta \, \beta^{s-1}\,
e^{-\varepsilon\beta}\right] \qquad .
\]
We neglect $I_2[n_0]$ and regularize the divergent integral
$I_1[n_0]$ for ${\rm Re}\, s\leq 0$ by assigning to it the value of
the analytic continuation of $I^R[n_0]=\frac{n_0}{b^s s\Gamma(s)}$,
valid for ${\rm Re}\, s> 0$, to the whole complex $s$-plane.

\subsection{One-loop kink mass shift formula}
We shall now perform the renormalizations needed to tame the
ultraviolet divergences, part of which we have already regularized
using the zeta function regularization method.

\subsubsection{Kink Casimir energy: zero-point
renormalization}

The first renormalization consists of subtracting the vacuum
zero-point energy. By analogy, we shall call this quantity the kink
Casimir energy. In the Casimir effect, the vacuum fluctuations are
subtracted from the zero-point energy around some geometrical
set-up, e.g., two infinite impenetrable plates.

The regularized kink Casimir energy is:
\[\bigtriangleup E_K^C(s)=\bigtriangleup E_K(s)-\bigtriangleup
E_{K_0}(s)=2^{s-1}\hbar
\left(\frac{\mu^2}{m^2}\right)^s\mu\left(\zeta_K(s)-\zeta_{K_0}(s)\right)\quad
.
\]
The proper kink Casimir energy is the value of this quantity at the
physical point $s=-\frac{1}{2}$:
\[
\bigtriangleup E_K^C=\lim_{s\to -\frac{1}{2}}\,\bigtriangleup
E_K^C(s)=\frac{\hbar
m}{2\sqrt{2}}\left(\zeta_K(-\frac{1}{2})-\zeta_{K_0}(-\frac{1}{2})\right)\qquad
.
\]
Because $s=-\frac{1}{2}$ is a pole of both $\zeta_K(s)$ and
$\zeta_{K_0}(s)$, with different residua in $(1+1)$-dimensional
models, the kink Casimir energy is still a divergent quantity.

\subsubsection{Kink energy engendered by mass renormalization}

There are other ultraviolet divergences due to one-loop graphs that
are controlled by the normal-ordered of the quantum Hamiltonian. The
contraction of two fields at the same point in Minkowski space is
the sum of a normal ordered product, all the creation operators
placed to the left of the annihilation operators, and a divergent
integral that in the normalization 1D \lq\lq box" becomes a familiar
divergent series:
\[
\hat{\phi}_a^2(x^\mu)=:\hat{\phi}_a^2(x^\mu):+ \delta\mu_a^2 \qquad
, \qquad \delta\mu_a^2=\int \, \frac{dk}{4\pi}\,
\frac{1}{\sqrt{k^2+\mu_a^2}}=\frac{1}{2l}\sum_{n\in{\mathbb
Z}}\,\frac{1}{(\frac{4\pi^2}{l^2}n^2+\mu_a^2)^{1/2}} \qquad .
\]
The quantum Hamiltonian
\[
\hat{\cal H}(\hat{\phi_1}(x^\mu),\cdots ,
\hat{\phi_N}(x^\mu))=\sum_{a=1}^N \,
\frac{1}{2}\left(\frac{\partial\hat{\phi}_a}{\partial
t}\frac{\partial\hat{\phi}_a}{\partial t}+
\frac{\partial\hat{\phi}_a}{\partial
x}\frac{\partial\hat{\phi}_a}{\partial
x}\right)+\hat{V}(\hat{\phi_1}(x^\mu),\cdots ,
\hat{\phi_N}(x^\mu))
\]
is normal-ordered by applying Wick's theorem:
\[
:\hat{\cal H}:=\hat{\cal H}+ :\left(1-{\rm
exp}\widehat{[-\hbar\sum_{a=1}^N\,
\delta\mu_a^2\frac{\delta^2}{\delta\phi_a^2}]V}\right):=\hat{\cal
H}+\hbar \sum_{a=1}^N\, \delta\mu_a^2 :\widehat{\frac{\delta^2
V}{\delta\phi_a^2}}:+{\cal O}(\hbar^2) \quad ,
\]
a process that is equivalent, at one-loop order, to adding quadratic
counter-terms to the Hamiltonian. Regularizing the divergent
coefficients of these counter-terms by means of the spectral zeta
function of $K_0$
\[
\delta\mu_a^2(s)=-\frac{1}{l}\frac{\Gamma(s+1)}{\Gamma(s)}\sum_{a=1}^N\zeta_{K_{0aa}}(s+1)
\]
we obtain at a regular point of $\zeta_{K_0}(s+1)$ the regularized
contribution to the kink energy due to the mass renormalization
counter-terms measured with respect to the mass renormalization
vacuum energy:
\begin{eqnarray*} \bigtriangleup E_K^R(s)&=&\hbar m\sum_{a=1}^N \,
\delta\mu_a^2\, \int \, dx \, \left( \langle 0; K|
:\widehat{\frac{\delta^2V}{\delta\phi_a^2}}:|K;0 \rangle -\langle 0;
V| :\widehat{\frac{\delta^2V}{\delta\phi_a^2}}:|V;0 \rangle\right)\\
&=& -\lim_{l\to\infty}\frac{\hbar
m}{2l}\left(\frac{2\mu^2}{m^2}\right)^{s+1/2}
\frac{\Gamma(s+1)}{\Gamma(s)}
\sum_{a=1}^N\zeta_{K_{0aa}}(s+1)\int_{-\frac{l}{2}}^\frac{l}{2}\, dx
\, U_{aa}(x)
\end{eqnarray*}
The expression in the second line of the formula is derived from the
fact that normal-ordered products of field operators acting on
coherent states select the ordinary product of the field
configuration characterizing the state, see \cite{SFC1}. Note also
that we have regularized the divergent graphs using
$\zeta_{K_0}(s+1)$ instead of $\zeta_{K_0}(s)$. The reason for this
is the convenience of comparing the residua of $\bigtriangleup
E^C_K(s)$ and $\bigtriangleup E^R_K(s)$ at the same pole:
$s=-\frac{1}{2}$.

Finally, the one-loop mass shift formula is found by seeking the
physical value of the $s$ parameter:
\begin{equation}
\begin{tabular}{|c|}
\hline  \\  $ \bigtriangleup M_K =\displaystyle \lim_{s\to
-1/2}\left(\bigtriangleup E_K^C(s)+\bigtriangleup E_K^R(s)\right)$
 \\ \\
\hline
\end{tabular}
\label{dos} \qquad .
\end{equation}

\subsection{N=1: the $\lambda(\phi^4)_2$ model}

We briefly describe the standard $\lambda(\phi^4)_2$ model. There is
only one real scalar field and no deformation parameters. The action
is:
\[
S[\phi]=\frac{m^2}{2\lambda}\int \, d^2x \,
\left\{\frac{\partial\phi}{\partial
x^\mu}\frac{\partial\phi}{\partial x_\mu}-(\phi^2(t,x)-1)^2\right\}
\qquad , \qquad \sigma_{11}=\sigma_1^2=0 \quad.
\]
The very well known vacua and kinks of this model, as well as the
differential operators governing the small vacuum and kink
fluctuations are written below:
\[
\phi^V=\pm 1 \quad , \quad K_0=-\frac{d^2}{dx^2}+4 \quad ; \quad
\phi^K(x)=\pm{\rm tanh}(x-x_0) \quad , \quad
K=-\frac{d^2}{dx^2}+4-\frac{6}{{\rm cosh}^2x} \quad .
\]

\begin{figure}[htbp]
\centerline{\includegraphics[height=4cm]{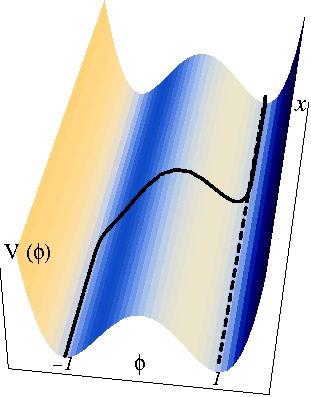}\hspace{4cm}
\includegraphics[height=3.7cm]{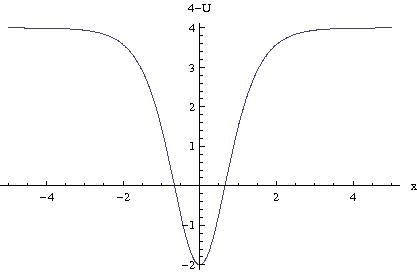}} \caption{ (left) 3D plot
of $V$ as a function of both $\phi$ and $x$. The kink is also shown
as a black line crossing from one vacuum valley to the other.
(right) Graphic of the potential well $4-U(x)$ in the Schr$\ddot{\rm
o}$dinger operator $K$.}
\end{figure}

\subsubsection{ Kink spectral heat and zeta functions}

Because the spectrum of $K$  is fully known, both the kink heat
function and all the Seeley coefficients can be computed exactly
{\footnote{In \cite{AMAJW}, these coefficients are obtained by
integration of the densities solving the recurrence relations.}}:
 \begin{eqnarray*}
{\rm Tr}e^{-\beta
K}&=&\frac{e^{-4\beta}}{\sqrt{4\pi}}\left(\frac{l}{\sqrt{\beta}}+\sqrt{4\pi}(e^\beta{\rm
Erf}[\sqrt{\beta}]+e^{4\beta}{\rm
Erf}[2\sqrt{\beta}])\right)\\&=&\frac{e^{-4\beta}}{\sqrt{4\pi}}\sum_{n=0}^\infty
\, c_n(K)\beta^{n-1/2} \quad ; \qquad c_0(K)=l \, \, \, \,  , \quad
c_n(K)=\frac{2^{n+1}(1+2^{2n-1})}{(2n-1)!!} \quad , \quad n\geq 1
\quad ,
\end{eqnarray*}
where ${\rm Erf}[z]$ is the error function. The exact kink zeta
function and the truncated theta functions, with the zero mode
subtracted, are respectively:
\begin{eqnarray} \zeta_K^*(s)&=&\frac{1}{\Gamma(s)}\int_0^\infty \,
d\beta \, \beta^{s-1}\left({\rm Tr}e^{-\beta K}-1\right)\nonumber
\\&=&\frac{1}{\sqrt{4\pi}\Gamma(s)}\left[\frac{l}{4^s}\Gamma(s-\frac{1}{2})+
\left(\frac{2}{3^{s+\frac{1}{2}}}{}_2F_1[\frac{1}{2},\frac{1}{2}+s,
\frac{3}{2};-\frac{1}{3}]-\frac{1}{4^ss}\right)
\Gamma(s+\frac{1}{2})\right]\\
\zeta_K^*(s;b,N_0)&=&\frac{1}{\Gamma(s)}\int_0^b \, d\beta \,
\beta^{s-1}\left(\frac{e^{-4\beta}}{\sqrt{4\pi}}\sum_{n=0}^{N_0} \,
c_n(K)\beta^{n-1/2}-1\right)\nonumber\\&=&
\frac{1}{\sqrt{4\pi}\Gamma(s)}\left(l\frac{\gamma[s-\frac{1}{2};4
b]}{(4b)^{s-\frac{1}{2}}}+\sum_{n=1}^{N_0}\,
\frac{2^{n+1}(1+2^{2n-1})}{(2n-1)!!}\frac{\gamma[s+n-\frac{1}{2};4
b]}{(4b)^{s+n-\frac{1}{2}}}-\frac{1}{b^s s}\right) \label{tzf}
\end{eqnarray}
$\gamma[z;c]$ are incomplete Euler Gamma functions and
${}_2F_1[a,b,c;z]$ are Gauss hypergeometric functions. We follow the
notation of \cite{ABRAM}.
\subsubsection{ One-loop $\lambda\phi^4$-kink mass shift}

The limit $s\to -\frac{1}{2}$ is very delicate: it is a pole of
$\zeta_K(s)$. To take this limit properly we look at the regularized
formulae for $s=-\frac{1}{2}+\varepsilon$ and then allow
$\varepsilon$ to go to zero:

\begin{eqnarray*}
\Delta E_K^C&=& \frac{\hbar m}{2\sqrt{2\pi}} \lim_{\varepsilon
\rightarrow 0}\left( \frac{2 \mu^2}{m^2} \right)^{\varepsilon}
\frac{\Gamma(\varepsilon)}{\Gamma(-\frac{1}{2}+\varepsilon)} \left[
\frac{2}{3^\varepsilon} \, {}_2
F_1[{\textstyle\frac{1}{2}},\varepsilon,
{\textstyle\frac{3}{2}},-{\textstyle\frac{1}{3}}]-
\frac{1}{(-\frac{1}{2}+\varepsilon)\,4^{-\frac{1}{2}+\varepsilon}}
\right]\\ &=& \frac{\hbar m}{2\sqrt{2}\pi} \lim_{\varepsilon
\rightarrow 0} \left[ -\frac{3}{\varepsilon}+2 +\ln \frac{3}{4}-3
\ln \frac{2 \mu^2}{m^2}
-{}_2F_1^{(0,1,0,0)}[{\textstyle\frac{1}{2}},0,{\textstyle\frac{3}{2}},-{\textstyle\frac{1}{3}}]
+o(\varepsilon) \right]\\&=&-\frac{\hbar
m}{2\sqrt{2}\pi}\lim_{\varepsilon\rightarrow
0}\left[{3\over\varepsilon}+3\ln\frac{2\mu^2}{m^2}-\frac{\pi}{\sqrt{3}}\right]
\qquad ,
\end{eqnarray*}
where ${}_2F_1^{(0,1,0,0)}[a,b,c;z ]$ is the derivative of the
hypergeometric function with respect to the second argument. The
same strategy is used with $\bigtriangleup E_K^R(s)$ to find:
\begin{eqnarray*}
\Delta E_K^R &=&-\frac{3 \hbar m}{\sqrt{2\pi}} \lim_{\varepsilon
\rightarrow 0} \left( \frac{2 \mu^2}{m^2} \right)^\varepsilon
\frac{4^{-\varepsilon}
\Gamma(\varepsilon)}{\Gamma(-\frac{1}{2}+\varepsilon)}\\&=&\frac{3\hbar
m}{2 \sqrt{2}\pi} \lim_{\varepsilon \rightarrow 0} \left[
\frac{1}{\varepsilon}+ \ln \frac{2 \mu^2}{m^2} -\ln
4+(\psi(1)-\psi(-{1\over 2})) +o(\varepsilon)
\right]\\&=&\frac{\hbar m}{2\sqrt{2}\pi}\lim_{\varepsilon\rightarrow
0}\left[{3\over\varepsilon}+3\ln\frac{2\mu^2}{m^2}-2(2+1)\right]
\qquad ,
\end{eqnarray*}
where $\psi(z)$ is the digamma function: i. e., the logarithmic
derivative of the Euler Gamma Function. Therefore, the divergences
at the pole cancel exactly and we are left with the finite answer:
\begin{eqnarray*}
\bigtriangleup M_K&=&\lim_{s\to -1/2}\left(\bigtriangleup
E^C_K(s)+\bigtriangleup
E^R_K(s)\right)=\left(\frac{1}{2\sqrt{6}}-\frac{3}{\pi\sqrt{2}}\right)
\hbar m =-0.471113 \hbar m
\end{eqnarray*}
in perfect agreement with the Dashen-Hasslacher-Neveu result in
\cite{DHN}. We remark that the cancelations above and those implicit
 in (\ref{dos}) set finite renormalizations, such as the large mass
subtraction scheme, by imposing the condition that tadpoles vanish,
see Reference \cite{BGvNV}.

From the sign of the correction we learn a qualitative fact about
the global effect of the scalar boson fluctuations on the kink. The
kink energy density is the square of the derivative with respect to
$x$ of the kink profile, whereas the kink energy is the area
enclosed by this curve. The decrease in kink energy due to kink
fluctuations is tantamount to a small decrease in the area.
Therefore, the net effect is equivalent to a force from the right
and another from the left exerted by the scalar bosons on the kink
in such a way that the kink profile shrinks slightly. Exactly as in
the ideal Casimir effect with two plates of infinite area.

As a test to prove the quality of the high-temperature
approximation, we now write the result for $N_0=10$ and $b=1$. See
\cite{AMAJW} and \cite{AMAJJW} for full details:
\[
\bigtriangleup M_K \cong
-\left(0.19947+\sum_{n=2}^{10}c_n(K)\frac{\gamma[n-1,4]}{8\pi\sqrt{2}\,
\, 4^{n-1}}\right)\hbar m=-0.471371 \hbar m \qquad .
\]
The answer departs from the right result in the four-decimal figure;
a reasonably acceptable approximation. There are two elements,
however, in formula (\ref{tzf}) to play with: namely, the length $b$
of the integration interval in the Mellin transform and the number
$N_0$ of terms in the series remaining. Figure 2 shows that these
two variables are not independent. For a given value of $N_0$, there
is an optimum choice of $b$ after which the approximate result (it
comes from an asymptotic series) rapidly departs from the exact
result. There is no point in enlarging the integration interval.
Conversely, taking $b$ longer forces us to increase $N_0$ to achieve
an acceptable approximation. These remarks are particularly
important for light-mass (less than one) particles. The reason is
that in such cases we need $b$ to be large because, if not, too much
would be neglected in the entire parts.

\begin{figure}[htb]
\centerline{\includegraphics[height=4.3cm]{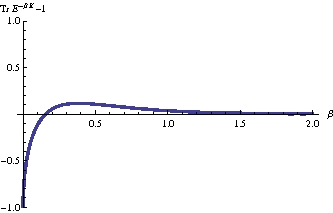}\hspace{1cm}
\includegraphics[height=4.3cm]{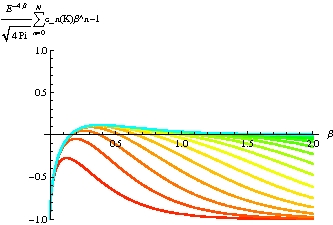}}\caption{ (left) Plot of the
exact heat trace as a function of $\beta$ with the zero mode
subtracted. (right) Graphics of the approximated formula, keeping
$N_0=2$, $N_0=3$, $\cdots $, $N_0=11$ terms.}
\end{figure}

\subsection{N=2: the BNRT model}
Next, we address a model with two real scalar fields. This field
theoretical system can be seen as the bosonic sector of an ${\cal
N}=1$ supersymmetric Wess-Zumino model of two chiral super-fields
dimensionally reduced to $(1+1)$-dimensions (plus a reality
condition in the fields), see \cite{SV}. Interactions are derived
from the simplest polynomial holomorphic super-potential
accommodating the second super-field. In fact, the super-potential
was discovered independently by Bazeia, Nascimento, Ribeiro, and
Toledo (BNRT) \cite{BNRT} directly in the $(1+1)$-dimensional field
theoretical model that we shall describe.

The following choice of parameters in the general linear ${\mathbb
O}(2)$-sigma model
\[
\sigma_1^2=0 \quad , \quad \sigma_2^2=1-\frac{\sigma}{2} \quad ,
\quad \sigma_{11}=\frac{3}{2} \quad , \quad
\sigma_{22}=\frac{\sigma^2-1}{2} \quad , \quad
1+\sigma_{12}=2\sigma(\sigma+1)
\]
leads to the one-parametric family of actions:
\begin{eqnarray*}
S[\phi_1,\phi_2]&=&\frac{m^2}{\lambda}\int \, d^2x \,
\left\{\frac{1}{2}\left(\frac{\partial\phi_1}{\partial
x^\mu}\frac{\partial\phi_1}{\partial
x_\mu}+\frac{\partial\phi_1}{\partial
x^\mu}\frac{\partial\phi_1}{\partial
x_\mu}\right)-\frac{1}{2}(\phi_1^2(t,x)+\phi_2^2(t,x)-1)^2\right.\\
&-&\left.\frac{3}{2}\phi_1^4+
\frac{1-\sigma^2}{2}\phi_2^4+(1-2\sigma(\sigma+1))\phi_1^2\phi_2^2+\frac{\sigma-2}{2}\phi_2^2\right\}
\quad .
\end{eqnarray*}
Quartic and quadratic anisotropies are induced by the positive real
parameter $\sigma\in {\mathbb R}^+$.
\subsubsection{Vacua and TK1 topological kinks}
The four classical minima are degenerated in energy for any
$\sigma\in {\mathbb R}^+$. Henceforth, all the classical minima are
ground states, and there are \lq\lq two points" in the vacuum moduli
space.
\begin{enumerate}

\item Vertical vacua:
\[
\phi_1^{V^{(2)}}=0 \,\,\, , \, \, \,
\phi_2^{V^{(2)}}=\pm\frac{1}{\sqrt{2\sigma}} \qquad  , \qquad
V(0,\pm\frac{1}{\sqrt{2\sigma}})=\frac{3}{8} \qquad ,
\]
such that the particle masses are
\[
\mu_1^2=2\left(\frac{1-\sigma_2^2}{1+2\sigma_{22}}(1+\sigma_{12})-(1-\sigma_1^2)\right)=2\sigma
\qquad , \qquad \mu_2^2=4(1-\sigma_2^2)=2\sigma \qquad .
\]

\item Horizontal vacua

\[
\phi_1^{V^{(1)}}=\pm\frac{1}{2} \,\,\, , \, \, \, \phi_2^{V^{(1)}}=0
\qquad , \qquad V(\pm\frac{1}{2},0)=\frac{3}{8}
\]
such that the particle masses are
\[
\mu_1^2=4(1-\sigma_1^2)=4 \qquad , \qquad
\mu_2^2=2\left(\frac{1-\sigma_1^2}{1+2\sigma_{11}}(1+\sigma_{12})-(1-\sigma_2^2)\right)=\sigma^2
\qquad .
\]

\end{enumerate}

Therefore, there are also two kinds of one-component topological
kinks, see Figure 5.

1. \lq\lq Vertical" topological kinks:
\[
\phi_1^K(x)=0 \qquad , \qquad \phi_2^K(x)=\pm
\frac{1}{\sqrt{2\sigma}}{\rm tanh}[\sqrt{\frac{\sigma}{2}}(x-x_0)]
\qquad .
\]
The vacuum and kink fluctuation operators are:
\[
K_0=\left(\begin{array}{cc}-\frac{d^2}{dx^2}+2\sigma & 0
\\ 0 & -\frac{d^2}{dx^2}+2\sigma \end{array}\right) \quad , \quad
K=\left(\begin{array}{cc}-\frac{d^2}{dx^2}+2\sigma
-\frac{2(1+\sigma)}{{\rm cosh}^2[\sqrt{\frac{\sigma}{2}}x]}& 0
\\ 0 & -\frac{d^2}{dx^2}+2\sigma -\frac{3\sigma}{{\rm
cosh}^2[\sqrt{\frac{\sigma}{2}}x]}\end{array}\right)\, \,  .
\]

2. \lq\lq Horizontal" topological kinks:
\[
\phi_1^K(x)=\pm \frac{1}{2}{\rm tanh}(x-x_0) \qquad , \qquad
\phi_2^K(x)=0 \qquad .
\]
The vacuum and kink fluctuation operators are:
\[
 K_0=\left(\begin{array}{cc}-\frac{d^2}{dx^2}+4 & 0
\\ 0 & -\frac{d^2}{dx^2}+\sigma^2 \end{array}\right) \qquad , \qquad
K=\left(\begin{array}{cc}-\frac{d^2}{dx^2}+4 -\frac{6}{{\rm
cosh}^2x}& 0
\\ 0 & -\frac{d^2}{dx^2}+\sigma^2 -\frac{\sigma(\sigma+1)}{{\rm
cosh}^2x}\end{array}\right) \, \, \, \,  .
\]

\subsubsection{Fluctuation spectrum of horizontal TK1 kinks}
We shall not discuss the vertical sector here; the main results can
be found in \cite{AMAJW2}. And it is more instructive to explain the
horizontal sector. In the kink orbit direction, we meet the
Schr$\ddot{\rm o}$dinger operator of the $\lambda\phi^4$-kink. There
is no need of re-compute again the effect of the fluctuations
parallel to the orbit; we simply use the results of the previous
model.

Fluctuations orthogonal to the orbit, however, are governed by the
Schr$\ddot{\rm o}$dinger operator shown above. We shall focus for a
while on the case in which $\sigma=J\in{\mathbb Z}^+$ is a positive
integer. The reason for this is that in this discrete set of models
transparent (reflection coefficient equal to zero) P$\ddot{\rm
o}$sch-Teller potentials arise:
\[ K_{22}=-\frac{d^2}{dx^2}+J^2-\frac{J(J+1)}{{\rm cosh}^2x} \qquad
.
\]
The previously  defined parameter $A$ determining the number of
bound states becomes $A=J+\frac{1}{2}$, and the threshold of the
continuous spectrum is: $4\frac{\mu_a^2}{\mu_c^2}=J^2$. The spectrum
of $K_{22}$ is summarized as follows:

\begin{enumerate}

\item  The eigenvalues and eigenfunctions of the discrete spectrum
are:
\begin{eqnarray*}
\varepsilon_2^2(j)&=&(2J-j)j \qquad , \qquad j=0,1,2, \cdots, J \\
f^2_0(x)&=&\frac{1}{{\rm cosh}^Jx} \qquad , \qquad
f_j^2(x)=\Pi_{r=0}^{j-1}\left(-\frac{d}{dx}+(J-r){\rm
tanh}x\right)\frac{1}{{\rm cosh}^{J-j}x} \quad , \quad j\geq 1
\qquad .
\end{eqnarray*}
There are $J$ bound states with eigenvalues below the threshold
starting from a zero mode. The highest eigenvalue bound state sits
immediately at the threshold of the continuous spectrum. These
eigenfunctions are termed half-bound states and they always
accompany zero reflection coefficients. In the remarkable array of
numbers that follows we have collected the bound state and
half-bound state eigenvalues up to $J=10$.

{\normalsize \[
\begin{array}{cccccccccc} J=1 & J=2 & J=3 & J=4 & J=5 & J=6 & J=7 &
J=8 & J=9 & J=10 \\ 0 & 0 & 0 & 0 & 0 & 0 & 0 & 0 & 0 & 0 \\ 1 & 3 &
5 & 7 & 9 & 11 & 13 & 15 & 17 & 19 \\  & 4 & 8 & 12 & 16 & 20 & 24 &
28 & 32 & 36 \\ & & 9 & 15 & 21 & 27 & 33 & 39 & 45 & 51 \\ & & & 16
& 24 & 32 & 40 & 48 & 56 & 64 \\ & & & & 25 & 35 & 45 & 55 & 65 & 75
\\ & & & & & 36 & 48 & 60 & 72 & 84 \\ & & & & & & 49 & 63 & 77 & 91
\\ & & & & & & & 64 & 80 & 96 \\ & & & & & & & & 81 & 99 \\ & & & &
& & & & & 100
\end{array}
\]}

\item The scattering states are also known and are listed below together
with the pase shifts and spectral densities.
\begin{eqnarray*}
\varepsilon_2^2(k)&=&k^2+J^2  \quad , \qquad
f^2_k(x)=\Pi_{p=1}^J\left(-\frac{d}{dx}+p {\rm tanh}x\right)e^{ikx}\\
\delta_2(k)&=&2{\rm arctan}\left[\frac{{\rm
Im}(\Pi_{p=1}^J(p-ik))}{{\rm Re}(\Pi_{p=1}^J(p-ik))}\right] \qquad ,
\qquad \rho_2(k)=\frac{l}{2\pi}-\frac{1}{\pi}\sum_{p=1}^J
\frac{p}{p^2+k^2} \qquad .
\end{eqnarray*}
\end{enumerate}

From this information, one derives the exact heat function:
\newpage
\begin{eqnarray*}
{\rm Tr}_{L^2} e^{-\beta K}&=&\sum_{j=0}^J \, e^{-\beta(2
J-j)}+\frac{l}{2\pi}\int_{-\infty}^\infty \, dk \, e^{-\beta
(k^2+J^2)}\left[1-\frac{2}{l}\sum_{p=1}^J \,
\frac{p}{p^2+k^2}\right]
\\ &=& \frac{e^{-\beta
N^2}}{\sqrt{4\pi}}\left(\frac{l}{\sqrt{\beta}}+\sqrt{4\pi}\sum_{p=1}^J
\, e^{\beta p^2}{\rm Erf}[p\sqrt{\beta}]\right)
\end{eqnarray*}
as a sum of error functions. The coefficients of the heat function
expansion are also easily calculated:
\begin{eqnarray*}
{\rm Tr}_{L^2} e^{-\beta K}&=& \frac{e^{-\beta
J^2}}{\sqrt{4\pi}}\sum_{n=0}^\infty \,
c_n(K)\beta^{n-\frac{1}{2}} \qquad , \qquad   c_0(K)=l \\
c_n(K)&=&\frac{2^{n+1}}{(2n-1)!!}\cdot\sum_{p=1}^J\,
p^{2n-1}=\frac{2^{n+1}}{(2n-1)!!}\left(J^{2n-1}+\frac{B_{2n}(J)-B_{2n}}{2n}\right)
\quad , \quad n\geq 1 \qquad .
\end{eqnarray*}
They can be expressed in terms of Bernouilli numbers $B_{2n}$ and
polynomials $B_{2n}(J)$.

These systems can be useful as patterns for other systems where no
analytical information about the spectrum is available and there is
no hope of finding the exact heat function. The coefficients of the
heat function must be similar for potentials with similar thresholds
and areas (number of bound states).  Thus, it is important to know
the behavior of the heat kernel coefficients in these \lq\lq
integrable" cases. Considering $n$ as a continuous variable $u$, the
function
\[
f_N(u)=\frac{2^{u+1}}{(2u-1)!!}\sum_{p=0}^J\,p^{2u-1}
\]
is plotted in Figure 3 for three values of $J$. The peaks, as they
should be, correspond to integer values $u=n$, and the coefficients
pass to zero when $n=12$ for $J=2$, $n=20$, if $J=3$, $n=30$ for
$J=4$, etcetera. These are the orders at which the double factorial
in the denominator dominates the numerator, ensuring the convergence
of the series despite the fact that the coefficients grow really
large before these orders for $J\geq 4$.

\begin{figure}[htbp]
\centerline{\includegraphics[height=2.7cm]{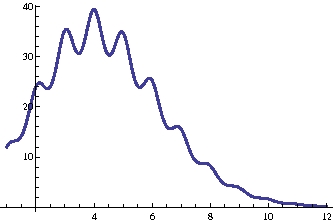}\hspace{1cm}
\includegraphics[height=2.7cm]{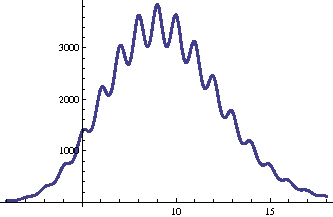}\hspace{1cm}\includegraphics[height=2.7cm]{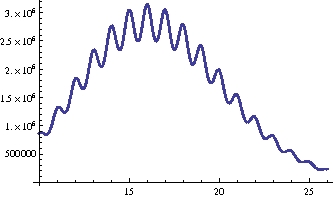}}
\caption{Plots of the function $f_J(u)$: (left) $J=2$ (middle) $J=3$
(right) $J=4$.}
\end{figure}

\subsubsection{Cancelation of divergences: one-loop mass shifts of horizontal TK1 kinks}
The consequences of the previous analyses can be seen in the
magnitude of the one-loop mass shifts. The spectral zeta function
\begin{eqnarray*}
\zeta^*_K(s)&=& \frac{1}{\Gamma(s)}\int_0^\infty \, d\beta \,
\beta^{s-1}\left({\rm Tr}_{L^2} e^{-\beta
K}-1\right)\\&=&\frac{1}{\sqrt{\pi}}\left[\frac{l}{2}\frac{\Gamma(s-\frac{1}{2})}{J^{2s-1}\Gamma(s)}
+\left(\sum_{j=1}^{J-1}\, \frac{2j}{(J^2-j^2)^{s+\frac{1}{2}}}\cdot
{}_2F_1[\frac{1}{2},\frac{1}{2}+s,\frac{3}{2};\frac{-j^2}{J^2-j^2}]-\frac{1}{J^{2s}
s}\right)\frac{\Gamma(s+\frac{1}{2})}{\Gamma(s)}\right]
\end{eqnarray*}
is essentially a sum of Gauss hypergeometric functions. For
instance, in the case $J=3$
\begin{eqnarray*}
\zeta^*_K(s)&=&\frac{1}{\sqrt{\pi}}\left[\frac{l}{2}\frac{\Gamma(s-\frac{1}{2})}{3^{2s-1}\Gamma(s)}
\right.\\ &+& \left.\left( \frac{2}{8^{s+\frac{1}{2}}}\cdot
{}_2F_1[\frac{1}{2},\frac{1}{2}+s,\frac{3}{2};-\frac{1}{8}]+\frac{4}{5^{s+\frac{1}{2}}}\cdot
{}_2F_1[\frac{1}{2},\frac{1}{2}+s,\frac{3}{2};-\frac{4}{5}]-\frac{1}{9^s
s}\right)\frac{\Gamma(s+\frac{1}{2})}{\Gamma(s)}\right]
\end{eqnarray*}
one can envisage the general behavior from the array of bound-state
energies. The ingredients are the bound-state eigenvalues, the
difference in the eigenvalues with respect to the thresholds, and
the thresholds themselves (in the subtraction of the zero modes). It
is amazing how cleanly the maths capture these fine physical data!

The kink Casimir energy in the direction orthogonal to the orbit for
generic $J$ is:
\begin{eqnarray*}
\bigtriangleup E_K^C&=&\lim_{s\to -\frac{1}{2}}\,
2^{s-1}\hbar\left(\frac{\mu^2}{m^2}\right)^s\mu\left(\zeta_{K_{22}}^*(s)-\zeta_{K_{022}}(s)\right)\\
&=&\frac{\hbar
m}{2\sqrt{2}}\lim_{\varepsilon\to 0}\,
\left(\frac{2\mu^2}{m^2}\right)^\varepsilon\left(\zeta_{K_{22}}^*(-\frac{1}{2}+\varepsilon)
-\zeta_{K_{022}}(-\frac{1}{2}+\varepsilon)\right)
\\ &=& \frac{\hbar
m}{2\sqrt{2\pi}}\lim_{\varepsilon\to 0}\,
\left(\frac{2\mu^2}{m^2}\right)^\varepsilon
\frac{\Gamma(\varepsilon)}{\Gamma(-\frac{1}{2}+\varepsilon)}\left(\sum_{j=1}^{J-1}
\frac{2j}{(J^2-j^2)^\varepsilon}\cdot
{}_2F_1[\frac{1}{2},\varepsilon
,\frac{3}{2};\frac{-j^2}{J^2-j^2}]-\frac{J}{J^{2\varepsilon}(-\frac{1}{2}+\varepsilon}\right)\,
.
\end{eqnarray*}
Simili modo, the kink energy due to the mass renormalization
counter-terms of the second particle reads:
\begin{eqnarray*}
\bigtriangleup E_{K_{22}}^R&=&-\frac{\hbar
m}{\sqrt{2}}\lim_{l\to\infty}\frac{1}{2l}\lim_{s\to -\frac{1}{2}}\,
\left(\frac{\mu^2}{m^2}\right)^{s+\frac{1}{2}}\frac{l}{\sqrt{4\pi}}\cdot
\frac{\Gamma(s+\frac{1}{2})}{J^{2s+1}\Gamma(s)}\cdot\int_{-\frac{l}{2}}^\frac{l}{2}\,
dx \, \frac{J(J+1)}{{\rm cosh^2x}} \\&=& -\frac{\hbar
m}{2\sqrt{2\pi}} \lim_{\varepsilon\to
0}\left(\frac{2\mu^2}{m^2}\right)^\varepsilon\,
\frac{J(J+1)}{J^{2\varepsilon}}\cdot\frac{\Gamma(\varepsilon)}{\Gamma(-\frac{1}{2}+\varepsilon)}\,
\, \, \, \, \, \, \, .
\end{eqnarray*}
For instance applying these formulae to the $J=3$ case, we find:
\begin{eqnarray*}
\bigtriangleup E_{K_{22}}^C&=& -\frac{\hbar
m}{\sqrt{2}\pi}\left(\lim_{\varepsilon\to
0}\frac{3}{\varepsilon}+3\log\frac{2\mu^2}{m^2}+\frac{1}{2}\left(
{}_2F_1^{(0,1,0,0)}[{\textstyle\frac{1}{2}},0,{\textstyle\frac{3}{2}},-{\textstyle\frac{1}{8}}]
-\log\frac{8}{4}\right)+\right. \\ &+& \left.
{}_2F_1^{(0,1,0,0)}[{\textstyle\frac{1}{2}},0,{\textstyle\frac{3}{2}},-{\textstyle\frac{4}{5}}]-\log\frac{5}{4}-3-
\frac{3}{2}\log\frac{9}{4} \right)\\ \bigtriangleup
E_{K_{22}}^R&=&\frac{3\hbar
m}{\sqrt{2}\pi}\left(\lim_{\varepsilon\to
0}\frac{1}{\varepsilon}+\log\frac{2\mu^2}{m^2}-\log\frac{9}{4}-2\right)\quad
.
\end{eqnarray*}
The divergent terms, as well as the terms depending on the auxiliary
parameter $\mu$ cancel each other exactly , to give:
\begin{equation}
\bigtriangleup M_{K_{22}}=-\frac{\hbar
m}{\sqrt{2}\pi}\left[\frac{1}{2}\left(
{}_2F_1^{(0,1,0,0)}[{\textstyle\frac{1}{2}},0,{\textstyle\frac{3}{2}},-{\textstyle\frac{1}{8}}]+
\log\frac{9}{8}\right)+
{}_2F_1^{(0,1,0,0)}[{\textstyle\frac{1}{2}},0,{\textstyle\frac{3}{2}},-{\textstyle\frac{4}{5}}]
+\log\frac{9}{5}+3\right] \label{N2kmf}\quad .
\end{equation}
Using this result and the analogous one for $J=4$, we obtain the
exact answer for the $J=3$, and $J=4$ systems:
\begin{equation}
\bigtriangleup M_K^{J=3}=-(0.47113+0.766861)\hbar m \qquad , \qquad
\bigtriangleup M_K^{J=4}=-(0.47113+1.11725)\hbar m \label{olmsN}
\qquad .
\end{equation}

\begin{figure}[htbp]
\centerline{\includegraphics[height=5cm]{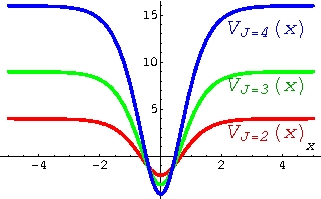}}
\caption{Potential wells for $J=2$, $J=3$, and $J=4$ plotted
together.}
\end{figure}
The higher the threshold and the broader the area, see Figure 4, the
more negative is the correction, because the attraction of the well
on the scattered and bound mesons is more intense. The
high-temperature asymptotic formula also provides very good
approximations to these exact results. We find:
\[
\bigtriangleup M_K^{J=3}=-(0.47137+0.76675)\hbar m \qquad , \qquad
\bigtriangleup M_K^{J=4}=-(0.47137+1.11723)\hbar m,
\]
taking $N_0=20$ coefficients for $J=3$ and $N_0=40$ coefficients if
$J=4$ in the asymptotic formulae of $\bigtriangleup
M_{K_{22}}^{J=3}$ and $\bigtriangleup M_{K_{22}}^{J=4}$. The
(optimal) truncations in the series have been suggested by the plots
in Figure 3. For the same reason, we consider $N_0=10$ in the
truncated series of $\bigtriangleup M_{K_{11}^{J=2}}$.

Even if $\sigma$ is not an integer the one-loop mass shift is still
known. The spectrum is more complex but the main differences passing
through $J$ are the lack of the (half) bound state, sitting at the
threshold, one more bound state and a different spectral density.
Because the vacuum half-bound state, (the constant function) is not
compensated by a kink half-bound state it must be accounted for in
the vacuum spectral function. The half-bound state weight is
$\frac{1}{2}$ (hence the name) owing to the one-dimensional Levinson
theorem \cite{BAR}, \cite{LJB}. The jump in the number of bound
states miraculously conspires with the change in spectral density to
produce a quantum correction that is a continuous function of
$\sigma$. Interested readers can find full details about this
problem in \cite{AMAJW2}.

\subsubsection{TK2 topological kinks}

The zero-mode fluctuation orthogonal to the kink orbit is a sign of
the existence of other topological kinks degenerated in energy with
the TK1 kink, which must have the two components of the field
different from zero, thus being TK2 kinks. Fortunately, a fair
knowledge of the main features and the structure of the TK2 kinks in
this model is available. The reason is that the potential energy
density can be written in terms of a polynomial \lq\lq
superpotential":
\[
W(\phi_1, \phi_2)=2\left({1\over 3}\phi_1^3-{1\over
4}\phi_1+{\sigma\over 2}\phi_1\phi_2^2 \right) \qquad , \qquad
V(\phi_1,\phi_2)-\frac{3}{8}={1\over 2}\sum_{a=1}^2\, \frac{\partial
W}{\partial\phi_a}\cdot\frac{\partial W}{\partial\phi_a} \qquad .
\]
The energy, arranged à la Bogomolny,
\begin{equation}
E={4m^3\over\lambda\sqrt{2}}\int \, dx \, {1\over
2}\left[\sum_{a=1}^2\left(\frac{d\phi_a}{dx}\mp\frac{\partial
W}{\partial\phi_a}\right)\left(\frac{d\phi_a}{dx}\mp\frac{\partial
W}{\partial\phi_a}\right)\right] \pm {4m^3\over\lambda\sqrt{2}}\cdot
\left\{W(\phi_a(+\infty))-W(\phi_a(-\infty))\right\} \label{kebb}
\end{equation}
shows that the solutions of the first-order ODE system
\begin{equation}
{d\phi_a\over dx}=\pm\frac{\partial W}{\partial\phi_a}\quad \equiv
\quad \left\{\begin{array}{c} \qquad
\displaystyle\frac{d\phi_1}{dx}=(-1)^\alpha
(2\phi_1^2+\sigma\phi_2^2-{1\over 2}) \\
\hspace{-1cm}\displaystyle\frac{d\phi_2}{dx}=(-1)^\beta
2\sigma\phi_1\phi_2
\end{array}\right. \quad , \quad \alpha,\beta=0,1 \label{kffoe}
\end{equation}
are absolute minima of the energy.

In \cite{SV}, the kink orbits solving this system of equations were
identified for the first time. We found the flow lines of ${{\rm
grad}\, W}$ between the top and the bottom of $W$, the kink orbits,
in \cite{AMAJ} simply by taking the quotient of the first equation
by the second in (\ref{kffoe}) to find:
\[
(-1)^\alpha\frac{d\phi_1}{2\phi_1^2+\sigma\phi_2^2-{1\over
2}}+(-1)^\beta\frac{d\phi_2}{2\sigma\phi_1\phi_2}=0 \qquad .
\]
Integration of this first-order equation is achieved using the
integrating factor $|\phi_2|^{-{2\over\sigma}}$. The kink orbits
are:
\[
\phi_1^2 + \frac{\sigma}{2(1-\sigma)} \phi_2^2 = \frac{1}{4}+
\frac{c}{2\sigma} |\phi_2|^{\frac{2}{\sigma}} \hspace{1cm} , \,
\sigma\neq 1 \, , \hspace{1cm} c\in (-\infty, c^s=
\frac{1}{4}\frac{\sigma}{1-\sigma}(2\sigma)^{\frac{1+\sigma}{\sigma}})
\qquad .
\]
Note that the integration constant $c$ must be lower than the
critical integration function $c^s$. The reason for this is easy to
understand just by looking at Figure 5. The $c^s$ kink orbit joins
one horizontal vacuum with the other two vertical vacua. Beyond that
value the orbits escape to infinity giving infinite energy to the
associated static-field theoretical solutions.

\begin{figure}[htbp]
\centerline{\includegraphics[height=3.5cm]{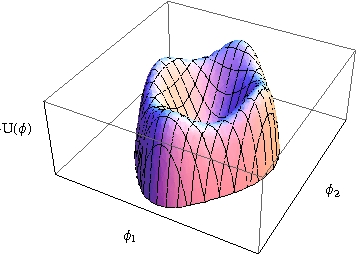}\hspace{1cm}
\includegraphics[height=3cm]{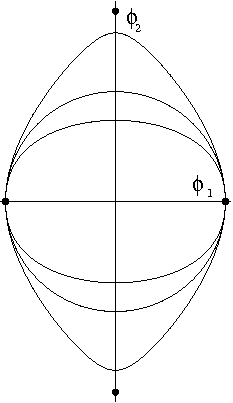}
\hspace{1cm}
\includegraphics[height=3cm]{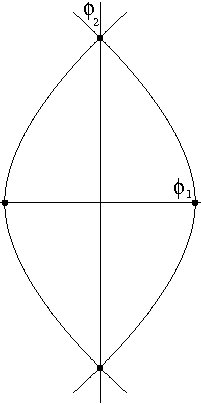} \hspace{1cm}
\includegraphics[height=3cm]{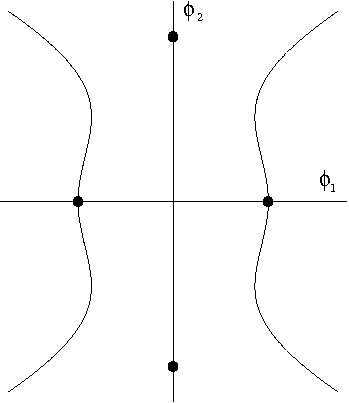}} \caption{3D graph of $-V(\phi_1,\phi_2)$ as a function of $\phi_1$ and $\phi_2$
(left) Flow-lines: in the ranges $c \in (-\infty,c^s)$ (middle
left), $c=c^s$ (middle right), and $c \in (c^s ,\infty)$ (right)}
\end{figure}
The energy of all these TK2 kink orbits is the same:
\[
E(\phi_1^K,\phi_2^K)=\frac{4m^3}{\lambda\sqrt{2}}|W(\pm
\frac{2}{2},0)-W(\mp\frac{1}{2},0)| =\frac{4m^3}{3\lambda\sqrt{2}}
\qquad .
\]
The zero-energy fluctuations orthogonal to the TK1 kink orbit obeys
 this fact: there is no cost in energy fluctuating from one kink in
this family to another.

For generic $\sigma$, it is not possible go further than this: i.e.,
knowledge of the TK2 kink orbits and their energy. There are two
exceptions, however. If $c=-\infty$, we again  find the TK1 kink
profile:
\[
\phi_1^{\rm 1K}(x)=(-1)^{\alpha}{1\over 2}{\rm tanh}(x-x_0)\\
\quad , \quad \phi_2^{\rm 1K}(x)=0 \qquad , \qquad \alpha=0,1 \qquad
.
\]
For $c=0$, the profiles of the corresponding TK2 kinks are also
 known analytically:
\[
\phi_1^{\rm 2K}(x)=(-1)^{\alpha_1}{1\over 2}{\rm
tanh}[2(1-\sigma)(x-x_0)] \quad , \quad \phi_2^{\rm
2K}(x)=(-1)^{\alpha_2}\sqrt{{1-\sigma\over \sigma}}{\rm
sech}[2(1-\sigma)(x-x_0)] \qquad ,
\]
where $\alpha_1,\alpha_2=0,1$. The two kink orbits are
half-ellipses.

The generic form of the kink profiles can be inferred from the
$\sigma=\frac{1}{2}$ case. Equations (\ref{kffoe}) are separable in
parabolic coordinates if $\sigma=\frac{1}{2}$, and we can give the
analytic expressions of the kink profiles \cite{AMAJ}:
\begin{eqnarray*}
\phi_1^{{\rm 2K}(d)} [x;x_0,d]&=&(-1)^{{\alpha_1}}\left(\frac{1}{2}
\frac{\sinh\left(
 (x-x_0)\right)}{\cosh\left( (x-x_0)\right)+d^2}\right)\\ \phi_2^{{\rm 2K}(d)}
[x;x_0,d]&=&(-1)^{{\alpha_2}}\left(\frac{d}{\sqrt{d^2+\cosh\left((x-x_0)\right)}}\right)
\qquad , \qquad d=\pm\sqrt{\frac{1}{\sqrt{1-4c}}} \qquad .
\end{eqnarray*}
There are two integration constants that are the parameters of this
TK2 family with a clear physical meaning: $x_0$ sets the kink center
as in TK1 kinks. The parameter $d$, related to the kink orbit label
$c$, can be loosely interpreted as the relative coordinate between
two basic kinks.
\begin{figure}[htbp]
\centerline{ \includegraphics[height=2.3cm]{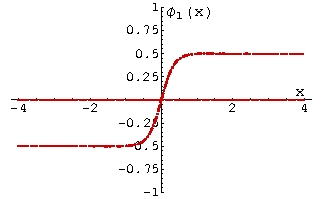}
\includegraphics[height=2.3cm]{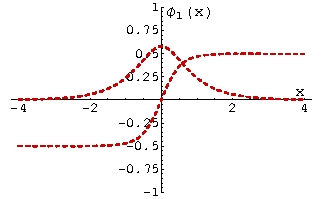} \includegraphics[height=2.3cm]{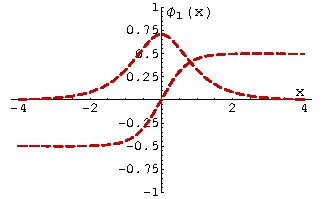}
\includegraphics[height=2.3cm]{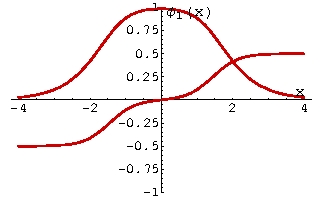}} \caption{Kink profiles corresponding to: {\it (a)} $d=0$,
{\it (b)} $d=\sqrt{0.5}$, {\it (c)} $d=1$ and {\it (d)}
$d=\sqrt{30}$.}
\end{figure}

In Figure 6, the kink profiles for four values of $d$ are shown. If
$d=0$ only one component is kink-shaped and non-null. For $d>0$, all
the profiles are two-component; one of the components is kink-shaped
and the other bell-shaped up to $d=1$. If $d>1$ one component has
the shape of two kinks and the other one tends to show two lumps.
\begin{figure}[htbp]
\centerline{ \includegraphics[height=2.3cm]{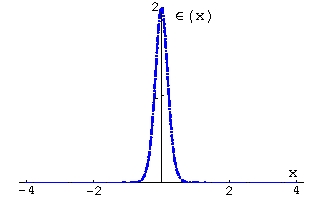}
\includegraphics[height=2.3cm]{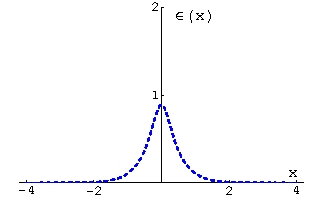}
\includegraphics[height=2.3cm]{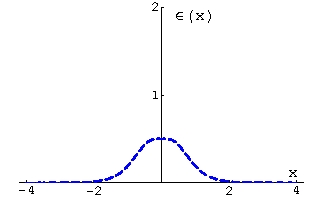} \includegraphics[height=2.3cm]{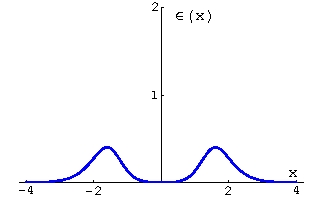}}
\caption{Energy density ${\cal E}^K[x;0,d]$ for {\it (a)}
$d=0$, {\it (b)} $d=\sqrt{0.5}$, {\it (c)} $d=1$ and {\it (d)}
$d=\sqrt{30}$.}
\end{figure}
This Figure is correlated with Figure 7, where the energy density is
plotted for the same values of $d$. It is possible to prove
analytically that the energy density as a function of $x$ has only
one maximum if $d\leq 1$ and two maxima if $d>1$. Thus, for $d>1$
this parameter is a relative coordinate between the two peaks,
although the distance between maxima is a transcendent function of
$d$. The meaning of the peaks is also clear: these values of $d$
correspond to orbits close to the critical orbit where two TK2 kinks
joining the horizontal vacuum with different vertical vacua live.
These are the basic kinks. All this is explained in detail in
Reference \cite{AMAJ}.

For generic values of $\sigma$, one must rely on numerical
integration methods to find the profiles of the kink solutions. We
do this by solving the first-order equations by standard numerical
methods with the \lq\lq initial" conditions:
\[
\phi_1(0)=0 \qquad , \qquad \frac{\sigma}{2(1-\sigma)}\phi_2^2(0)
-\frac{c}{2\sigma}|\phi_2(0)|^{\frac{2}{\sigma}}=\frac{1}{4} \qquad
.
\]
The reasons for this choice are two fold: (1) For any kink solution,
$\phi_1(x)$ always has a zero. Translational invariance allows us to
set the zero at $x=0$; (2) To ensure that we will find a numerical
kink solution, we fix $\phi_2(0)$ on a kink orbit for a given value
of $\sigma$ and arbitrary choices of $c$.

The numerical method provides us with a succession of points of the
kink solution generated by an interpolation polynomial. The plots of
the numerical results show that the behavior derived analytically
for the kink profiles when $\sigma={1\over 2}$ is generic. For any
value of $\sigma$, the kink profiles are composed of two kinks. The
parameter $c$ giving the orbit is related to the separation between
basic kinks. In some range of $c$, the two basic kinks completely
melt into a composite kink. The precise value of $c$ at which this
happens depends on the value of $\sigma$. $\sigma={1\over 2}$ is
singled out, because in this case $c=0$ is the value where two kinks
fuse into a composite kink or, viceversa, a composite kink splits
into two basic kinks.

\subsubsection{ TK2 kink small fluctuations}

We consider now small fluctuations of TK2 kinks:
$\phi_a(x)=\bar{\phi}_a(x;c)+\eta_a(x)$, where we denote the TK2
kink profiles in the form  $\phi_a^{\rm
TK2(c)}(x)=\bar{\phi}_a(x;c)$. The second-order kink fluctuation
operator is a non-diagonal Schr$\ddot{\rm o}$dinger operator:
\begin{eqnarray*}
K(c)&=&\left(\begin{array}{cc} -\frac{d^2}{dx^2}+4-U_{11}(x;c) &
U_{12}(x;c)\\ U_{21}(x;c) &
-\frac{d^2}{dx^2}+4-U_{22}(x;c)\end{array}\right) \quad , \quad
\\
U_{11}(x;c)&=&-(24\bar{\phi}_1^2(x;c)+4\sigma(\sigma+1)\bar{\phi}_2^2(x;c)-6)\\
U_{12}(x;c)&=&U_{21}(x;c)=-8\sigma(\sigma+1)\bar{\phi}_1(x;c)\bar{\phi}_2(x;c)
\\
U_{22}(x;c)&=&-(4\sigma(\sigma+1)\bar{\phi}_1^2(x;c)+6\sigma^2\bar{\phi}_2^2(x;c)-\sigma(\sigma+1))\qquad
.
\end{eqnarray*}
In Figure 8 the potential wells (or barriers) in the diagonal
components of $K$ are shown for the integrable case
$\sigma=\frac{1}{2}$ for several values of $c$. The correlation with
the kink profiles is clear. Therefore, the non-integrable cases have
similar wells because the kink profiles obtained numerically are
similar to the analytic ones.
\begin{figure}[htbp] \centerline{
\includegraphics[height=3cm]{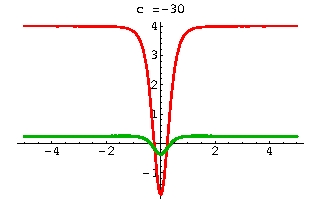}\quad
\includegraphics[height=3cm]{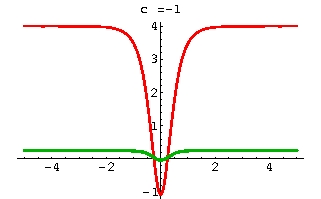}}
\centerline{\includegraphics[height=3cm]{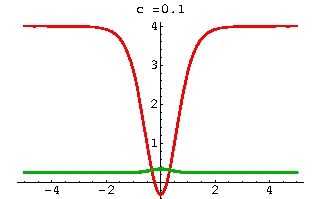}\quad
\includegraphics[height=3cm]{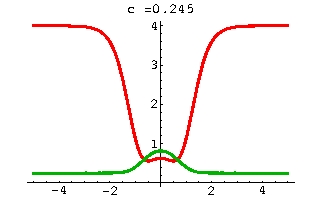}\quad
\includegraphics[height=3cm]{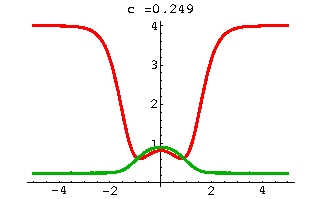}} \caption{ Diagonal components $-U_{11}(x)$ (red) and
$-U_{22}(x)$ (green) of the $\sigma=\frac{1}{2}$ potential for
$c=-30, c=-1, c=0.1, c=0.245$ and $c=0.249$.}
\end{figure}

The one-loop mass shifts can be computed using the high-temperature
formula.  We use the first-order equations to express field
derivatives as functions of the fields (avoiding problems with the
discreteness of space required in numerical methods) and write the
Seeley densities. Then, we plug the numerically generated kink
profiles into these densities and perform numerical integration over
the real line, to finally find the coefficients that enter the
formula.

The results are shown in the following Tables for three values of
$\sigma$: $1.5$, $2$, and $2.5$, and represented in Figure 9.
\newpage
\begin{table}[h]
{\tiny
\hspace{1cm}{\begin{tabular}{cc} $\sigma=1.5$ & \\ \hline $c$ & $\Delta M$ \\
\hline $-30$ & $-1.16009$ \\ $-27.5$ & $-1.16017$ \\ $-25$ &
$-1.16128$ \\ $-22.5$ & $-1.16042$ \\ $-20$ & $-1.16061$ \\
$-17.5$ & $-1.16088$ \\ $-15$ & $-1.16128$ \\ $-12.5$ & $-1.16193$
\\ $-10$ & $-1.16313$ \\ $-7.5$ & $-1.16597$ \\ $-5$  & $-1.18205$
\\ $-4.6801886$ & $-1.24345$ \\ $-4.68018860186678332$ &
$-1.25103$ \\ \hline
 & \\
 & \\
\end{tabular}} \hspace{1.cm} {\begin{tabular}{cc} $\sigma=2.0$ & \\ \hline $c$ & $\Delta M$ \\
\hline $-30$ & $-1.33281$ \\ $-27.5$ & $-1.33281$ \\ $-25$ &
$-1.33281$ \\ $-22.5$ & $-1.33281$ \\ $-20$ & $-1.33281$ \\
$-17.5$ & $-1.33281$ \\ $-15$ & $-1.33281$ \\ $-12.5$ & $-1.33281$
\\ $-10$ & $-1.33281$ \\ $-7.5$ & $-1.33281$ \\ $-5$  & $-1.33280$
\\ $-4.001$  & $-1.33280$ \\ $-4.00001$  & $-1.33280$ \\ \hline &
\\ & \\
\end{tabular}}
\hspace{1.cm} {\begin{tabular}{cc} $\sigma=2.5$ & \\ \hline
 $c$ & $\Delta M$ \\ \hline
$-30$ & $-1.52784$ \\ $-27.5$ & $-1.52782$ \\ $-25$ & $-1.52780$
\\ $-22.5$ & $-1.52778$ \\ $-20$ & $-1.52774$ \\ $-17.5$ &
$-1.52769$
\\ $-15$ & $-1.52760$ \\ $-12.5$ & $-1.52744$ \\ $-10$ &
$-1.52711$ \\ $-7.5$ & $-1.52626$ \\ $-5$  & $-1.52285$ \\ $-4$  &
$-1.52168$ \\ $-3.97$ & $-1.52915$ \\ $-3.96594571$ & $-1.55402$
\\ $-3.96594570565808127$ & $-1.56127$ \\ \hline
\end{tabular}}}
\caption{Quantum Mass Corrections to the TK2 family in the BNRT model with $\sigma=1.5,2$, and $2.5$}
\end{table}

\begin{figure}[h] \centerline{
\includegraphics[height=3.5cm]{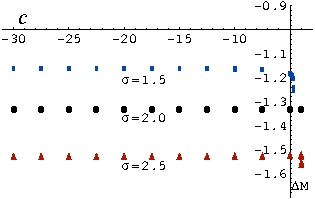}}
\caption{ The one-loop mass correction in the cases $\sigma=1.5$,
$\sigma=2.0$, and $\sigma=2.5$}
\end{figure}

In the $\sigma=1.5$ and $\sigma=2.5$ cases, the pattern is similar.
The one-loop correction is equal in the zone where the two basic
kinks are fused into a single kink and the degeneracy is not broken
at one-loop level. For values of $c$ in the zone where the two basic
kinks are split, the one-loop corrections become more and more
negative. The classical degeneracy is broken by quantum fluctuations
that press the two basic kinks apart from each other. This
conclusion could be reached by arguing qualitatively from Figure 8.
For small $c$, the potential wells both in the parallel and
orthogonal directions to the kink orbit are attractive. After $c=1$
the orthogonal wells start to become weakly repulsive. For larger
$c$, the parallel wells develop two peaks whereas the orthogonal
wells become strongly repulsive. This is the explanation of the
repulsion between the two basic kinks and the reason why the less
energetic TK2 kink corresponds to maximum separation of the basic
kinks, even though the classical energy is independent of the
distance between centers. The case $\sigma=2$ is special. The
classical degeneracy does not disappear because a simple change of
variables shows that this case is equivalent to two independent
$\lambda\phi^4$ models, such that the one-loop correction is twice
the correction of the $\lambda\phi^4$ kink.

\subsection{ N=3: the massive non-linear ${\mathbb S}^2$-sigma
model}

The last scalar field theoretical model that we shall discuss is the
massive non-linear ${\mathbb S}^2$-sigma model studied in Reference
\cite{AMAJ2}. This is the formal $\lambda\to\infty$ limit of the
deformed linear ${\mathbb O}(3)$-sigma model, a limit that is only
meaningful if the fields satisfy the constraint
\[
\chi_1^2(t,x)+\chi_2^2(t,x)+\chi_3^3(t,x)=R^2 \qquad, \qquad
R^2=\frac{m^2}{\lambda}
\]
such that the target space is a ${\mathbb S}^2$-sphere of radius
$R$. The action becomes
\[
S[\chi_1,\chi_2,\chi_3]={1\over 2}\int \, dx^2 \,
\left\{\sum_{a=1}^3\frac{\partial\chi_a}{\partial
x_\mu}\cdot\frac{\partial\chi_a}{\partial
x^\mu}-\alpha_1^2\chi_1^2(t,x)-\alpha_2^2\chi_2^2(t,x)-\alpha_3^2\chi_3^2(t,x)\right\}\quad
,
\]
where $\alpha_1^2=2\sigma_1^2 \, \, , \, \, \alpha_2^2=2\sigma_2^2
\, \,  , \, \,  \alpha_3^2=2\sigma_3^2$.

Despite appearances this is a highly non-linear system due to the
constraint between the fields. This statement is made evident by
solving $\chi_3$ in terms of $\chi_1$ and $\chi_2$. The action
becomes:
\[
S={1\over 2}\int \, dtdx \,
\left\{\partial_\mu\chi_1\partial^\mu\chi_1+\partial_\mu\chi_2\partial^\mu\chi_2
+ \frac{(\chi_1\partial_\mu\chi_1+\chi_2\partial_\mu\chi_2
)^2}{R^2-\chi_1^2-\chi_2^2}
-\chi_1^2(t,x)-\sigma^2\cdot\chi_2^2(t,x)\right\} \qquad .
\]
The vacua are the North and South poles: $\chi_3=\pm 1$. The
parameter
$0<\sigma^2=\frac{\sigma_2^2-\sigma_3^2}{\sigma_1^2-\sigma_3^2}< 1$
is the mass of the pseudo-Nambu-Goldstone boson, the quantum of the
$\chi_2$-field. The mass of the other pseudo-Nambu-Goldstone boson
is 1.

Interactions, however, come from the geometry:
\begin{eqnarray*}
&&\frac{(\chi_1\partial_\mu\chi_1+\chi_2\partial_\mu\chi_2
)(\chi_1\partial^\mu\chi_1+\chi_2\partial^\mu\chi_2
)}{R^2-\chi_1^2-\chi_2^2}\simeq \\
&\simeq& {1\over R^2}\left(1+{1\over R^2}(\chi_1^2+\chi_2^2)+{1\over
R^4}(\chi_1^2+\chi_2)^2+\cdots\right)\cdot\left(
\chi_1\partial_\mu\chi_1+\chi_2\partial_\mu\chi_2\right)\left(
\chi_1\partial^\mu\chi_1+\chi_2\partial^\mu\chi_2\right)
\end{eqnarray*}
which shows that $\frac{1}{R^2}$ is a non-dimensional coupling
constant.

\subsubsection{TK1 topological ${\mathbb S}^2$-kinks}

Using spherical coordinates in the target space
\[
\chi_1(t,x)=R\sin\theta(t,x)\cos\varphi(t,x)\quad , \quad
\chi_2(t,x)=R\sin\theta(t,x)\sin\varphi(t,x)\quad , \quad
\chi_3(t,x)=R\cos\theta(t,x) \, \, ,
\]
the field equations become:
\[
\Box \theta-{1\over 2}{\rm sin}2\theta\left(\partial^\mu\varphi
\partial_\mu\varphi-\cos^2\varphi
-\sigma^2\sin^2\varphi\right)= 0\quad , \quad  \partial^\mu
(\sin^2\theta
\partial_\mu\varphi)-{1\over 2}\bar{\sigma}^2\sin^2\theta \sin
2\varphi= 0 \quad .
\]
On the half-meridians $\varphi=\frac{\pi}{2}$,
$\varphi=\frac{3\pi}{2}$ and $\varphi=0$, $\varphi=\pi$ the field
equations reduce to the sine-Gordon equation. Therefore, there are
two types of two-component topological kinks joining the North and
South poles:
\begin{enumerate}
\item $K1$ kinks
\begin{eqnarray*}
\varphi_{K_1}(x)&=&{\pi\over 2} \, \, , \, \,
\varphi_{K_1^*}(x)={3\pi\over 2} \quad ; \quad
\theta_{K_1}(x)=2\arctan \, e^{\pm \sigma(x-x_0)}\\
\chi_1^{K_1}(x)&=&0 \quad , \quad \chi_2^{K_1}(x)=\pm\frac{R}{
\cosh[\sigma(x-x_0)]}\quad , \quad \chi_3^{K_1}(x)=\pm R \, \tanh
[\sigma(x-x_0)]
\end{eqnarray*}

\item $K2$ kinks
\begin{eqnarray*}
\varphi_{K_2}(x)&=&0 \, \, , \, \, \varphi_{K_1^*}(x)=\pi \quad ;
\quad \theta_{K_2}(x)=2\arctan \, e^{\pm (x-x_0)}\\
\chi_1^{K_2}(x)&=&\pm\frac{R}{ \cosh[(x-x_0)]}\quad , \quad
\chi_2^{K_2}(x)=0 \quad , \quad \chi_3^{K_2}(x)=\pm R \, \tanh
[(x-x_0)]
\end{eqnarray*}
\end{enumerate}
\begin{figure}[htbp]
\centerline{\includegraphics[height=3.5cm]{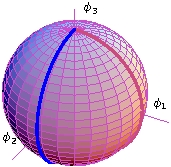}
\quad \qquad
\includegraphics[height=3.5cm]{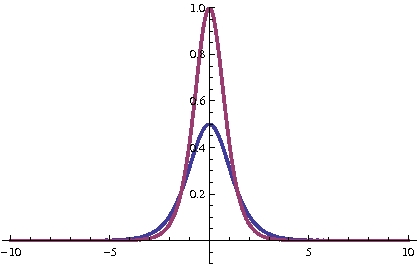}}\caption{a) $K_1$ and $K_2$
($\sigma^2=\frac{1}{2}$) kink orbits. b) $K_1$ (blue) and $K_2$
(red) kink energy densities}
\end{figure}
In Figure 10 the topological kink orbits are plotted, together with
the corresponding Kink energy densities. The classical kink energies
are:
\[
E_{K_1}^C= m \sqrt{2} R^2 \sigma \quad \qquad , \quad \qquad
E_{K_2}^C= m \sqrt{2} R^2
\]
and the $K_2$ kinks are heavier than the $K_1$-kinks.
\subsubsection{Kink fluctuations and one-loop mass
shifts}

The geodesic deviation from the $K_1$ kink orbit plus the Hessian of
the potential for the $K_1$ kink (in a parallel frame to the kink
orbit) reads, see \cite{AMAJ3}:
\[
K=\left(\begin{array}{cc}
-\frac{d^2}{dx^2}+\sigma^2-\frac{2\sigma^2}{{\rm cosh}^2\sigma x} &
0
\\ 0 & -\frac{d^2}{dx^2}+1-\frac{2\sigma^2}{{\rm cosh}^2\sigma
x}\end{array}\right)\qquad .
\]
As might be expected, for parallel fluctuations to the orbit we find
the same Schr$\ddot{\rm o}$dinger operator as for the sine-Gordon
kink. What comes as a surprise is that the orthogonal fluctuations
are also governed by an operator, which is of the transparent
P$\ddot{\rm o}$sch-Teller type of the $N=1$ class with slightly
different threshold.

The one-loop $K_1$ mass shift is immediately obtained from the
Cahill-Comtet-Glauber formula:
\begin{eqnarray*}
E_{K_1}^{OL}(\sigma)&=& E_{K_1}(\sigma)+\Delta M_{K_1}(\sigma)+{\cal
O}(\frac{\hbar^2}{R^2}) \\&=&\sqrt{2}m R^2\sigma-\frac{\hbar
m\sigma}{\sqrt{2}\pi}
[\sin\nu_1+{1\over\sigma}\sin\nu_2-\nu_1\cos\nu_1 -
{1\over\sigma}\nu_2\cos\nu_2]+{\cal O}(\frac{\hbar^2}{R^2}) \qquad .
\end{eqnarray*}
Here, $\nu_1= \arccos(\frac{0}{\sigma})={\pi\over 2}$ and
$\nu_2=\arccos (\frac{\bar{\sigma}}{1})$. The arguments are the
square root of the value of the bound state eigenvalue
$\bar{\sigma}^2=1-\sigma^2$ divided by the threshold of the
continuous spectrum. For instance, as a function of $\sigma$ the
correction reads:
\[ E_{K_1}^{OL}(\sigma) =\sqrt{2}m R^2\sigma-\frac{\hbar
m\sigma}{\sqrt{2}\pi}\left[
2-\frac{\bar{\sigma}}{\sigma}\arccos(\bar{\sigma}) \right]+{\cal
O}(\frac{\hbar^2}{R^2})
\]
whereas for $\sigma=\frac{1}{2}$ we obtain :
\[
E_{K_1}^{OL}(\frac{1}{2})=\frac{m}{\sqrt{2}} R^2-\frac{3\hbar
m}{2\sqrt{2}\pi}\left(\frac{2}{3}-\frac{\pi}{6\sqrt{3}}\right)+{\cal
O}(\frac{\hbar^2}{R^2}) \qquad .
\]

The second-order fluctuation operator for the $K_2$ kink (in a
 parallel frame to the kink orbit) is:
\[
K=\left(\begin{array}{cc} -\frac{d^2}{dx^2}+1-\frac{2}{{\rm cosh}^2
x} & 0
\\ 0 & -\frac{d^2}{dx^2}+\sigma^2-\frac{2}{{\rm cosh}^2
x}\end{array}\right) \qquad .
\]

The bound state eigenvalue $-\bar{\sigma}^2$ in the orthogonal
direction to the orbit is negative, telling us that the $K_2$ kinks
are unstable. The CCG formula applied to calculate the one-loop
$K_2$ mass shift captures this fact, because $\nu_2$ ceases to be an
angle and becomes ${\rm arccosh}\bar{\sigma}$:
\begin{eqnarray*}
E_{K_2}^{OL}(\sigma)&=& E_{K_2}(\sigma)+\Delta M_{K_2}(\sigma)+{\cal
O}(\frac{\hbar^2}{R^2})\qquad ; \qquad \nu_1= \arccos(0)={\pi\over
2} \, \, \, ,  \, \, \, \nu_2=\arccos (i\bar{\sigma})\\&=&\sqrt{2}m
R^2-\frac{\hbar m\sigma}{\sqrt{2}\pi}
[\sin\nu_1+{1\over\sigma}\sin\nu_2-\nu_1\cos\nu_1 -
{1\over\sigma}\nu_2\cos\nu_2]+{\cal O}(\frac{\hbar}{R^2}) \qquad .
\end{eqnarray*}
Therefore,
\[
E_{K_2}^{OL}(\sigma) =\sqrt{2}m R^2-\frac{\hbar
m\sigma}{\sqrt{2}\pi}\left[
\frac{1}{\sigma}+\sqrt{2-\sigma^2}-i\frac{\pi}{2}\bar\sigma+\bar\sigma\log\left(\sqrt{2-\sigma^2}-\bar\sigma\right)
\right]+{\cal O}(\frac{\hbar^2}{R^2})
\]
has an imaginary part, as corresponds to the energy of a resonant
state in quantum physics. We end this first part by computing:
\[
E_{K_2}^{OL}(\frac{1}{2})=\frac{m}{\sqrt{2}} R^2-\frac{\hbar
m}{4\sqrt{2}\pi}\left(4+\sqrt{7}+\sqrt{3}\log\left(\frac{\sqrt{7}-\sqrt{3}}{2}\right)-i\frac{\pi\sqrt{3}}{2}\right)
+{\cal O}(\frac{\hbar^2}{R^2}) \qquad .
\]
\newpage
\section{Abelian Higgs models in a plane}
The second part will be devoted to discussing the quantum
fluctuations of the topological solitons that arise in gauge
theoretical models in $(2+1)$-dimensions with spontaneous breaking
of the gauge symmetry. The prototype of these solitons is the
Abrikosov-Nielsen-Olesen vortex, see \cite{AAA}, \cite{NO}. Because
the spatial components of the gauge field form a purely vorticial
vector field, the planar soliton is made from lumps of magnetic
field. By extending these objects to the third spatial dimension
magnetic flux lines arise like those existing in Type II
superconductors, predicted by the phenomenological Ginzburg-Landau
theory. In the relativistic Abelian Higgs model the Nielsen-Olesen
vortices form strings that might confine magnetic monopoles. The
one-loop mass shift due to quantum fluctuations of ANO vortices has
been computed at the critical point between Type I and Type II
superconductors in \cite{AMJW}, \cite{AMJW1}, and \cite{AMJW2} using
heat kernernel/zeta function regularization methods.

It was discovered in \cite{AV} that a generalization of the Higgs
model, the so called semi-local Abelian Higgs model containing a
doublet of complex scalar fields, also admits planar topological
solitons similar to the ANO vortices. The moduli space of these
solitons is, however, richer, encompassing both ${\mathbb
CP}^1$-lumps and ANO vortices as limiting cases with very
complicated mixtures of these two classes in between, see
\cite{GORS}. The one-loop fluctuations of self-dual semi-local
topological solitons have been analyzed in \cite{AMJW3} and
\cite{AMJW4}, with some surprising results. Our aim in this part is
the description of these results. We remark that the semi-local
Abelian Higgs model is the bosonic sector of the electro-weak theory
when the Weinberg angle is $\frac{\pi}{2}$ and the $Z$, $W^\pm$
gauge fields decouple.

\subsection{The planar semi-local Abelian Higgs model}

\subsubsection{Action and field equations}
In the semi-local Abelian Higgs model there is a Higgs doublet
\[
\Phi(x^\mu)=\left(\begin{array}{c}\Phi_1(x^\mu)\\
\Phi_2(x^\mu)\end{array}\right)=
\left(\begin{array}{c}\phi_1(x^\mu)+i\phi_2(x^\mu)\\
\phi_3(x^\mu)+i\phi_4(x^\mu)\end{array}\right) \,\, : \quad {\mathbb
R}^{1,2} \longrightarrow \quad {\mathbb C}^2
\]
and an Abelian gauge field:
\[
A_\mu(x^\mu)\frac{\partial}{\partial x^\mu}: \quad  T{\mathbb
R}^{1,2}\longrightarrow \quad Lie {\mathbb U}(1) \qquad .
\]
The action is built from the electromagnetic tensor, and
interactions between the gauge and scalar fields determined from a
minimal coupling principle by means of the covariant derivative and
self-interactions of the scalar field are induced by a quartic
potential energy density:
\[
F_{\mu\nu}=\partial_\mu A_\nu-\partial_\nu A_\mu \qquad , \qquad
D_\mu \Phi=\partial_\mu \Phi- i A_\mu \Phi \qquad , \qquad
U(\Phi)=\frac{\lambda}{8 e^2} (\Phi^\dagger \Phi-1)^2 \qquad .
\]
The action governing the dynamics of the system is:
\[
S= \frac{v}{e}\int dx^3 \left\{ -\frac{1}{4} F_{\mu \nu} F^{\mu
\nu}+\frac{1}{2} (D_\mu \Phi)^\dagger D^\mu \Phi - \frac{\kappa}{8}
(\Phi^\dagger \Phi-1)^2  \right\} \qquad .
\]
There are the following (dimensional) parameters: $[v^2]=M$, the
vacuum expectation value of the Higgs field, the scalar-vector and
scalar-scalar couplings, $[e^2]=[\lambda]=M^{-1}L^{-2}$, and the
Higgs and vector particle masses: \hspace{-0.2cm} $m^2=e^2v^2 $,
$\mu^2=\lambda v^2$. The ratio between masses
$\kappa^2=\frac{\lambda}{e^2}$ is a very important parameter that
determines whether we are in a Type I or Type II
superconductivity-like regime. Needless to say, we continue working
with non-dimensional fields, $\Phi$, $A_\mu$, and Minkowski
coordinates $x_\mu$, $\mu=0,1,2$. The metric and volume integration
are chosen such that:
\begin{eqnarray*}
g_{\mu\nu}={\rm diag}(1,-1,-1) \qquad &,& \qquad dx^3=dx_0dx_1dx_2
\\  x_\mu
x^\mu=g_{\mu\nu}x^\mu x^\nu=x_0^2-x_1^2-x_2^2 \qquad &,& \qquad
\partial_\mu\partial^\mu=g_{\mu\nu}\partial^\mu\partial^\nu=\frac{\partial^2}{\partial
x_0^2}-\frac{\partial^2}{\partial x_1^2}-\frac{\partial^2}{\partial
x_2^2} \qquad .
\end{eqnarray*}
The field equations are:
\[
\partial_\mu
F^{\mu\nu}=i\left[(D^\nu\Phi)^\dagger\Phi-\Phi^\dagger
D^\nu\Phi\right]\qquad , \qquad D_\mu D^\mu\Phi
=\frac{\kappa^2}{4}\Phi(1-\Phi^\dagger \Phi) \qquad .
\]
\subsubsection{Global and local symmetries: vacuum orbit structure}
The ${\mathbb S}{\mathbb U}(2)$ global weak iso-spin transformations
\[
\Phi(x^\mu)\longrightarrow \Phi^\prime(x^\mu)=
\exp(-\frac{i}{2}\vec{\theta}\cdot\vec{\sigma}) \Phi(x^\mu)
\]
as well as the ${\mathbb U}(1)$-gauge local transformations
\[
\Phi(x^\mu)\longrightarrow \Phi^\prime(x^\mu)=e^{i\alpha
(x^\mu)}\Phi(x^\mu) \hspace{1cm} , \hspace{1cm}
A_\mu(x^\mu)\longrightarrow
A^\prime_\mu(x^\mu)=A_\mu(x^\mu)+\frac{\partial\alpha}{\partial
x^\mu}(x^\mu)
\]
are symmetries of this system suggesting the slightly deceptive name
for this model. The vacuum orbit has a very subtle structure due to
the combined action of these two symmetries.

The manifold of zero energy configurations, $\Phi^V(x^\mu)=\Phi^V$,
$A_\mu^V(x^\mu)=0_\mu$, is:
\[
(\Phi^V)^\dagger \Phi^V=(\Phi_1^V)^*\Phi_1^V+(\Phi_2^V)^*\Phi_2^V
=1=(\phi_1^V)^2+(\phi_2^V)^2 +(\phi_3^V)^2+(\phi_4^V)^2
\]
The Higgs vacuum orbit is the ${\mathbb S}^3$  unit sphere in
${\mathbb C}^2$: the orbit of the point $\Phi^V_0
=\left(\begin{array}{c}1\\0
\end{array}\right)$, the north pole of the ${\mathbb S}^3$ sphere, under the global
${\mathbb S}{\mathbb U}(2)$ action, i.e., the Hopf fibre bundle:
\[
{\mathbb S}^1\, \, \, \longrightarrow \, \, \, {\mathbb S}^3 \, \,
\, \longrightarrow \, \, \, {\mathbb S}^2
\]
given by the action of a ${\mathbb U}(1)$ subgroup on each point of
the ${\mathbb S}^2$ sphere.

Use of the Hopf coordinates $\theta_1,\theta_2\in [0,2\pi]$,
$\psi\in[0,{\pi\over 2}]$ in the ${\mathbb S}^3$ sphere allows us to
write the ${\mathbb S}{\mathbb U}(2)$ action in the form:
\[
G\Phi^V_0=\left(\begin{array}{cc} e^{i\theta_1}{\rm sin}\psi &
e^{i\theta_2}{\rm cos}\psi \\ -e^{-i\theta_2}{\rm cos}\psi &
e^{-i\theta_1}{\rm sin}\psi
\end{array}\right)\left(\begin{array}{c} 1 \\ 0
\end{array}\right)=\left(\begin{array}{c} e^{i\theta_1}{\rm sin}\psi \\
-e^{-i\theta_2}{\rm cos}\psi  \end{array}\right)
\]
such that the Higgs vacuum orbit is parametrized as follows:
\[
\phi_1^V={\rm cos}\theta_1{\rm sin}\psi \quad , \quad \phi_2^V={\rm
sin}\theta_1{\rm sin}\psi \quad , \quad \phi_3^V=-{\rm
cos}\theta_2{\rm cos}\psi \quad , \quad \phi_4^V={\rm
sin}\theta_2{\rm cos}\psi \qquad .
\]
Let us consider the one form $\omega={1\over
\pi}\left(\phi_1^Vd\phi_2^V+\phi_3^Vd\phi_4^V\right)\in
\Omega^1({\mathbb S}^3)$. The Hopf index, labeling the homotopy
class of the third homotopy group of the 2-sphere $\Pi_3({\mathbb
S}^2)={\mathbb Z}$, is:
\[
h={1\over\pi^2}\int_{S^3}\, \omega \wedge d\omega
={2\over\pi^2}\int_{S^3}\, \phi_1^V d\phi_2^V\wedge d\phi_3^V\wedge
d\phi_4^V \qquad .
\]
Thus,
\[
\phi_1^V d\phi_2^V \wedge d\phi_3^V \wedge d\phi_4^V=-{\rm
cos}^2\theta_1{\rm sin}^3\psi {\rm cos}\psi d\psi\wedge
d\theta_1\wedge d\theta_2
\]
and the Hopf index of the vacuum orbit is:
\[
h={2\over\pi^2}\left[2\pi \cdot \int_0^{2\pi} \, d\theta_1 \, {\rm
cos}^2\theta_1 \cdot \int_0^{\pi\over 2} \, d\psi \, {\rm
sin}^3\psi{\rm cos}\psi\right]={2\over\pi^2}\left[2\pi \cdot \pi
\cdot {1\over 4} \right]=1 \qquad .
\]
\subsubsection{Higgs mechanism and Feynman rules}
The choice of a point in the vacuum orbit ${\mathbb S}^3$
spontaneously breaks the ${\mathbb S}{\mathbb U}(2)$-global symmetry
and one would expect three Goldstone bosons. One of the three
Goldstone bosons, however, will undergo the Higgs mechanism. We
shift the scalar field away from the vacuum in $H(x^\mu)$ -real
Higgs-, $G(x^\mu)$ -real Higgs ghost-, and $\varphi(x^\mu)$ -complex
Goldstone- fields:
\[
\Phi(x^\mu)=\left(\begin{array}{c} 1+H(x^\mu)+i G(x^\mu) \\
\sqrt{2}\varphi(x^\mu)\end{array}\right)\qquad .
\]
The choice of the Feynman-'t Hooft R-gauge
\[
R(A_\mu , G)=\partial_\mu A^\mu(x^\mu) + G(x^\mu) \qquad , \qquad
S_{{\rm g.f.}}=-{1\over 2}\int \, d^3x \, \left(\partial_\mu
A^\mu(x^\mu) + G(x^\mu)\right)^2
\]
needs a Faddeev-Popov determinant to restore unitarity which amounts
to introducing a complex ghost field:
\begin{eqnarray*}
&&R(A^\prime_\mu,G^\prime)\simeq
R(A_\mu,G)+\left(\Box-1-H(x^\mu)\right)\cdot
\delta\alpha(x^\mu)\\&&{\rm Det}\frac{\delta R}{\delta\alpha}=\int
[d\chi^*(x^\mu)][d\chi(x^\mu)] {\rm exp}\left\{i\int dx^3
\chi^*(x^\mu)\left(\Box-1-H(x^\mu)\right)\chi(x^\mu)\right\}\qquad .
\end{eqnarray*}
The action becomes:
\begin{eqnarray} S&=&{v\over e}\int \, d^3x \,
\left[ -\frac{1}{2} A_\mu
[-g^{\mu\nu}(\Box +1)]A_\nu +\partial_\mu\chi^*\partial^\mu \chi- \chi^*\chi\right.\nonumber\\
&+&\frac{1}{2}\partial_\mu G\partial^\mu G-\frac{1}{2} G^2+
\frac{1}{2}\partial_\mu H\partial^\mu H-\frac{\kappa^2}{2}  H^2+\partial_\mu\varphi^* \partial^\mu \varphi\nonumber\\
&-& {\kappa^2\over 2}H (H^2+G^2)+ A_\mu (\partial^\mu H
G-\partial^\mu G H)+H (A_\mu A^\mu -\chi^* \chi) +i A_\mu(\varphi^*
\partial^\mu \varphi-\varphi \partial^\mu \varphi^*)\nonumber\\&+& \left.
A_\mu A^\mu |\varphi|^2 -\frac{\kappa^2}{8} (H^2+G^2)^2
+\frac{1}{2}(G^2+H^2) A_\mu A^\mu
-\frac{\kappa^2}{2}|\varphi|^2(|\varphi|^2+H^2+G^2+2H)\right]\label{fpha}
\, .
\end{eqnarray}
The Feynman rules are summarized in the next two Tables.
\begin{table}[h]
\begin{center}
\begin{tabular}{lccc} \\ \hline
\textit{Particle} & \textit{Field} & \textit{Propagator} & \textit{Diagram} \\
\hline \\ Higgs & $H(x)$ & $\displaystyle\frac{i e \hbar
}{v(k^2-\kappa^2+i\varepsilon)}$ &
\parbox{3.0cm}{\includegraphics[width=3.2cm]{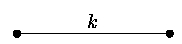}} \\[0.5cm]
Higgs Ghost & $G(x)$ & $\displaystyle\frac{ie
\hbar}{v(k^2-1+i\varepsilon)}$ &
\parbox{3.0cm}{\includegraphics[width=3.2cm]{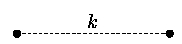}} \\[0.5cm]
Complex Goldstone & $\varphi(x)$ & $\displaystyle\frac{ie
\hbar}{v(k^2+i\varepsilon)}$ & \hspace{0.05cm}
\parbox{3.0cm}{\includegraphics[width=3.15cm]{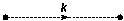}} \\[0.5cm]
Ghost & $\chi(x)$ & $\displaystyle\frac{ie
\hbar}{v(k^2-1+i\varepsilon)}$ &
\parbox{3.0cm}{\includegraphics[width=3.2cm]{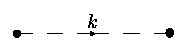}} \\[0.5cm]
Vector Boson & $A_\mu(x)$  &  {\Large $\frac{-ie \hbar g^{\mu\nu}
}{v(k^2-1+i\varepsilon)}$} &
\parbox{3.0cm}{\includegraphics[width=3.5cm]{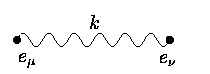}}
\\[0.5cm] \hline
\end{tabular}
\caption{Propagators}
\end{center}
\begin{center}
\begin{tabular}{llllllll} \\ \hline
\textit{Vertex} & \textit{Weight} & \textit{Vertex} &
\textit{Weight}
& \textit{Vertex} & \textit{Weight} & \textit{Vertex} & \textit{Weight} \\
\hline \\
\parbox{1.2cm}{\includegraphics[width=1.5cm]{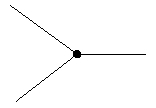}} &
$-3i\kappa^2\frac{v}{\hbar e} $  &
\parbox{1.2cm}{\includegraphics[width=1.2cm]{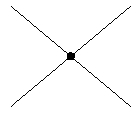}} &
$ -3i\kappa^2\frac{v}{\hbar e}$ &
\parbox{1.2cm}{\includegraphics[width=1.5cm]{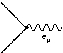}} &
$ i(k^\mu+q^\mu)\frac{v}{\hbar e} $  &
\parbox{1.5cm}{\includegraphics[width=1.0cm]{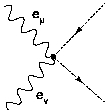}} &
$ 2 i g^{\mu\nu}\frac{v}{\hbar e}$ \\[0.5cm]
\parbox{1.2cm}{\includegraphics[width=1.5cm]{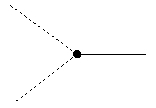}} &
$ -i\kappa^2\frac{v}{\hbar e}$  &
\parbox{1.2cm}{\includegraphics[width=1.2cm]{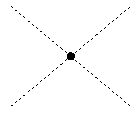}} &
$-3i\kappa^2\frac{v}{\hbar e}$ &
\parbox{1.2cm}{\includegraphics[width=1.5cm]{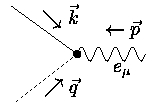}} & $
(k^\mu-q^\mu)\frac{v}{\hbar e}$  &
\parbox{1.2cm}{\includegraphics[width=1.5cm]{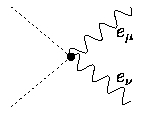}} &
$ 2 i \frac{v}{\hbar e} g^{\mu \nu}$ \\[0.5cm]
 \parbox{1.2cm}{\includegraphics[width=1.5cm]{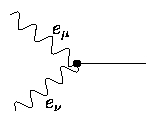}} &
$ 2i\frac{v}{\hbar e} g^{\mu \nu}$  &
\parbox{1.2cm}{\includegraphics[width=1.2cm]{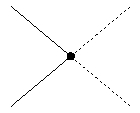}} &
$ -i\kappa^2\frac{v}{\hbar e}$ & \hspace{0.1cm}
\parbox{1.2cm}{\includegraphics[width=1.2cm]{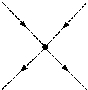}} &
$ -2i\kappa^2\frac{v}{\hbar e} $  &
\parbox{1.2cm}{\includegraphics[width=1.2cm]{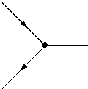}} &
$ -i\kappa^2\frac{v}{\hbar e}$
\\[0.5cm]
 \parbox{1.2cm}{\includegraphics[width=1.5cm]{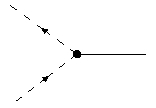}} &
$-i\frac{v}{\hbar e}$ &
\parbox{1.2cm}{\includegraphics[width=1.5cm]{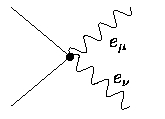}} &
$ 2 i\frac{v}{\hbar e}g^{\mu \nu}$ &\hspace{0.1cm}
\parbox{1.2cm}{\includegraphics[width=1.2cm]{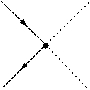}} &
$ -i \kappa^2\frac{v}{\hbar e}$ &
\parbox{1.2cm}{\includegraphics[width=1.2cm]{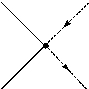}} &
$ -i \kappa^2\frac{v}{\hbar e}$  \\[0.8cm]\hline
\end{tabular}
\caption{Third- and fourth-valent vertices }
\end{center}
\end{table}
Table 4 gives the propagators in the $R$-gauge. There is a Higgs
field propagating with mass $\kappa^2$, a Higgs ghost of mass $1$,
and a vector particle with mass also $1$. A Faddeev-Popov
(anti-commuting) ghost of mass $1$ must be included to compensate
the non-physical Higgs ghost. Finally, there is a complex Goldstone
boson, as expected from this partial Higgs mechanism. The vertices
are read from the cubic and quartic terms in action (\ref{fpha}) and
shown in Table 5. There are accordingly three-valent and four-valent
vertices. The cubic terms with derivatives in (\ref{fpha}) provide
weights proportional not only to the coupling constants but also
dependent on the momenta.

\newpage
\subsection{Topological solitons}

The search for time-independent finite energy solutions requires the
use of the Weyl gauge: $A_0(x)=0$; otherwise, time-dependent gauge
transformations would spoil the time independence. We thus look for
solutions of
\begin{equation}
1)\, -\partial_0^2 A_i+ \partial_i F_{ij} = \frac{i}{2} \left(
\Phi^\dagger D_i \Phi - (D_i \Phi)^\dagger \, \Phi \right) \quad ,
\quad 2)\,  -\frac{1}{2} \partial_0^2 \Phi+\frac{1}{2} D_i D_i \Phi=
\frac{\kappa}{4}\Phi(\Phi^\dagger\Phi-1) \label{spde}
\end{equation}
such that their energy
\begin{equation}
E(\Phi,A_i)=\int \, d^2x \, \left\{{1\over 4}F_{ij}F_{ij}+{1\over
2}(D_i\Phi)^\dagger D_i\Phi+{\kappa\over
8}(\Phi^\dagger\Phi-1)^2\right\} \label{ssfe}
\end{equation}
is finite.
\subsubsection{Topology of the configuration space}

The configuration space
\[
{\cal C}=\left\{\Phi(\vec{x})\in Maps({\mathbb R}^2,{\mathbb
C}^2),A_i(\vec{x})\in Maps({\mathbb R}^2,T{\mathbb
R}^2)/E(\Phi,A_i)<+\infty \right\}
\]
is the set of all the field configurations of finite energy.
Consider polar coordinates in the plane: $r=+\sqrt{x_1^2+x_2^2}$,
$\theta={\rm arctan}\frac{x_2}{x_1}$. The equations
\begin{equation}
1) \, \, \,  \Phi^{\dagger} \Phi |_{{\mathbb S}_\infty^1} =1
\hspace{2cm} , \hspace{2cm} 2) \, \, \, D_i \Phi |_{{\mathbb
S}_\infty^1} =(\partial_i \Phi - i A_i \Phi)|_{{\mathbb
S}_\infty^1}=0 \label{abfc}
\end{equation}
are the necessary conditions to be satisfied by finite energy static
field configurations at the boundary of ${\mathbb R}^2$ at infinity:
${\mathbb S}^1_\infty \equiv
\left\{x_1,x_2/\lim_{r\rightarrow\infty}(x_1^2+x_2^2=r^2)\right\}\equiv\partial{\mathbb
R}^2$. Therefore, equation (\ref{abfc}(1)) determines a map from
${\mathbb S}^1_\infty$ in ${\mathbb S}^3$. If $\Phi_2|_{{\mathbb
S}^1_\infty}=0$
\[
\Phi|_{{\mathbb S}_\infty^1}=\left(\begin{array}{c}\Phi_1|_{{\mathbb S}_\infty^1}\\
\Phi_2|_{{\mathbb
S}_\infty^1}\end{array}\right)=\left(\begin{array}{c}\phi_1|_{{\mathbb
S}_\infty^1}
+i\phi_2|_{{\mathbb S}_\infty^1}\\
\phi_3|_{{\mathbb S}_\infty^1}+i\phi_4|_{{\mathbb
S}_\infty^1}\end{array}\right) = \left(\begin{array}{c}e^{il
\theta}\\0\end{array}\right)=\Phi_l^V \, \,\qquad , \, \, \qquad
l\in{\mathbb Z}
\]
are all the single-valued maps complying with (\ref{abfc}(1)). Acting
with the ${\mathbb S}{\mathbb U}(2)$ matrices parametrized by Hopf
coordinates we obtain the general solution by changing the base
point in ${\mathbb S}^2$ at which ${\mathbb S}^1$ is fibred:
\[
G\Phi_l^V=\left(\begin{array}{cc} e^{i\theta_1}{\rm sin}\psi &
e^{i\theta_2}{\rm cos}\psi \\ -e^{-i\theta_2}{\rm cos}\psi &
e^{-i\theta_1}{\rm sin}\psi
\end{array}\right)\left(\begin{array}{c} e^{il\theta} \\ 0
\end{array}\right)=\left(\begin{array}{c} e^{i(\theta_1+l\theta)}{\rm sin}\psi \\
-e^{-i(\theta_2-l\theta)}{\rm cos}\psi  \end{array}\right) \qquad .
\]
Equation (\ref{abfc}(2)) is consequently solved by:
\begin{eqnarray}
A_i |_{{\mathbb S}_\infty^1} &=&- i \Phi^\dagger\partial_i
\Phi|_{{\mathbb S}_\infty^1}
= l\frac{\partial\theta}{\partial x_i} \label{qmf}\\
&=&\phi_1(\vec{x})\frac{\partial\phi_2}{\partial x_i}(\vec{x})\left
|_{{\mathbb
S}^1_\infty}\right.-\phi_2(\vec{x})\frac{\partial\phi_1}{\partial
x_i}(\vec{x})\left |_{{\mathbb
S}^1_\infty}\right.+\phi_3(\vec{x})\frac{\partial\phi_4}{\partial
x_i}(\vec{x})\left |_{{\mathbb
S}^1_\infty}\right.-\phi_4(\vec{x})\frac{\partial\phi_3}{\partial
x_i}(\vec{x})\left |_{{\mathbb S}^1_\infty}\right. \nonumber \quad ,
\end{eqnarray}
showing that the vector field $A_i(x_1,x_2)$ is asymptotically
vorticial. The integer number $l$ has correlated mathematical and
physical meanings:
\begin{enumerate}
\item It is the winding number of the map provided by the Higgs field at
infinity:
\[
{\mathbb S}^1_\infty \quad  \longrightarrow \quad {\mathbb S}^1_1
\qquad , \qquad {\cal C}=\sqcup_{l\in{\mathbb Z}}{\cal C}_l \qquad ,
\]
where ${\mathbb S}^1_1$ is the fiber at the north pole of ${\mathbb
S}^2$.

\item It is the magnetic flux of the field configuration:
\[
g=\oint_{{\mathbb S}^1_\infty}\,
\left(A_1(\vec{x})dx^1+A_2(\vec{x})dx^2\right)=l\oint_{{\mathbb
S}^1_\infty}\, \left(\frac{\partial\theta}{\partial
x_1}dx^1+\frac{\partial\theta}{\partial
x_2}dx^2\right)=l\int_0^{2\pi}\, d\theta =2\pi l \qquad  .
\]
\end{enumerate}
Because the first homotopy group of a circle is non-trivial,
$\Pi_1({\mathbb S}^1_1)={\mathbb Z}$, the configuration space is the
union of numerable infinite disconnected sectors distinguished by
the integer $l$: ${\cal C}=\bigsqcup_l\, {\cal C}_l$.

\subsubsection{Self-dual semi-local topological solitons}
At the critical point between Type II and Type I superconductivity
where the masses of Higgs and vector particles are equal,
$\kappa^2=1$, it is possible to write the energy, up to a total
derivative, as a Bogomolny splitting:
\[
E= \int \frac{d^2 x}{2} \left( (D_1 \Phi \pm i D_2 \Phi)^\dagger
(D_1 \Phi \pm i D_2 \Phi) + [ F_{12} \pm {\textstyle\frac{1}{2}}
(\Phi^\dagger \Phi-1) ]^2 \right)+\frac{1}{2}|g| \qquad .
\]
The solutions of the first-order system of partial differential
equations (\ref{sdfopd})
\begin{equation}
D_1 \Phi \pm i D_2 \Phi=0 \hspace{1.5cm},\hspace{1.5cm} F_{12} \pm
\frac{1}{2} (\Phi^\dagger\Phi-1) =0 \label{sdfopd}
\end{equation}
are absolute minima of the energy and saturate the topological bound
proportional to the quantized magnetic flux in each topological
sector. Because the first-order vortex equations can be obtained as
a dimensional reduction of the self-duality equations of Euclidean
Yang-Mills theory in four dimensions, solutions of the PDE system
(\ref{sdfopd}) -that also solve the second order PDE system
(\ref{spde})- are usually called self-dual.

The structure of the moduli space of solutions of (\ref{sdfopd}) has
been completely unveiled in \cite{GORS}, see also \cite{AV1}. The
parameters underlying the $4 l$ dimensional moduli space of
topological solitons are the coordinates of the $l$ zeroes of
$\Phi_1$, the coordinates of the $l-1$ zeroes of $\Phi_2$, and the
scale and phase of $\Phi_2$. Full details can also be found in
\cite{AMJW3}.

\subsubsection{Self-dual semi-local topological solitons with
mixed circle-symmetry}

We shall restrict ourselves to study solutions enjoying symmetry
with respect to combined circle transformations in the ${\mathbb
R}^2$ plane and the internal space ${\mathbb C}^2$. This symmetry is
materialized by means of the ansatz:
\[
\Phi(x_1,x_2) = \left(\begin{array}{c}f(r) e^{il\theta}\\
|h(r)|e^{i(\omega+ m\theta)}\end{array}\right) \quad , \quad l,
m\in{\mathbb Z}^+\,\, , \,\, \omega\in{\mathbb R} \quad , \quad
A_i(x_1,x_2)=-l \varepsilon_{ij} \frac{\alpha(r)}{r^2}x_j \quad.
\]
Note that by taking some care with the behavior of $\alpha$ at the
origin the vector field $A_i(x_1,x_2)$ is divergence-free (purely
rotational) in the whole plane. The PDE system (\ref{sdfopd})
reduces to another first-order non-linear ODE system linking $f(r)$,
$\alpha(r)$, and $h(r)$:
\begin{equation}
{1\over r} {d \alpha \over d r}= - \frac{1}{2 l}
(f^2(r)+|h(r)|^2-1)\quad , \quad {d f\over d r}=  \frac{l}{r}
f(r)[1-\alpha(r)]  \quad , \quad  {d |h|\over dr}={l\over
r}|h|(r)({m\over l}-\alpha(r)) \label{sdfood} \quad .
\end{equation}
Finite energy solutions, regular at the origin where the (multi)
vortex sits, require us to solve (\ref{sdfood}) with asymptotic and
core behavior:
\begin{eqnarray}
&&{\displaystyle \lim_{r\rightarrow\infty}} f(r) = 1 \hspace{1cm} ,
\hspace{1cm} {\displaystyle \lim_{r\rightarrow\infty}}h(r) = 0
\hspace{1cm},\hspace{1cm} {\displaystyle
\lim_{r\rightarrow\infty}}\alpha(r) = 1  \label{bca}\\ &&f(0) =0
 \hspace{1.5cm}
, \hspace{1.3cm}|h(0)|=|h_0|
\delta_{m,0}\hspace{0.6cm},\hspace{0.8cm}\alpha(0)=0
\hspace{0.8cm},\hspace{0.8cm} m<l  \label{bco}
\end{eqnarray}
We stress that the contribution to the vorticity of the $\Phi_2$
field $m$ must be always smaller than the vorticity $l$ of the
$\Phi_1$ for the topological solution to have finite energy. For
later use, we give the magnetic field $B(r)=\frac{l}{
2r}\frac{d\alpha}{dr}$ and the energy density of these circle
configurations:
\[
 {\cal E}(r)={1\over 8}({1\over
l^2}+1)(1-f^2(r)-|h(r)|^2)^2+{l^2f^2(r)\over
r^2}(1-\alpha(r))^2+\frac{l^2|h(r)|^2}{r^2}({m\over l}-\alpha(r))^2
\qquad .
\]
\subsubsection{Topological solutions of one quantum of magnetic
flux} We now go on to the most elementary solutions that carry a
quantum of magnetic flux, or, $l=1=m+1$ . We are guided by the
procedure developed in \cite{dVS} to solve the non-linear ODE system
(\ref{sdfood}) with boundary conditions (\ref{bca})-(\ref{bco}).
First, we consider small values of $r$ and in the first-order
differential equations we test the power series
\begin{eqnarray}
f(r)&\equiv& f_1\cdot r+f_2\cdot r^2+f_3\cdot r^3+f_4 r^4+ \cdots \label{eq:sdex1}\\
\alpha(r) & \equiv & \alpha_1\cdot r+\alpha_2\cdot r^2+\alpha_3\cdot
r^3+\alpha_4\cdot r^4+ \cdots \label{eq:sdex2}\\ h(r)&\equiv&
h_0+h_1\cdot r+h_2\cdot r^2+h_3\cdot r^3+ h_4\cdot r^4+\cdots \qquad
, \label{eq:sdex3}
\end{eqnarray}
where $f_j$ and $\alpha_j$, $j=1,2,3, \cdots $, are real, whereas
$h_j$, $j=0,1,2, \cdots $, are complex coefficients. The coupled
first-order ODE's are solved at this limit by
(\ref{eq:sdex1})-(\ref{eq:sdex2})-(\ref{eq:sdex3}) if
\begin{eqnarray*}
f(r) & \simeq & f_1\cdot r  +\frac{f_1}{8} (|h_0|^2-1)\cdot  r^3 +
\frac{f_1}{128} \left[ (|h_0|^2-1)(2|h_0|^2-1)+4 f_1^2 \right]\cdot
r^5
+\dots \\
\alpha(r) & \simeq &  \frac{1}{4}(1-|h_{0}|^2)\cdot r^2- \left[
\frac{1}{32} |h_0|^2 (|h_0|^2-1)+\frac{1}{8} f_1^2 \right]\cdot r^4 - \\
&-&  \left[ \frac{1}{768} |h_0|^2
(|h_0|^2-1)(3|h_0|^2-2)+\frac{1}{192} f_1^2 (5|h_0|^2-4)
\right]\cdot r^6
+\dots \\
h(r) & \simeq & h_0 + \frac{h_0}{8}(|h_0|^2-1)\cdot r^2+
\frac{h_0}{128}  \left[(|h_0|^2-1)(2|h_0|^2-1)+4 f_1^2\right]\cdot
r^4 +\dots \qquad .
\end{eqnarray*}
We stress that $h_0\in [0,1]$ is determined by the behavior of the
solution at the origin such that only a free parameter, $f_1$, is
left. Second, a numerical scheme is implemented by setting a
boundary condition at a non-singular point of the ODE system, which
is obtained from the power series for a small value of $r$
($r=0.001$ in our case). This scheme prompts a shooting procedure by
varying $f_1$, where the correct asymptotic behavior of the
solutions is obtained setting a optimal value for $f_1$ for a given
value of $h_0$. Finally, the first-order ODE system is solved for
long $r$ by means of a power series in $\frac{1}{r}$:
\[
f(r) = \sum_{j=0} f^j\cdot r^{-j} \hspace{1cm} \alpha(r) =
\sum_{j=0} \alpha^j\cdot r^{-j} \hspace{1cm} h(r) = \sum_{j=1}
h^j\cdot r^{-j}
\]
with the result that
\begin{eqnarray*}
f(r) & \simeq & 1-\frac{|h^1|^2}{2}\cdot  r^{-2} + (-2|h^1|^2+
\frac{3}{8}|h^1|^4)\cdot
r^{-4}+|h^1|^2(-32+5|h^1|^2-\frac{5}{16}|h^1|^4)\cdot r^{-6}+\\&+&
|h^1|^2(-1152+158|h^1|^2-\frac{35}{4}|h^1|^4+\frac{35}{128}|h^1|^6)\cdot r^{-8}+\dots \\
\alpha(r) & \simeq &  1- |h^1|^2\cdot r^{-2}+|h^1|^2 (
-8+|h^1|^2)\cdot r^{-4} +|h^1|^2(-192+24|h^1|^2-|h^1|^4)\cdot r^{-6} \\
&& + |h^1|^2 (9216+1120|h^1|^2-48|h^1|^4+|h^1|^6)\cdot r^{-8}
+\dots \\
|h(r)| & \simeq & |h|^1\cdot r^{-1}+-\frac{|h^1|^3}{2}\cdot
r^{-3}+|h^1|^3(-2+{3\over 8}|h^1|^2)\cdot r^{-5}+\\&+&
|h^1|^3(-32+5|h^1|^2-\frac{5}{16}|h^1|^4)\cdot r^{-7} +\dots \qquad
.
\end{eqnarray*}
Again, only one free parameter $h^1$ is left. The value of $h^1$ is
fixed by demanding the continuity of the solution at intermediate
distances ($r=15$ in our case) obtained by gluing the short-$r$ and
long-$r$ approximations. In particular, this has the important
implication that $|h_0|=0  \quad \Rightarrow \quad |h^1|=0$ linking
the null value of $|h_0|$, which gives the embedded ANO vortex, with
the null value of the constant $|h^1|$ setting the behavior of the
solution for very long $r$. Another important remark is that the
long $r$ behavior of self-dual semi-local defects differs from the
long $r$ behavior of self-dual ANO vortices that decay
exponentially.

The following Figures show the results obtained with this procedure
for several values of $h_0$. Note that $h_0=0$ for the ANO vortices,
and $h_0=1$ for the ${\rm CP}^1$-lumps. It may be observed in the
graphics that the field profiles reach their vacuum values at
distances of the order of $r=15$. Consequently, practically
identical numerical solutions would be generated by sewing the
numerical to the asymptotic solution at $r$ greater than $15$.

\begin{figure}[h]
\centerline{\includegraphics[height=3.0cm]{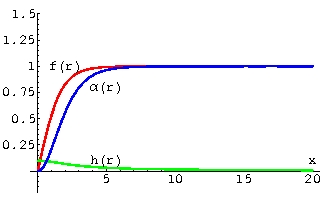}\hspace{0.3cm}
\includegraphics[height=3.0cm]{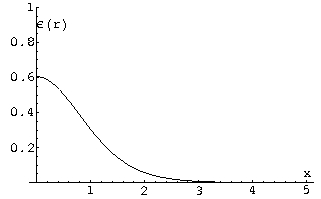}\hspace{0.3cm}
\includegraphics[height=3.0cm]{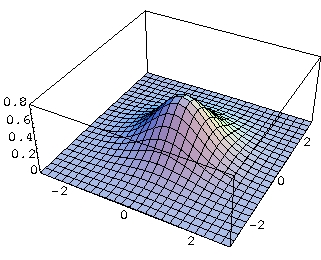}}
\centerline{\small \textit{a) Functions $f(r)$, $h(r)$ and
$\alpha(r)$ and b), c) Energy density for $h_0=0.1$}} \vspace{0.8cm}
\centerline{\includegraphics[height=3.0cm]{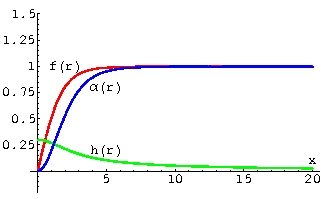}\hspace{0.3cm}
\includegraphics[height=3.0cm]{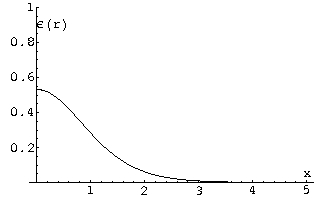}\hspace{0.3cm}
\includegraphics[height=3.0cm]{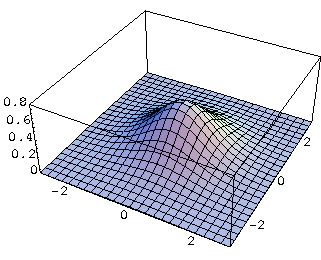}}
\centerline{\small \textit{a) Functions $f(r)$, $h(r)$ and
$\alpha(r)$ and b), c) Energy density for $h_0=0.3$}} \vspace{0.8cm}
\end{figure}
\begin{figure}[h]
\centerline{\includegraphics[height=3.3cm]{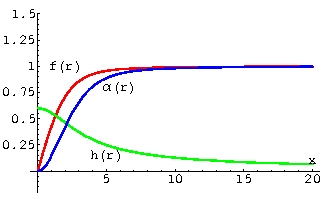}\hspace{0.3cm}
\includegraphics[height=3.3cm]{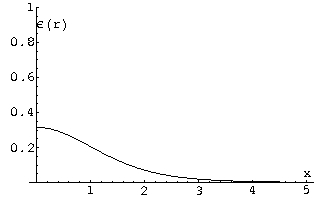}\hspace{0.3cm}
\includegraphics[height=3.3cm]{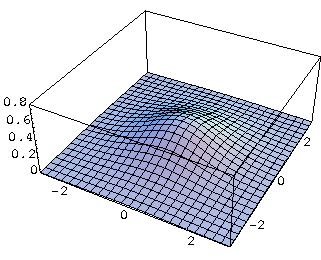}}
\centerline{\small \textit{a) Functions $f(r)$, $h(r)$ and
$\alpha(r)$ and b), c) Energy density for $h_0=0.6$}} \vspace{0.8cm}
\centerline{\includegraphics[height=3.3cm]{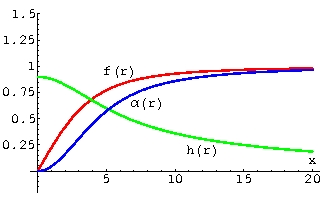}\hspace{0.3cm}
\includegraphics[height=3.3cm]{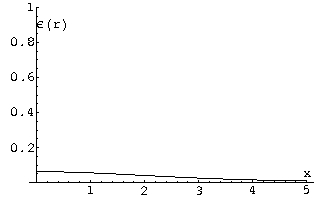}\hspace{0.3cm}
\includegraphics[height=3.3cm]{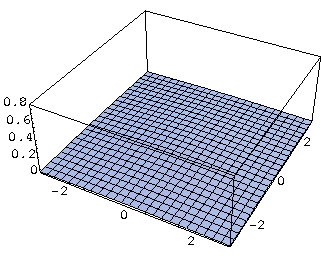}}
\centerline{\small \textit{a) Functions $f(r)$, $h(r)$ and
$\alpha(r)$ and b), c) Energy density for $h_0=0.9$}} \caption{Field
Profiles and Energy Densities for Semi-local Topological Defects}
\end{figure}
\par
\newpage

\subsubsection{Deformation of the first-order equations}
It is interesting at this point to consider small deformations of
the solutions of the (\ref{sdfopd}) system
\[
\Phi(\vec{x})=S(\vec{x})+\delta S(\vec{x}) \hspace{1.5cm} ,
\hspace{1.5cm} A_j(\vec{x})=V_j(\vec{x})+\delta a_j(\vec{x})\qquad ,
\]
which are still solutions of the same PDE system. The necessary and
sufficient conditions for this are tantamount to the linear PDE
system:
\begin{eqnarray}
&& -\partial_2\delta a_1+\partial_1\delta a_2+\frac{1}{2}(S^\dagger
\delta S+\delta S^\dagger S)=0 \nonumber \\ &&
(\frac{\partial}{\partial
x_1}-iV_1(\vec{x})+i\frac{\partial}{\partial
x_2}+V_2(\vec{x}))\delta S-i(\delta a_1+i\delta a_2) S=0
\label{slsdm} \qquad .
\end{eqnarray}
Pure gauge fluctuations are discarded from the solutions of
(\ref{slsdm}) by setting the background gauge:
\[
\partial_j\delta a_j(\vec{x})+\frac{i}{2}(S^\dagger \delta S-\delta S^\dagger
S)=0 \qquad .
\]
The tangent space to the moduli space of self-dual topological
solitons with a given magnetic charge $2\pi l$ is therefore the
kernel of the first-order deformation operator ${\cal D}$:
\[
{\cal D}\xi(\vec{x})= \left(\begin{array}{cccccc}
-\partial_2 &\partial_1 & S_1^1 & S_1^2&S_2^1&S_2^2 \\
 -\partial_1 & -\partial_2 & -S_1^2 & S_1^1&-S_2^2&S_2^1\\
S_1^1 & -S_1^2 & -\partial_2+V_1 & -\partial_1-V_2 &0&0\\
S_1^2 &S_1^1&\partial_1+V_2 & -\partial_2+V_1&0&0\\
S_2^1 & -S_2^2&0&0 & -\partial_2+V_1 & -\partial_1-V_2\\
S_2^2 &S_2^1&0&0&\partial_1+V_2 & -\partial_2+V_1\\
 \end{array}\right)\left(\begin{array}{c} \delta a_1(\vec{x})\\
\delta a_2(\vec{x})\\
\delta S_1^1(\vec{x})\\ \delta S_1^2(\vec{x})\\\delta
S_2^1(\vec{x})\\ \delta S_2^2(\vec{x}) \end{array}\right)=0 \qquad .
\]
Accordingly, $\bigtriangleup_+={\cal D}^\dagger{\cal D}$ is the
following $6\times 6$ matrix partial differential operator:
{\small\begin{displaymath}
\bigtriangleup_+=\left(\begin{array}{cccccc}
A&0&-2\nabla_1 S_1^2 &2\nabla_1 S_1^1 & -2\nabla_1 S_2^2 &2\nabla_1 S_2^1  \\
0&A&-2\nabla_2 S_1^2 & 2\nabla_2 S_1^1 & -2\nabla_2 S_2^2 & 2\nabla_2 S_2^1 \\
-2\nabla_1 S_1^2 &-2\nabla_2 S_1^2 &B&-2 V_k\partial_k&S_1^1S_2^1+S_1^2S_2^2 &S_1^1S_2^2-S_1^2S_2^1 \\
2\nabla_1 S_1^1 & 2\nabla_2 S_1^1&2 V_k \partial_k&B&-S_1^1S_2^2+S_1^2S_2^1 & S_1^1S_2^1+S_1^2S_2^2 \\
-2\nabla_1 S_2^2 &-2\nabla_2 S_2^2  &S_1^1S_2^1+S_1^2S_2^2  & -S_1^1S_2^2+S_1^2S_2^1 &C&-2 V_k\partial_k\\
2\nabla_1 S_2^1 & 2\nabla_2 S_2^1 &S_1^1S_2^2-S_1^2S_2^1  &
S_1^1S_2^1+S_1^2S_2^2 &2 V_k\partial_k&C
\end{array}\right) \nonumber
\end{displaymath}}
\begin{eqnarray*}
&& A=-\partial_k \partial_k+|S_1|^2+|S_2|^2 \, , \, \, \, \, j,k=1,2
\quad , \quad \partial_k\partial_k=\frac{\partial^2}{\partial
x_1^2}+\frac{\partial^2}{\partial x_2^2} \quad , \quad
V_k(\vec{x})V_k(\vec{x})=V_1^2(\vec{x})+V_2^2(\vec{x})
\\ && B=-\partial_k \partial_k+\frac{1}{2}(3|S_1|^2+|S_2|^2+2V_k
V_k-1) \quad , \quad
C=-\partial_k\partial_k+\frac{1}{2}(|S_1|^2+3|S_2|^2+2V_k V_k-1)  \\
&& \nabla_j S_M^A=\partial_j S_M^A+\varepsilon^{AB} V_jS_M^B \qquad
, \qquad M=1,2 \, \, \, , \, \, \, A,B=1,2 \quad , \quad
\varepsilon^{12}=-\varepsilon^{21}=1 \, \, , \, \,
\varepsilon^{11}=\varepsilon^{22}=0 \, .
\end{eqnarray*}
One easily checks that $\bigtriangleup_+$ has a supersymmetric
partner{\footnote{This is an example of hidden bosonic supersymmetry
\cite{PLYU}. }}: $\bigtriangleup_- ={\cal D}{\cal D}^\dagger$:
{\small
\begin{displaymath} \bigtriangleup_-=\left(\begin{array}{cccccc}
A&0&0&0&0&0\\
0&A&0&0&0&0\\
0&0&B_-&-2 V_k\partial_k&S_1^1S_2^1+S_1^2S_2^2 & S_1^1S_2^2-S_1^2S_2^1 \\
0&0&2 V_k \partial_k&B_-&-S_1^1S_2^2+S_1^2S_2^1 & S_1^1S_2^1+S_1^2S_2^2 \\
0&0&S_1^1S_2^1+S_1^2S_2^2  & -S_1^1S_2^2+S_1^2S_2^1 &C_-&-2 V_k\partial_k\\
0&0&S_1^1S_2^2-S_1^2S_2^1  & S_1^1S_2^1+S_1^2S_2^2 &2
V_k\partial_k&C_-
\end{array}\right) \quad , \quad V_k \partial_k =V_1(\vec{x})\frac{\partial}{\partial x_1}
+V_1(\vec{x})\frac{\partial}{\partial x_1} \nonumber
\end{displaymath}}
\[
B_-=-\partial_k \partial_k+\frac{1}{2}(|S_1|^2-|S_2|^2+2V_k
V_k+1)\quad  , \quad   C_-=-\partial_k
\partial_k+\frac{1}{2}(-|S_1|^2+|S_2|^2+2V_k V_k+1) \, \, \,  .
\]
The index of the deformation operator - ${\rm ind}\,{\cal D}={\rm
dim} {\rm Ker} {\cal D}- {\rm dim} {\rm Ker} {\cal D}^\dagger$ - is
in this case equal to the dimension of ${\rm Ker}\bigtriangleup_+$
because ${\rm dim} {\rm Ker} {\cal D}^\dagger=0$, $\bigtriangleup_-$
being definite positive.

Because $\bigtriangleup_+$ and $\bigtriangleup_-$ are iso-spectral
up to zero modes the index of ${\cal D}$ can be regularized in the
following form:
\[
{\rm index}{\cal D}=\lim_{\beta\to 0}\left[{\rm
Tr}_{L^2}e^{-\beta\bigtriangleup_+}-{\rm
Tr}_{L^2}e^{-\beta\bigtriangleup_-}\right].
\]
Let us use the heat trace expansion: ${\rm
Tr}_{L^2}e^{-\beta\bigtriangleup_\pm}={\rm
Tr}_{L^2}e^{-\beta\bigtriangleup}\sum_{n=0}^\infty\,
c_n[\bigtriangleup_\pm]\beta^n $,
\[
\bigtriangleup=\left(\begin{array}{cccccc}-\partial_k\partial_k+1 &
0 & 0 & 0 & 0 & 0 \\ 0 & -\partial_k\partial_k+1 & 0 & 0 & 0 & 0
\\ 0 & 0 & -\partial_k\partial_k+1 & 0 & 0 & 0 \\
 0 & 0 & 0 & -\partial_k\partial_k+1 & 0 & 0 \\ 0 & 0 & 0 & 0 & -\partial_k\partial_k
 & 0 \\ 0 & 0 & 0 & 0 & 0 & -\partial_k\partial_k\end{array}\right)
 \qquad .
\]
Because
\[
{\rm
Tr}_{L^2}e^{-\beta\bigtriangleup}=\frac{e^{-\beta}}{\pi}+\frac{1}{2\pi\beta}
\quad , \quad c_0[\bigtriangleup_\pm]=6 l^2 , \quad , \quad
c_1[\bigtriangleup_\pm]=\int \, dx^2 \, {\rm
tr}\left(\bigtriangleup_\pm(\vec{x})-\bigtriangleup\right)
\]
we obtain:
\begin{equation}
{\rm ind}{\cal D}=-\frac{1}{\pi}\int \, dx^2 \, {\rm
tr}\left(\bigtriangleup_+(\vec{x})-\bigtriangleup_-(\vec{x})\right)=
\frac{2}{\pi}\int \, dx^2 \,
\left(|S_1|^2+|S_2|^2-1\right)=\frac{2}{\pi}\int \, dx^2 \,
F_{12}=4l \label{ind} \, \, \, .
\end{equation}
In the derivation of (\ref{ind}) we have used the vortex equation
(\ref{sdfopd}) and ${\rm tr}$ means trace in the matrix sense. We
find that the number of zero modes is twice the magnetic flux (mod
$\pi$), in perfect agreement with the number of parameters of the
self-dual topological solitons.
\subsection{One-loop correction to the masses of  semi-local
self-dual topological solitons (SSTS)}

\subsubsection{SSTS fluctuations}
Let us now consider time-dependent small fluctuations of the
self-dual topological solitons:
\[
\Phi(\vec{x})=S(\vec{x})+\delta S(x_0,\vec{x}) \hspace{1cm} ,
\hspace{1cm} A_j(\vec{x})=V_j(\vec{x})+\delta
a_j(x_0,\vec{x})\hspace{1cm} , \hspace{1cm} \chi(\vec{x})=\delta
\chi(\vec{x}) \qquad .
\]
In order to discard pure gauge fluctuations we impose the
Weyl/background gauge condition (the R gauge in the topological
sector of magnetic flux 1):
\[
A_0(x_0,\vec{x})=0 \qquad \qquad , \qquad \qquad \partial_j\delta
a_j(x_0,\vec{x})+ \frac{i}{2}(S^\dagger(\vec{x}) \delta
S(x_0,\vec{x})-\delta S^\dagger(x_0,\vec{x}) S(\vec{x}))=0 \qquad .
\]
The classical energy up to ${\cal O}(\delta^3)$ order (one-loop) of
the SSTS fluctuations is:
\[
H^{(2)}=H^{(2)}_B+H^{(2)}_F={v^2\over 2}\int \,
d^2x\left\{\frac{\partial\delta\xi^T}{\partial
x_0}\frac{\partial\delta\xi}{\partial x_0}+\delta \xi^T K \delta\xi+
\delta\chi^* K^G \delta\chi\right\} \quad ,
\]
where
\[
\delta\xi^T(x_0,\vec{x})=\left(\delta a_1(x_0,\vec{x}) \, \, \, \,
\delta a_2(x_0,\vec{x}) \, \, \, \, \delta S_1^1(x_0,\vec{x}) \, \,
\, \, \delta S_1^2(x_0,\vec{x}) \, \, \, \, \delta
S_2^1(x_0,\vec{x}) \, \, \, \, \delta S_2^2(x_0,\vec{x})\right)
\]
is a file vector assembling the fluctuations of the two
polarizations of vector particles, the Higgs and Higgs ghost
fluctuations, and Goldstone fluctuations around the topological
soliton solution, i.e., the bosonic fluctuations and the
second-order operator determining the small fluctuations of these
extended objects is precisely $K=\bigtriangleup_+$.

We shall impose periodic boundary conditions on the fluctuations
$\delta\xi_a(x_1,x_2)=\delta\xi_a(x_1+l,x_2+l)$, $a=1,2, \cdots ,6$,
$l=mL$. Therefore, $K$ acts on the Hilbert space
$L^2=\bigoplus_{a=1}^6 \, L^2_a({\mathbb S}^1\bigotimes {\mathbb
S}^1)$. Assuming the ortho-normality and completeness of the
eigenfunctions of $K$, $K\xi_n(\vec{x})=\varepsilon^2_n
\xi_n(\vec{x})$, in the sub-space orthogonal to its kernel, one
finds the quantum Hamiltonian of the one-loop bosonic fluctuations:
\begin{equation}
\hat{H}^{(2)}_B=\hbar m\, \sum_{{\rm Spec}_+\, K}\,
\varepsilon_n(\hat{B}_n^\dagger\hat{B}_n+\frac{1}{2}) \qquad ,
\end{equation}
where $[\hat{B}_n^\dagger,\hat{B}_m]=\delta_{nm}$ are the expansion
coefficients promoted to creation and annihilation operators. The
index theorem argument in the previous Section shows that the number
of normalizable zero modes is: ${\rm dim}{\rm Ker}{\cal D}=4 l$
where ${\cal D}$ is the partial differential operator arising in the
deformation of the first-order equation. Because $K={\cal D}^\dagger
{\cal D}$, standard supersymmetry strategies allow us to conclude
that the spectrum of $K$ is formed by $4l$ zero modes and positive
eigenvalues giving rise to ${\rm Spec}_+ K$.

The topological soliton ground state is a coherent state annihilated
by all destruction operators:
\[
\hat{B}_n|0;TS \rangle =0, \, \forall n \quad \Rightarrow \quad
\begin{array}{c} \hat{\Phi}(x_0,\vec{x})|0;TS \rangle = S(\vec{x})|0;TS \rangle
\\ \hat{A}_i(x_0,\vec{x})|0;TS \rangle = V_i(\vec{x})|0;TS
\rangle\end{array} \qquad .
\]
Therefore, the bosonic energy of the ground state in a topological
sector of magnetic charge $l\neq 0$ is:
\begin{equation}
\langle 0; TS|\hat{H}_B^{(2)}|TS; 0 \rangle=\frac{\hbar m}{2}{\rm
Tr}_{L^2}\, K^\frac{1}{2} \label{boe} \qquad .
\end{equation}

In the Figure 12 we show the diagonal potential wells/barriers in
the differential operator $K$ for magnetic flux $2\pi$. The vector
bosons \lq\lq essentially" feel -there are also non-diagonal terms-
the potential $A$; Higgs bosons and Higgs ghosts feel the potential
$B$, and the Goldstone bosons move along the potential $C$. Leaving
apart the non-diagonal exchange interactions, all three types of
particles move through attractive potentials, exponentially decaying
to their vacuum values if the background is the ANO vortex, $h_0=0$.
If the background is a $h_0=0.3$ topological soliton, both vector
and Goldstone bosons move in less attractive potential wells that,
moreover, only decay as some negative power of $r$ near infinity.
For $h_0=0.9$, a value giving almost a ${\mathbb C}{\mathbb
P}^1$-lump, the vector bosons pass through the background feeling a
extremely weak attraction, the Higgs boson and ghost note a
considerably weaker attraction as compared to the attraction of the
previous backgrounds, and the Goldstone bosons are repelled by the
topological soliton. The decay at infinity is extremely slow. The
analysis of these physical features will give qualitative support to
our results on the one-loop corrections to be presented later.

\begin{figure}[h]
\centerline{\includegraphics[height=3.0cm]{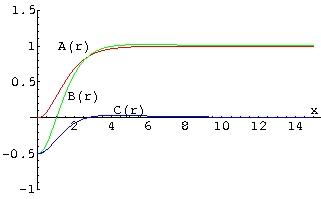}\hspace{0.3cm}
\includegraphics[height=3.0cm]{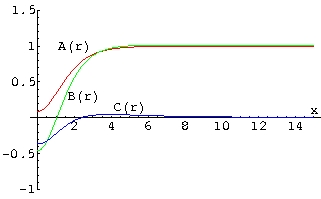}\hspace{0.3cm}
\includegraphics[height=3.0cm]{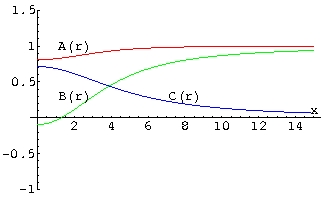}}
\caption{Potential wells for $l=1$ $V_A(r)$, $V_B(r)$, $V_C(r)$: a)
$h_0=0.0$, b) $h_0=0.3$, c) $h_0=0.9$ }
\end{figure}

It remains to get rid off the contribution of the Higgs ghosts by
considering the ground state energy of the (fermionic) Faddeev-Popov
ghosts. The FP ghost fluctuations (with PBC) are determined by the
Sch$\ddot{\rm o}$dinger operator:
\[
K^G=-\partial_k\partial_k+|S_1|^2+|S_2|^2 \qquad , \qquad
\delta\chi(x_1,x_2)=\delta\chi(x_1+l,x_2+l) \qquad ,
\]
such that $K^G$ acts on $L^2_G=L^2({\mathbb S}^1\bigotimes {\mathbb
S}^1)$. Again assuming the ortho-normality and completeness of the
eigenfunctions of $K^G$, $K^G\chi_n(\vec{x})=\varepsilon^2_n
\chi_n(\vec{x})$, in the sub-space orthogonal to its kernel, one
finds the quantum Hamiltonian of the one-loop fermionic
fluctuations:
\begin{equation}
\hat{H}^{(2)}_F=\hbar m\, \sum_{{\rm Spec}_+\, K^G}\,
\varepsilon_n(\hat{C}_n^\dagger\hat{C}_n-\frac{1}{2}) \qquad ,
\end{equation}
where $\{\hat{C}_n^\dagger,\hat{C}_m\}=\delta_{nm}$ are the
expansion coefficients promoted to fermionic creation and
annihilation operators. There are no fermionic ghosts in the
topological soliton ground state. Henceforth, $\hat{C}_n |0;TS
\rangle =0, \forall n$ and the ground state energy of the
topological soliton is:
\begin{equation}
\langle 0;TS|\hat{H}^{(2)}_B|0;TS \rangle + \langle
0;TS|\hat{H}^{(2)}_F|0;TS \rangle =\frac{\hbar m}{2}\left({\rm
Tr}_{L^2}\, K^\frac{1}{2}-{\rm Tr}_{L^2}\,
(K^G)^\frac{1}{2}\right)=\frac{\hbar m}{2}{\rm STr}_{L^2}\,
K^\frac{1}{2} \qquad . \label{tgse}
\end{equation}
\subsubsection{Vacuum fluctuations}
It is instructive to specify this analysis for the vacuum
fluctuations:
\[
S^V(\vec{x})=\left(\begin{array}{c}1 \\ 0 \end{array}\right)\qquad ,
\qquad \qquad V_i^V(\vec{x})=0_i \qquad .
\]
In this case the second-order operator fluctuations are
$K_0=\bigtriangleup$ and $K^G_0=-\partial_k\partial_k+1$ such that
the spectrum is completely known. In a normalization square of area
$l^2$, periodic boundary conditions on the fluctuations plus
canonical quantization produce the following bosonic and fermionic
free Hamiltonians:
\begin{eqnarray*}
&&\hat{H}^{(2)}_B=\hbar m\, \sum_{a=1}^6\,\sum_{n_1\in{\mathbb
Z}}\,\sum_{n_2\in\mathbb Z}\,\,
\omega_a(n_1,n_2)\left(\hat{b}_{n_1}^{a\dagger}\hat{b}_{n_1}^a+
\hat{b}_{n_2}^{a\dagger}\hat{b}_{n_2}^a+1\right)\quad ,
\quad [\hat{b}_{n_\alpha}^{a\dagger},\hat{b}_{m_\beta}^c]=\delta^{ac}\delta_{n_\alpha m_\beta} \\
&&\hat{H}^{(2)}_F=\hbar m\,\sum_{n_1\in{\mathbb
Z}}\,\sum_{n_2\in\mathbb Z}\,\,
\omega(n_1,n_2)\left(\hat{c}_{n_1}^{\dagger}\hat{c}_{n_1}+\hat{c}_{n_2}^{\dagger}\hat{c}_{n_2}-1\right)
\quad ,
\quad \{\hat{c}_{n_\alpha}^{\dagger},\hat{c}_{m_\beta}\}=\delta_{n_\alpha m_\beta} \\
&&\omega(n_1,n_2)=\omega_a(n_1,n_2)=\frac{4\pi^2}{l^2}(n_1^2+n_2^2)+1\,
, \, \, \, a=1,2,3,4\, \\ &&
\omega_a(n_1^2,n_2^2)=\frac{4\pi^2}{l^2}(n_1^2+n_2^2)\, , \, \, \,
a=5,6 \qquad .
\end{eqnarray*}
Therefore, the ground state in the vacuum sector is also a coherent
state:
\[
\hat{b}_{n_\alpha}^a |V; 0\rangle = \hat{c}_{n_\alpha}|V; 0\rangle=0
\, \, , \quad \forall a \, \, , \, \, \forall n_\alpha \quad  \equiv
\quad \left\{\begin{array}{c}\hat{\Phi}(t,\vec{x})|V; 0\rangle
=\left(\begin{array}{c}|V; 0\rangle
\\ 0 \end{array}\right)\\ \hat{A}_i(t,\vec{x})|V; 0\rangle=0_i\end{array}\right.
\]
such that the ground state energy follows immediately:
\[
\langle 0;V| \left(\hat{H}^{(2)}_B+\hat{H}_F^{(2)}\right)|V;0
\rangle={\rm Tr}_{L^2} K_0^\frac{1}{2}-{\rm
Tr}_{L^2}\left(K_0^G\right)^\frac{1}{2}={\rm
STr}_{L^2}\left(K_0\right)^\frac{1}{2} \qquad .
\]
It is clear that the ghosts cancel the contribution to the vacuum
energy of the non-physical Higgs ghosts and render the theory
unitary.
\subsection{Zero point energy and mass renormalizations}
\subsubsection{Topological soliton Casimir energy}
Subtracting the ground state vacuum energy from the ground state
energy of the topological solitons
\begin{eqnarray}
\bigtriangleup E_{\rm TS}^C &=&\langle 0;
TS|\left(\hat{H}_B^{(2)}+\hat{H}_F^{(2)}\right)|TS; 0 |\rangle
-\langle 0; V|\left(\hat{H}_B^{(2)}+\hat{H}_F^{(2)}\right)|V; 0
|\rangle \nonumber\\ &=& \bigtriangleup E_{\rm TS}-\bigtriangleup
E_0 ={\hbar m\over 2}\left({\rm STr}\, K^{{1\over 2}}- {\rm STr}\,
K_0^{{1\over 2}}\right) \label{cev}
\end{eqnarray}
one formally measures the semi-local self-dual topological soliton
Casimir energy.

\subsubsection{Mass renormalization counter-terms} In $(2+1)$-dimensions, the
semi-local Abelian Higgs model is super-renormalizable. This means
that there are a finite number of divergent graphs. We list the
one-loop graphs that diverge.
\begin{enumerate}
\item Higgs boson tadpole:
\vspace{0.4cm}

\centerline{
\parbox{2.5cm}{\includegraphics[width=2.2cm]{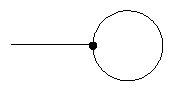}} $+$
\parbox{2.5cm}{\includegraphics[width=2.2cm]{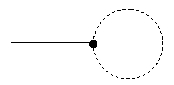}} $+$
\parbox{2.5cm}{\includegraphics[width=2.2cm]{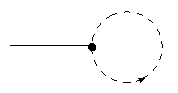}} $+$
\parbox{2.5cm}{\includegraphics[width=2.1cm]{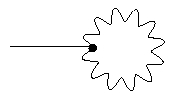}} $+$
\parbox{2.5cm}{\includegraphics[width=2.0cm]{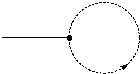}}$=$}
\vspace{-0.4cm}
\[
= - 2 i (\kappa^2 + 1)\cdot\, I(1)-i\kappa^2\cdot\, I(0) +
\mbox{finite part}
\]

\item Higgs boson self-energy:
\vspace{0.4cm}

\centerline{
\parbox{2.5cm}{\includegraphics[width=2.2cm]{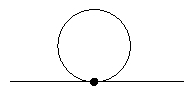}} $+$
\parbox{2.5cm}{\includegraphics[width=2.2cm]{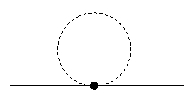}} $+$
\parbox{2.5cm}{\includegraphics[width=2.2cm]{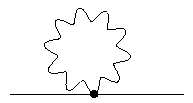}} $+$
\parbox{2.5cm}{\includegraphics[width=2.2cm]{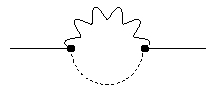}}$+$
\parbox{2.5cm}{\includegraphics[width=2.15cm]{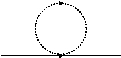}} $=$}
\vspace{-0.2cm}
\[
= - 2 i (\kappa^2+ 1)\cdot\, I(1)-i\kappa^2\cdot\, I(0) +
\mbox{finite part}
\]

\item Higgs ghost self-energy:
\vspace{0.4cm}

\centerline{
\parbox{2.5cm}{\includegraphics[width=2.2cm]{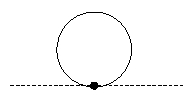}} $+$
\parbox{2.5cm}{\includegraphics[width=2.2cm]{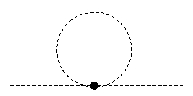}} $+$
\parbox{2.5cm}{\includegraphics[width=2.2cm]{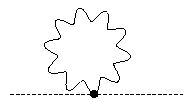}} $+$
\parbox{2.5cm}{\includegraphics[width=2.2cm]{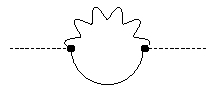}}$+$
\parbox{2.5cm}{\includegraphics[width=2.15cm]{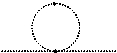}}$=$}
\vspace{-0.2cm}
\[
= - 2 i (\kappa^2+ 1)\cdot\, I(1)-i\kappa^2\cdot\, I(0) +
\mbox{finite part}
\]
\item Goldstone boson self-energy:
\vspace{0.4cm}

\centerline{
\parbox{2.5cm}{\includegraphics[width=2.2cm]{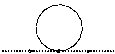}} $+$
\parbox{2.5cm}{\includegraphics[width=2.2cm]{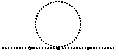}} $+$
\parbox{2.5cm}{\includegraphics[width=2.2cm]{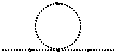}} $+$
\parbox{2.5cm}{\includegraphics[width=2.2cm]{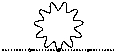}}
$+$
\parbox{2.5cm}{\includegraphics[width=2.8cm]{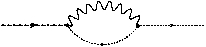}}
\,  $=$} \vspace{-0.2cm}
\[
= - 4i (\kappa^2+ 1)\cdot\, I(1)-2i\kappa^2\cdot\, I(0) +
\mbox{finite part}
\]

\item  Vector boson self-energy:
\vspace{0.4cm}

\centerline{
\parbox{2.5cm}{\includegraphics[width=2.2cm]{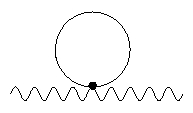}} $+$
\parbox{2.5cm}{\includegraphics[width=2.2cm]{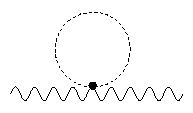}} $+$
\parbox{2.6cm}{\includegraphics[width=2.3cm]{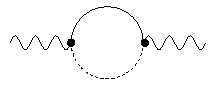}} $+$
\parbox{2.5cm}{\includegraphics[width=2.2cm]{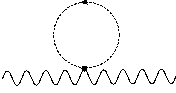}}$+$
\parbox{2.7cm}{\includegraphics[width=2.5cm]{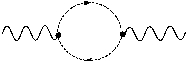}} \, $=$}
\vspace{-0.4cm}
\[
=  2i\cdot\, [I(1)+I(0)]g^{\mu\nu} + \mbox{finite part} \qquad .
\]

\end{enumerate}
Note that the ultraviolet divergences come from integrals of the
form:
\[
I(c^2)=\int \, \frac{d^3k}{(2\pi)^3} \cdot
\frac{i}{k^2-c^2+i\varepsilon}\qquad ,
\]
and that, unlike in $(1+1)$-dimensional scalar field theory, normal
ordering is not enough.

We add the following counter-terms to cancel these divergences:
\begin{eqnarray*}
{\cal L}_{c.t.}^S &=& {\hbar\over 2}\left[2(\kappa^2+1)\cdot\,
I(1)+\kappa^2\cdot\, I(0)\right]\cdot\,
\left[\Phi^*_1(x^\mu)\Phi_1(x^\mu)+\Phi^*_2(x^\mu)\Phi_2(x^\mu)-1
\right]\\{\cal L}_{c.t.}^A &=&-\hbar [I(1)+I(0)]\cdot\, A_\mu(x^\mu)
A^\mu(x^\mu)
\end{eqnarray*}

\begin{table}[hbt]
\begin{center}
\begin{tabular}{ccc} \\ \hline
\textit{Diagram} & \hspace{0.3cm} & \textit{Weight} \\
\hline \\[-0.3cm]
\parbox{4.5cm}{\includegraphics[width=2.5cm]{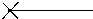}} & &
$\displaystyle i[2(\kappa^2+1)I(1)+\kappa^2 I(0)] $
\\
\parbox{4.5cm}{\includegraphics[width=2.5cm]{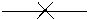}} & &
$\displaystyle i[2(\kappa^2+1) I(1)+\kappa^2 I(0)]$ \\
 \parbox{4.5cm}{\includegraphics[width=2.5cm]{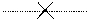}} & &
$\displaystyle i[2(\kappa^2+1) I(1)+\kappa^2 I(0)]$
\\
 \parbox{4.5cm}{\includegraphics[width=2.5cm]{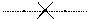}} & &
$\displaystyle i[4(\kappa^2+1) I(1)+2\kappa^2 I(0)]$
\\
 \parbox{4.5cm}{\includegraphics[width=2.5cm]{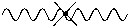}} & &
$\displaystyle -2i [I(1)+I(0)]g^{\mu\nu}$ \\[0.3cm] \hline
\end{tabular}
\caption{Counter-term vertices}
\end{center}
\end{table}
We have used a minimal subtraction prescription and the main
criteria to set finite renormalizations have been: 1) The divergence
due to the tadpole graph is exactly canceled in the self-dual limit
$\kappa^2=1$. 2) The global ${\mathbb S}{\mathbb U}(2)$ symmetry
remains unbroken after one-loop renormalizations. Interested readers
can find a fully detailed description of our renormalization
conventions in \cite{AMJW4}.

\subsubsection{Mass renormalization counter-term energies}

Therefore, the topological soliton energy due to mass
renormalization counter-terms in the self-dual limit $\kappa^2=1$
receives the following contribution from the scalar and vector
fields:
\[
\Delta E_{CT}^S=\frac{\hbar\, m}{2}[4I(1)+I(0)]\,\int \, d^2x
\,(1-|S_1|^2-|S_2|^2)\, \,  , \quad \Delta E_{CT}^A=-\hbar \, m \,
[I(1)+I(0)]\int \, d^2x \, V_kV_k \quad .
\]
We reshuffle the sum of these two quantities into two pieces
$\bigtriangleup E_{CT}^{\rm I(1)}= \frac{\hbar\, m}{2}\,I(1)
\,\Sigma^{(1)}(S,V_k)$ and $\bigtriangleup E_{CT}^{\rm I(0)}=
\frac{\hbar\, m}{2}\,I(0) \,\Sigma^{(0)}(S,V_k)$, respectively
proportional to $I(1)$ and $I(0)$:
\[
\Sigma^{(1)}(S,V_k)=4\int \, d^2x \, (1-|S_1|^2-|S_2|^2-\frac{1}{
2}V_kV_k ) \, \, \, , \, \, \, \Sigma^{(0)}(S,V_k)= \int \, d^2x
\,\left(1-|S_1|^2-|S_2|^2-2V_kV_k\right) \, \, .
\]
As in the zero point renormalization one must subtract the energy
induced by the counter-terms in the vacuum from the same quantity
for the topological soliton:
\begin{equation}
\bigtriangleup E_{TS}^R=\bigtriangleup E_{CT}^{\rm
I(1)}(TS)-\bigtriangleup E_{CT}^{\rm I(1)}(V)+\bigtriangleup
E_{CT}^{\rm I(0)}(TS)-\bigtriangleup E_{CT}^{\rm I(0)}(V) \quad .
\label{mrev}
\end{equation}
\subsection{High-temperature one-loop mass shift formula for
self-dual semilocal topological solitons}

\subsubsection{Spectral zeta-function regularization} Both
$\bigtriangleup E_{TS}^C$ and $\bigtriangleup E_{TS}^R$ are
divergent quantities that we shall regularize by means of the zeta
function procedure before being added. We recall that the spectral
zeta functions of elliptic operators $A$ (with positive definite and
discrete spectrum) are formally defined as infinite sums of complex
powers of their eigenvalues:
\[
\zeta_A(s)=\sum_{{\rm Spec}\,A}\,\lambda_n^{-s} \qquad \qquad ,
\qquad \qquad s\in{\mathbb C} \qquad .
\]
These sums are usually convergent for ${\rm Re}\, s>s_0>0$, where
$s_0$ is a positive real constant, but are susceptible to being
analytically continued to the complex $s$-plane. In many favorable
cases, their analytic continuations are meromorphic functions of $s$
and we shall regularize: 1) the ground state energy in topological
sectors, 2) the ground state energy in the vacuum sector, and 3) the
SSTS Casimir energy, by assigning to these quantities the values
\begin{eqnarray*}
\Delta E_{TS}(s)&=&\frac{\hbar\mu}{2}\left({\mu^2\over
m^2}\right)^s\left\{\zeta_{K}(s)-\zeta_{K^G}(s)\right\}\hspace{1cm},
\hspace{1cm} \Delta E_0(s)=\frac{\hbar\mu}{2}\left({\mu^2\over
m^2}\right)^s\left\{ \zeta_{K_0}(s)-\zeta_{K^G_0}(s)\right\}
\\
\Delta E_{TS}^C(s)&=& \Delta E_{TS}(s)- \Delta E_0(s)\quad\qquad
\hspace{1.8cm} , \quad\qquad \Delta E_{TS}^C=\lim_{s\rightarrow
-\frac{1}{2}}\Delta M_{TS}^C(s)
\end{eqnarray*}
at a regular point of the spectral zeta functions in the complex
$s$-plane. The spectral zeta functions of $K_0$ and $K_0^G$ with
periodic boundary conditions on the edges of a square of area $l^2$
are given by meromorphic Euler Gamma functions:
\[
\zeta_{K_0}(s)=\frac{l^2}{\pi}\cdot \frac{\Gamma[s-1]}{\Gamma(s)}+
\frac{l^2}{2\pi}\cdot \frac{1}{(s-1)\Gamma(s)}\hspace{1cm} ,
\hspace{1cm} \zeta_{K^G_0}(s)=\frac{l^2}{4\pi}\cdot
\frac{\Gamma[s-1]}{\Gamma(s)} \qquad .
\]
$\Delta E_{TS}^R$ can be regularized in a similar vein. On a square
of area $l^2$ $I(1)$ and $I(0)$ become infinite sums over discrete
momenta:
\[
I(c)={1 \over 2}\int {d^2k\over (2 \pi)^2} {1\over \sqrt{\vec k
\cdot \vec k +c^2}}={1\over 2}{1\over l^2}\sum_{\vec{k}\in{\mathbb
Z}^2}\frac{1}{\sqrt{\vec{k}\cdot\vec{k}+c^2}} \qquad .
\]
Therefore, we regularize $\Delta E_{TS}^R$ in the same form:
\[
\Delta E_{TS}^R(s) = {\hbar\over 2 l^2}\left({\mu^2\over
m^2}\right)^s \left(\zeta_{-\bigtriangleup+1} (s) \Sigma^{(1)}
(S,V_k)+\zeta_{-\bigtriangleup}(s) \Sigma^{(0)} (S,V_k)\right)
\qquad , \quad \Delta E_{TS}^R=\lim_{s\rightarrow{1\over 2}} \Delta
E_{TS}^R(s) \, \, \, ,
\]
knowing that:
\[
I(1)={1\over 2 l^2}\zeta_{-\bigtriangleup+1}({1\over 2})={1 \over 8
\pi} {\Gamma (-{1\over 2}) \over \Gamma({1\over 2})}=-{1\over 4\pi}
\qquad , \qquad I(0)={1\over 2 l^2}\zeta_{-\bigtriangleup}({1\over
2})=-\frac{1}{4\pi\sqrt{\pi}} \quad .
\]

\subsubsection{The heat kernel expansion of elliptic differential operators}
Because the $SSTS$ solutions are not known analytically, it is not
possible to compute the spectral functions of $\zeta_{K}(s)$ and
$\zeta_{K^G}(s)$. One possible loophole is to rely on the
high-temperature asymptotic expansion of the heat traces and to
build approximations to the spectral zeta functions from the Mellin
transform of these approximated heat traces.

This construction is particularly complicated for the second-order
fluctuation operator $K$. The Hessian is of the general form:
\[
K=K_0 +Q_k(\vec{x})\partial_k+V(\vec{x}) \qquad .
\]
The logical sequence is as follows:  the $K$-heat equation kernel is
the solution of the $K$-heat equation complying with the delta
function infinite temperature condition:
\begin{equation}
\left(\frac{\partial}{\partial\beta}{\mathbb I}+K \right)K_{
K}(\vec{x},\vec{y};\beta )=0 \qquad , \qquad K_{
K}(\vec{x},\vec{y};0)={\mathbb I}\cdot
\delta^{(2)}(\vec{x}-\vec{y})\qquad . \label{heatE}
\end{equation}
In our problem ${\mathbb I}$ is the ${\rm six}\times {\rm six}$ unit
matrix. The $K$-heat trace, or $K$-partition function, is the
integral over the whole ${\mathbb R}^2$-plane of the $K$-heat
equation kernel at the diagonal of ${\mathbb R}^2\bigotimes{\mathbb
R}^2$, whereas the spectral $K$-zeta function is the Mellin
transform of the $K$-heat trace:
\[
{\rm Tr}\,e^{-\beta K}={\rm tr}\int_{{\mathbb R}^2} \, d^2\vec{x} \,
K_{K}(\vec{x},\vec{x};\beta) \qquad , \qquad \zeta_{
K}(s)=\frac{1}{\Gamma(s)}\cdot \int_0^\infty \, d\beta
\,\beta^{s-1}{\rm Tr}e^{-\beta K} \qquad .
\]
Given the structure of $K$,
\[
K_{K}(\vec{x},\vec{y};\beta)=C_{
K}(\vec{x},\vec{y};\beta)K_{K_0}(\vec{x},\vec{y};\beta)
\]
is a good ansatz to solve (\ref{heatE}). Plugging the ansatz in
(\ref{heatE}), one finds the transfer PDE and the high-temperature
condition that $C(\vec{x},\vec{y};\beta)$ must satisfy:
\begin{equation}
\left\{ {\partial\over\partial\beta}{\mathbb
I}+{x_k-y_k\over\beta}(\partial_k{\mathbb I}-{1\over
2}Q_k)-\bigtriangleup{\mathbb I}+Q_k\partial_k+V
\rule[-0.1in]{0.in}{0.2in} \right\}C_{ K}(\vec{x},\vec{y};\beta)=0
\qquad , \qquad C_{ K}(\vec{x},\vec{y};0)={\mathbb I} \label{heatT}
\qquad .
\end{equation}
The solution of (\ref{heatT}) by means of a high-temperature power
series expansion
\[
C_K(\vec{x},\vec{y};\beta)=\sum_{n=0}^\infty
c_n(\vec{x},\vec{y};K)\beta^n \qquad , \qquad
c_0(\vec{x},\vec{y};K)={\mathbb I}
\]
is tantamount to the solving of the recurrence relations:
\begin{equation}
[n{\mathbb I}+(x_k-y_k)(\partial_k{\mathbb I}-{1\over
2}Q_k)]c_n(\vec{x},\vec{y};K) =[\bigtriangleup{\mathbb I}
-Q_k\partial_k-V]c_{n-1}(\vec{x},\vec{y};K)\hspace{1cm} .
\label{recE}
\end{equation}
It is easy to find the first Seeley density,
$c_1(\vec{x},\vec{x};K)=-V(\vec{x})$, a result used in the
subsection above addressing the index theorem. Higher-order Seeley
densities are harder to find. It is convenient to introduce the
notation
\[
{}^{(\alpha_1,\alpha_2)}C_n^{ab}(\vec{x})=\lim_{\vec{y}\rightarrow
\vec{x}} \frac{\partial^{\alpha_1+\alpha_2}
[c_n]_{ab}(\vec{x},\vec{y};K)}{\partial x_1^{\alpha_1}\partial
x_2^{\alpha_2}} \hspace{1cm} , \hspace{1cm}
[{c}_n]_{ab}(\vec{x},\vec{x};K)={}^{(0,0)}C_n^{ab}(\vec{x})
\]
because the recurrence relations between derivatives of the Seeley
densities can be written in the compact form
\clearpage
\begin{eqnarray*} &&(k+\alpha_1+\alpha_2+1)
{}^{(\alpha_1,\alpha_2)}C_{k+1}^{ab}(\vec{x})=
{}^{(\alpha_1+2,\alpha_2)}C_{k}^{ab}(\vec{x})+
{}^{(\alpha_1,\alpha_2+2)}C_{k}^{ab}(\vec{x})-  \\
&&-\sum_{d=1}^6 \sum_{r=0}^{\alpha_1}\sum_{t=0}^{\alpha_2} {\alpha_1
\choose r} {\alpha_2 \choose t} \left[ \frac{\partial^{r+t}
Q^{ad}_1}{\partial x_1^r\partial x_2^t}
{}^{(\alpha_1-r+1,\alpha_2-t)}C_{k}^{db}(\vec{x})\right.+\\&&+\left.
\frac{\partial^{r+t} Q^{ad}_2}{\partial x_1^r\partial x_2^t}
{}^{(\alpha_1-r,\alpha_2-t+1)}C_{k}^{db}(\vec{x}) \right]+  \\&&
+\frac{1}{2}\sum_{d=1}^6
\sum_{r=0}^{\alpha_1-1}\sum_{t=0}^{\alpha_2} \alpha_1{\alpha_1-1
\choose r} {\alpha_2 \choose t} \frac{\partial^{r+t}
Q^{ad}_1}{\partial x_1^r\partial x_2^t}
{}^{(\alpha_1-1-r,\alpha_2-t)}C_{k+1}^{db}(\vec{x})+ \\
&&+\frac{1}{2}\sum_{d=1}^6
\sum_{r=0}^{\alpha_2-1}\sum_{t=0}^{\alpha_1} \alpha_2{\alpha_2-1
\choose r} {\alpha_1 \choose t}  \frac{\partial^{r+t}
Q^{ad}_2}{\partial x_1^t\partial x_2^r}
{}^{(\alpha_1-t,\alpha_2-1-r)}C_{k+1}^{db}(\vec{x})-
\\&&-\sum_{d=1}^6
\sum_{r=0}^{\alpha_2}\sum_{t=0}^{\alpha_1}{\alpha_1 \choose
t}{\alpha_2\choose r}
 \frac{\partial^{r+t}
V^{ad}}{\partial x_1^t\partial x_2^r}
{}^{(\alpha_1-t,\alpha_2-r)}C_k^{db}(\vec{x})
\end{eqnarray*}
\[
c_0(\vec{x},\vec{x};K)={\mathbb I} \Rightarrow\left\{
\begin{array}{c} {}^{(\alpha,\beta)}C_0^{ab}(\vec{x})=0 , \, {\rm
if}\,
\alpha \neq 0, {{\rm and}/{\rm or}}\, \beta \neq 0 \\
{}^{(0,0)}C_0^{aa}(\vec{x})=1 , \, a=1,2, \cdots ,6
\end{array}\right.
\]
susceptible to being solved with the help of a symbolic programm
implemented in Mathematica.

\subsubsection{Mellin transform of the
heat trace asymptotic expansion}

The spectral zeta functions of of both $K$ and $K^G$ are obtained
from the high-temperature expansion of the heat traces via the
Mellin transform:
\begin{eqnarray*}
\zeta_{K}(s)&=&{1\over\Gamma(s)}\int_0^1 \, d\beta \,
\beta^{s-1}\left\{{1\over 4\pi\beta}
\sum_{n=0}^\infty\beta^n\cdot\left(e^{-\beta}\sum_{a=1}^4
[c_n^I(K)]_{aa}+ \sum_{a=5}^6
[c^O_n(K)]_{aa}\right)\right\}\\&+&{1\over\Gamma(s)}\int_1^\infty \,
d\beta \, \beta^{s-1} \, {\rm Tr} \, e^{-\beta K}
\\
\zeta_{K^G}(s)&=& {1\over\Gamma(s)}\int_0^1 \, d\beta \, \beta^{s-1}
\left\{{e^{-\beta}\over 4\pi\beta} \sum_{n=0}^\infty\beta^n
c_n(K^{\rm G})\right\}+{1\over\Gamma(s)}\int_1^\infty \, d\beta \,
\beta^{s-1} \, {\rm Tr} \, e^{-\beta K^G} \, \, \,  .
\end{eqnarray*}
Here the Seeley coefficients $[c_n^I(K)]_{aa}$, $[c_n^O(K)]_{aa}$,
and $c_n(K^G)$ are obtained through integration over the whole
${\mathbb R}^2$-plane of the Seeley densities:
\begin{eqnarray*}
[c_n^I(K)]_{aa}&=&\int \, d^2x \,[c_n]_{aa}(\vec{x},\vec{x};K)\quad,\quad a=1,2,3,4\\
{[c_n^O(K)]}_{aa}&=&\int \, d^2x
\,[c_n]_{aa}(\vec{x},\vec{x};K)\quad,\quad a=5,6\quad , \quad
{[c_n(K^G)]}=\int \, d^2x \,[c_n](\vec{x},\vec{x};K^G) \,\, .
\end{eqnarray*}
Finally, we find the spectral zeta functions as sums of meromorphic
and entire parts:
\begin{eqnarray}
\zeta_{K}(s)&=&\sum_{n=0}^\infty
\left\{\sum_{a=1}^4[c_n^I(K)]_{aa}\frac{\gamma[s+n-1,1]}{4\pi\Gamma(s)}+
\sum_{a=5}^6[c_n^O(K)]_{aa}\frac{1}{4\pi
\Gamma(s)(s+n-1)}\right\}+{1\over\Gamma(s)}B_{K}(s) \nonumber
\\
\zeta_{K^G}(s)&=&\sum_{n=0}^\infty
c_n(K^G)\frac{\gamma[s+n-1,1]}{4\pi\Gamma(s)}+{1\over\Gamma(s)}B_{K^G}(s)\label{zetTS}\,
\, \, .
\end{eqnarray}

An important warning: the physical point $s=-\frac{1}{2}$ is a
regular point -not a pole- of $\zeta_K(s)$, $\zeta_{K^G}(s)$,
$\zeta_{K_0}(s)$, and $\zeta_{K_0^G}(s)$. Unlike in kink cases,
after zeta function regularization is performed only finite
renormalizations are left in $(2+1)$-dimensional field theories.

\subsubsection{High-temperature one-loop mass
shift formula for self-dual semi-local topological solitons}

Neglecting the entire parts and truncating the zeta functions at a
finite number of summands $N_0$, the high-temperature one-loop
formula giving the semi-classical shift to the masses of semi-local
self-dual topological solitons is obtained:
\begin{eqnarray} \Delta M_{TS}&=&\bigtriangleup E^C_{TS}+\bigtriangleup
E^R_{Ts}= \nonumber\\&=& -{\hbar m \over 16\pi\sqrt{\pi}} \left[
\sum_{n=2}^{N_0}\,\left\{ [\sum_{a=1}^4[c_n(K)]_{aa}-c_n(K^{\rm G})]
\cdot\gamma[n-\frac{3}{2},1]+
\sum_{a=5}^6\frac{[c_n(K)]_{aa}}{n-\frac{3}{2}}\right\}+4 l\cdot
8\pi \right] \nonumber\\&-&\frac{\hbar m}{8\pi\sqrt{\pi}}\cdot \int
\, d^2x \, \left|S_2\right|^2(x_1,x_2)\cdot \left(\gamma[-{1\over
2},1]-2\right) \label{mfqc} \qquad \qquad .
\end{eqnarray}
The following remarks are meaningful:

\begin{enumerate}

\item The factor $-2l\frac{\hbar m}{\sqrt{\pi}}$ is due to the
subtraction of the $4l$ zero modes by exactly the same procedure as
the regularization method used for kinks.

\item Unlike in theories with only massive particles, the criterion
of the exact cancelation of tadpole graphs is not completely
equivalent to the cancelation of the contributions of the
first-order diagonal Seeley coefficient. The term in the second row
of formula (\ref{mfqc}) is the mismatch between these two criteria
due to the massless particles in the semi-local Abelian Higgs model.

\item The ANO vortices correspond to the SSTS solitons with
$S_2(\vec{x})=0$. Freezing the $\delta S_2(x_0,\vec{x})$
fluctuations and dropping away the associated $2l$ zero modes, the
one-loop mass shifts for ANO vortices are attained.

\end{enumerate}

\subsection{Mathematica calculations}
In this last Section we shall solve the recurrence relations to find
the Seeley densities and their associated Seeley coefficients by a
mixture of symbolic and numerical Programs implemented in
Mathematica. We shall focus on circle symmetric SSTS solutions,
\begin{eqnarray*}
S_1^1(x_1,x_2)&=&f(r)\cos\theta \hspace{2cm}S_1^2(x_1,x_2)=f(r)\sin\theta \\
S_2^1(x_1,x_2)&=&h(r)\hspace{2.9cm}S_2^2(x_1,x_2)=0 \\
V_1(x_1,x_2)&=&-\frac{\alpha(r)}{r}\sin\theta \hspace{1.7cm}
V_2(x_1,x_2)=\frac{\alpha(r)}{r}\cos\theta \qquad ,
\end{eqnarray*}
the only ones at our disposal, although $f(r)$, $h(r)$, and
$\alpha(r)$ have been found numerically.

\subsubsection{Seeley densities for circle symmetric semi-local
vortices}

The first step is to use the first-order rotationally symmetric ODE
(\ref{sdfood}) to write the field derivatives in terms of the field
profiles themselves: {\small \begin{eqnarray*} \frac{\partial
S_1^1}{\partial x_1}&=&\frac{
f(r)}{r}\left[1-\alpha(r)\cos^2\theta\right]\hspace{1cm}
\frac{\partial S_1^2}{\partial
x_2}=\frac{f(r)}{r}\left[1-\alpha(r)\sin^2\theta
\right]\hspace{1cm}\frac{\partial S_2^1}{\partial x_2}=-\frac{
h(r)}{r}\alpha(r)
\cos\theta \\
\frac{\partial S_1^1}{\partial x_2}&=&-\frac{ f(r)}{r}\alpha(r)
\cos\theta \sin \theta\hspace{1.2cm}\frac{\partial S_1^2}{\partial
x_1}=-\frac{ f(r)}{r}\alpha(r)) \sin\theta \cos
\theta\hspace{1cm}\frac{\partial S_2^1}{\partial x_1}=-\frac{
h(r)}{r}\alpha(r)) \sin\theta
\\\frac{\partial V_1}{\partial x_1}&=&{\cos
2\theta\over 2}\left[\frac{2 f(r)\alpha(r)}{r}+{f^2(r)+h^2(r)-1\over
2}\right] \hspace{0.4cm} \quad \frac{\partial V_1}{\partial x_2}=-
\cos
2\theta\frac{\alpha(r)}{r^2}+{\sin^2\theta\over 2}(f^2(r)+h^2(r)-1)\\
\frac{\partial V_2}{\partial x_1}&=&- \cos
2\theta\frac{\alpha(r)}{r^2}- {\cos^2\theta\over 2}
(f^2(r)+h^2(r)-1)\hspace{0.3cm}\hspace{0.5cm} \frac{\partial
V_2}{\partial x_2}=-{\cos 2\theta\over 2}\left[\frac{2
f(r)\alpha(r)}{r}+{f^2(r)+h^2(r)-1\over 2}\right] \, \,  .
\end{eqnarray*}}
This step is important because numerical calculations on field
derivatives cause considerable errors. The next step is the
Mathematica solution of the rotationally symmetric recurrence
relations. Below we list the three first-order circle symmetric
Seeley densities:

{\footnotesize \begin{eqnarray*}
{\rm tr}\,c_1^I(r) &=&5 - \frac{2\,{\alpha(r)}^2}{r^2} - 5\,{f(r)}^2 - 3\,{h(r)}^2 \\
{\rm tr}\,c_2^I(r)&=&\frac{1}{12\,r^4} \left[4\,{\alpha(r)}^4 +
27\,r^4\,{f(r)}^4 -
    8\,r^2\,\alpha(r)\,\left( -1 + 14\,{f(r)}^2 + {h(r)}^2 \right)  + \right. \\
    & & \hspace{0.8cm} +
    8\,{\alpha(r)}^2\,\left( -2 - 3\,r^2 + 9\,r^2\,{f(r)}^2 + 3\,r^2\,{h(r)}^2 \right)
    + \\ & & \hspace{0.8cm} + \left.
    {f(r)}^2\,\left( 56\,r^2 - 64\,r^4 + 34\,r^4\,{h(r)}^2 \right)  +
    r^4\,\left( 37 - 32\,{h(r)}^2 + 7\,{h(r)}^4 \right) \right] \\
{\rm tr}\,c_3^I(r) &=& \frac{1}{120\,r^6} \left\{-4\,{\alpha(r)}^6 -
4\,r^2\,{\alpha(r)}^3\,\left( 14 + 35\,{f(r)}^2 - 36\,{h(r)}^2
\right) + \right.
\\ && \hspace{0.9cm} +
    4\,{\alpha(r)}^4\,\left( 20 + 9\,r^2 + 32\,r^2\,{f(r)}^2 + 26\,r^2\,{h(r)}^2 \right)
     - \\ && \hspace{0.3cm} -
    2\,r^2\,\alpha(r)\,\left[ 57\,r^2\,{f(r)}^4 +
       {f(r)}^2\,\left( 32 + 331\,r^2 - 75\,r^2\,{h(r)}^2 \right)  -
       4\,\left( -1 + {h(r)}^2 \right) \,\left( -16 - 9\,r^2 + r^2\,{h(r)}^2 \right)
       \right]
       + \\ && \hspace{0.9cm} +
     {\alpha(r)}^2\,\left[ -256 - 144\,r^2 - 117\,r^4 + 99\,r^4\,{f(r)}^4 - 16\,r^2\,{h(r)}^2 +
       94\,r^4\,{h(r)}^2 - 61\,r^4\,{h(r)}^4 + \right. \\ && \hspace{1.3cm}
       + \left.
       2\,r^2\,{f(r)}^2\,\left( 56 + 183\,r^2 + 19\,r^2\,{h(r)}^2 \right)  \right]
       + \\ && \hspace{0.9cm} +
    r^4\,\left[ -16 + 151\,r^2 - 29\,r^2\,{f(r)}^6 + \left( 32 - 135\,r^2 \right) \,{h(r)}^2 +
       \left( -16 + 23\,r^2 \right) \,{h(r)}^4 + r^2\,{h(r)}^6 + \right. \\
        &+& \left. \left.
       {f(r)}^4\,\left( -20 + 199\,r^2 - 57\,r^2\,{h(r)}^2 \right)  +
       {f(r)}^2\,\left( 392 - 321\,r^2 + 2\,\left( -68 + 111\,r^2 \right) \,{h(r)}^2 -
          27\,r^2\,{h(r)}^4 \right)  \right] \right\} \, \, \, .
\end{eqnarray*}}
Here, ${\rm tr}$ means that we have summed up to the fourth diagonal
density.

{\footnotesize\begin{eqnarray*} {\rm tr}\,c_1^O(r) &=& 1 -
\frac{2\,{\alpha(r)}^2}{r^2}
- {f(r)}^2 - 3\,{h(r)}^2 \\
{\rm tr}\,c_2^O(r) &=& \frac{1}{12\,r^4} [4\,{\alpha(r)}^4 -
r^4\,{f(r)}^4 + 8\,r^2\,\alpha(r)\,\left( 1 + 2\,{f(r)}^2 - {h(r)}^2
\right)  - \\ && \hspace{0.8cm} -
    8\,{\alpha(r)}^2\,\left( 2 + r^2 + r^2\,{f(r)}^2 - 5\,r^2\,{h(r)}^2 \right)  +
    2\,r^2\,{f(r)}^2\,\left( -4 + 9\,r^2\,{h(r)}^2 \right)  + \\ &&
    \hspace{0.8cm} +
    r^4\,\left( 1 - 8\,{h(r)}^2 + 19\,{h(r)}^4 \right) ]
\end{eqnarray*}}
\newpage
{\footnotesize\begin{eqnarray*} {\rm tr}\,c_3^O(r) &=&
\frac{-1}{120\,r^6}\left\{4\,{\alpha(r)}^6 -
4\,r^2\,{\alpha(r)}^3\,\left( -14 + 9\,{f(r)}^2 + 84\,{h(r)}^2
\right) - \right.
\\ & & \hspace{0.8cm} -
    4\,{\alpha(r)}^4\,\left( 20 + 3\,r^2 + 2\,r^2\,\left( {f(r)}^2 + 4\,{h(r)}^2 \right)  \right)
    + \\ && \hspace{0.8cm} +
    {\alpha(r)}^2\,\left[ 256 + 48\,r^2 - 3\,r^4 + 45\,r^4\,{f(r)}^4 +
       2\,r^2\,\left( -40 + 89\,r^2 \right) \,{h(r)}^2 - 115\,r^4\,{h(r)}^4
       + \right. \\ && \hspace{2.5cm} + \left.
       2\,r^2\,{f(r)}^2\,\left( 8 + 5\,r^2 - 35\,r^2\,{h(r)}^2 \right)
       \right]
       - \\ && \hspace{0.3cm} -
    2\,r^2\,\alpha(r)\,\left[ 53\,r^2\,{f(r)}^4 +
       4\,\left( -1 + {h(r)}^2 \right) \,\left( -16 - 3\,r^2 + 7\,r^2\,{h(r)}^2 \right)  -
       {f(r)}^2\,\left( 32 + 17\,r^2 + 47\,r^2\,{h(r)}^2 \right)
       \right]
       + \\ & & \hspace{0.8cm} +
    r^4\,\left[ 16 + 3\,r^2 + 3\,r^2\,{f(r)}^6 - \left( 32 + 19\,r^2 \right) \,{h(r)}^2 +
       \left( 16 + 23\,r^2 \right) \,{h(r)}^4 + 33\,r^2\,{h(r)}^6 +
       \right.
       \\ && \hspace{1.8cm} + \left. \left.
       {f(r)}^4\,\left( 52 - r^2 + 39\,r^2\,{h(r)}^2 \right)  +
       {f(r)}^2\,\left( -24 - 5\,r^2 + \left( -72 + 22\,r^2 \right) \,{h(r)}^2 +
          69\,r^2\,{h(r)}^4 \right)  \right] \right\} \, \, .
\end{eqnarray*}}
${\rm tr}$ now means that we have summed $[c_n^O]_{55}$ and
$[c_n^O]_{66}$.

{\footnotesize \begin{eqnarray*}
c_1^G (r) &=& 1 - {f(r)}^2 - {h(r)}^2 \\
c_2^G(r) &=& \frac{1}{6\,r^2} \left[3\,r^2 + 2\,r^2\,{f(r)}^4 -
\left( 5\,r^2 + 4\,{\alpha(r)}^2 \right) \,{h(r)}^2 +
2\,r^2\,{h(r)}^4 + \right. \\ & & \hspace{0.8cm} \left. +
{f(r)}^2\,\left( -4 - 5\,r^2 + 8\,\alpha(r) - 4\, {\alpha(r)}^2 +
4\,r^2\,{h(r)}^2 \right) \right] \\
c_3^G(r) &=& \frac{-1}{60\,r^4}\left\{-10\,r^4 + 4\,r^4\,{f(r)}^6 +
\left[ 23\,r^4 - 8\,r^2\,\alpha(r) + 16\,\left( 1 + r^2 \right)
\,{\alpha(r)}^2 + 32\,{\alpha(r)}^3 + 16\,{\alpha(r)}^4 \right]
\,{h(r)}^2 + \right.
\\ & &  \hspace{0.9cm} + r^2\,\left[ -17\,r^2 + 8\,\alpha(r) - 16\,{\alpha(r)}^2 \right] \,{h(r)}^4 +
4\,r^4\,{h(r)}^6 +
 \\  & &  \hspace{0.9cm} +    r^2\,{f(r)}^4\,\left[ -24 - 17\,r^2 + 40\,\alpha(r) -
 16\,{\alpha(r)}^2 + 12\,r^2\,{h(r)}^2 \right]
 +  \\ &&  \hspace{0.9cm} + {f(r)}^2\,\left[ -32\,{\alpha(r)}^3 + 16\,{\alpha(r)}^4 +
 8\,r^2\,\alpha(r)\,\left( -5 + 6\,{h(r)}^2 \right)  +
 16\,{\alpha(r)}^2\,\left( 1 + r^2 - 2\,r^2\,{h(r)}^2 \right) \right. + \\
& & \left. \hspace{0.9cm} + \left. r^2\,\left( 24 + 23\,r^2 -
2\,\left( 10 + 17\,r^2 \right) \,{h(r)}^2 + 12\,r^2\,{h(r)}^4
\right)  \right] \right\} \, \, \, .
\end{eqnarray*}}

\subsubsection{One-loop SSTS mass shifts}
In this sub Section we offer some Tables with Mathematica
calculations of the one-loop corrections to the masses of
topological solitons in the semi-local Abelian Higgs model. We
denote the Seeley coefficients, calculated by means of numerical
integration of the circle symmetric Seeley densities evaluated at
the numerical solutions for the field profiles, in the form:
\[
{\rm tr}\,c_n^I=2\pi \int_0^\infty \, dr \, r {\rm
tr}\,c_n^I(r)\quad , \quad {\rm tr}\,c_n^O=2\pi \int_0^\infty \, dr
\, r {\rm tr}\,c_n^O(r)\quad , \quad  c_n^G=2\pi \int_0^\infty \, dr
\, r c_n^G(r) \qquad .
\]
First, we focus on the embedded ANO vortices, the $h_0$ topological
solitons. Second, we give results for several values of $h_0$ up to
a value of $h_0$ close to $h_0=1$, which corresponds to the
${\mathbb C}{\mathbb P}^1$-lumps.
\newpage

1. One-loop vortex mass shifts in units of $\hbar m$:

Because the ANO vortex solutions were generated numerically,
integration over the whole plane of the Seeley densities can also
only be performed numerically. Therefore, we are forced to put a
cut-off in the area and replace the infinite plane by a discus of
radius $R$, which in the calculations shown below was chosen to be
$R=10.000$

\begin{table}[h]
\hspace{2cm} \begin{tabular}{|c|c|c|c|}
 \hline  & \multicolumn{3}{|c|}{$h_0=0.0$}  \\ \hline
 $n$ & ${\rm tr}\,c_n^I$ & ${\rm tr}\,c_n^O$ & $c_n^G$  \\ \hline
 1 & -41.4469  & -91.8429   & 12.599 \\ \hline
 2 & 30.3736  & 0.96286     & 2.61518  \\ \hline
 3 & 12.9447  & -0.0592415  & 0.32005 \\ \hline
 4 & 4.22603  & 0.001512548  & 0.0230445 \\ \hline
 5 & 1.05059  & 0.000758663 & 0.0013023 \\ \hline
 6 & 0.20900  & -0.00023912 & 0.0000698185 \\ \hline
\end{tabular}
\hspace{1cm}
\begin{tabular}{|c|c|}  \hline
$N_0$ & $\Delta M_V (N_0)$ \\
 & $l=1$
\\ \hline
2 & -1.61536  \\
3 & -1.66862 \\
4 & -1.67809 \\
5 & -1.67966 \\
6 & -1.67989 \\ \hline
\end{tabular}
\caption{Seeley coefficients for the (left) and Quantum Mass Correction (right) to the soliton in the semi-local Abelian Higgs model with $h_0=0.0$.}
\end{table}
In the Table on the left, the first six Seeley coefficients are
given. The next coefficients are very small and one expects that the
approximation to the exact value of the one-loop mass shift is quite
good, as shown in the Table on the right: the mass shift obtained by
counting five or six coefficients agrees up to the third decimal
figure. In fact, the thresholds in the wells are $\leq 1$. 1 is the
expectation value of the scalar field at the vacuum, and we find a
similar situation to sine-Gordon kinks (see the first Part), where
keeping six coefficients provides a fairly good approximation.

From the same Table one can read the one-loop mass shifts of ANO
vortices with $l=1$ in the Abelian Higgs model: just take ${\rm
tr}c_n^O(K)$ equal to zero and subtract two zero modes, not 4. The
ratio is:
\[
\frac{\Delta M_V^{\rm SLAHM}}{\Delta M_V^{\rm
AHM}}=\frac{1.67989}{1.09449}=1.53486 \qquad .
\]
A similar proportion exists between the ratios of kink mass shifts
in the $\lambda(\phi)^4_2$ model and the BNRT model, both treated in
the first Part:
\[
\frac{\Delta M_K^{\rm BNRTM}}{\Delta M_K^{\rm \lambda\Phi
M}}=\frac{0.693943}{0.471113}=1.47299 \, \, , \, \, \sigma=0.99
\quad , \quad \frac{\Delta M_K^{\rm BNRTM}}{\Delta M_K^{\rm
\lambda\Phi M}}=\frac{0.698445}{0.471113}=1.48254 \, \, , \, \,
\sigma=1.01 \, \, \, .
\]
2. SSTS one-loop mass shifts

Whereas the one-loop mass shift of ANO embedded vortices is always
negative and varies extremely slowly as the area increases towards
more negative values, one-loop mass shifts of genuine semilocal
topological solitons with $|h_0|>0$ become less negative, and even
positive, for larger areas, as is shown in the following
Table.\vspace{0.2cm}
\begin{table}[h]
\hspace{0.3cm} {\small \begin{tabular}{|c|c|c|c|c|c|}  \hline
$R$ & $\Delta M_V (N_0=6,R)$ & $\Delta M_V (N_0=6,R)$ & $\Delta M_V
(N_0=6,R)$
& $\Delta M_V (N_0=6,R)$ &$\Delta M_V (N_0=6,R)$ \\
 & $h_0=0.0$ & $h_0=0.1$ & $h_0=0.3$ & $h_0=0.6$ & $h_0=0.9$
\\ \hline
$10^2$ & -1.67955 & -1.61672 & -1.05000 & 2.10142 & 24.6066 \\
$10^3$ & -1.67971& -1.58311 & -0.626167 &4.5485 &  42.7747 \\
$10^4$ & -1.67989 & -1.55133 & -0.252586 &6.41655 & 60.9433 \\
$10^5$ & -1.68005 & -1.51957 &  0.12086  &8.5741  & 79.1116 \\
$10^6$ & -1.68026 & -1.48779 &  0.49433  &10.7203 & 97.2798 \\
\hline
\end{tabular}}\caption{One-loop mass shifts for semi-local
topological solitons: Five values of $h_0$, five values of $R$, and
fixed $N_0=6$.}
\end{table}

The classical degeneracy in energy between semi-local topological
defects seems to be broken by one-loop fluctuations, the ANO
embedded vortices becoming the ground states in the topological
sector of one quantum of magnetic flux. It is remarkable how strong
this effect becomes for topological solitons close to ${\mathbb
C}{\mathbb P}^1$-lumps, $|h_0|\simeq 1$.

\subsubsection{Infrared divergences: quantum fate of semi-local
topological solitons}

The origin of the degeneracy breaking is the slow decay
(non-exponential) to their vacuum values of genuine semilocal
topological solitons as compared with ANO vortices. Plugging the
asymptotic form of the circle symmetric topological soliton
solutions in the Seeley densities, we find the following behavior at
infinity in terms of the parameter $|h^1|$ (which sets the long $r$
behavior of the solutions): {\small\begin{eqnarray*} 2\pi r{\rm
tr}c_1^I(r)& \stackrel{r\rightarrow\infty}{\simeq} &
-\frac{4\pi}{r}(1-|h^1|^2)+\frac{4\pi}{r^3}(12|h^1|^2-|h^1|^4)+{\cal
O}({1\over r^5}) \\ 2\pi r{\rm tr}c_1^O(r)& \stackrel{r\rightarrow\infty}{\simeq} &
-\frac{4\pi}{r}(1+|h^1|^2)+\frac{4\pi}{r^3}(4|h^1|^2+|h^1|^4)+{\cal
O}({1\over r^5}) \\ 2\pi r c_1^G(r)& \stackrel{r\rightarrow\infty}{\simeq} &\,\, \frac{8\pi}{r^3}|h^1|^2+{\cal O}({1\over r^5})
\end{eqnarray*}}
{\small\begin{eqnarray*} 2\pi r{\rm tr}c_2^I(r)& \stackrel{r\rightarrow\infty}{\simeq} &
\,\,\frac{2\pi}{r}|h^1|^2+\frac{2\pi}{r^3}(-1+4|h^1|^2-|h^1|^4)+{\cal
O}({1\over r^5}) \\ 2\pi r{\rm tr}c_2^O(r)& \stackrel{r\rightarrow\infty}{\simeq} & \,\,
\frac{2\pi}{r}|h^1|^2+\frac{2\pi}{r^3}(-1+4|h^1|^2-|h^1|^4)+{\cal
O}({1\over r^5}) \\2\pi r c_2^G(r)& \stackrel{r\rightarrow\infty}{\simeq} & \,\,
\frac{64\pi}{3r^5}|h^1|^2+\frac{\pi}{r^7}(768|h^1|^2-80|h^1|^4)+{\cal
O}({1\over r^9})
\end{eqnarray*}}
{\small\begin{eqnarray*} 2\pi r{\rm tr}c_3^I(r)& \stackrel{r\rightarrow\infty}{\simeq} &
\,\,\frac{2\pi}{3r}|h^1|^2+\frac{2\pi}{3r^3}(3|h^1|^2-|h^1|^4)+{\cal
O}({1\over r^5}) \\ 2\pi r{\rm tr}c_3^O(r)& \stackrel{r\rightarrow\infty}{\simeq} & \,\,
-\frac{2\pi^2}{3r}|h^1|^2+\frac{2\pi}{3r^3}(-4|h^1|^2+|h^1|^4)+{\cal
O}({1\over r^5}) \\2\pi r c_3^G(r)& \stackrel{r\rightarrow\infty}{\simeq} & \,\, \frac{384\pi}{5r^7}|h^1|^2+{\cal O}({1\over r^8})
\end{eqnarray*}}
{\small\begin{eqnarray*} 2\pi r{\rm tr}c_4^I(r)& \stackrel{r\rightarrow\infty}{\simeq} &
\,\,\frac{\pi}{6r}|h^1|^2+\frac{\pi}{6r^3}({12\over
5}|h^1|^2-|h^1|^4)+{\cal O}({1\over r^5}) \\ 2\pi r{\rm
tr}c_4^O(r)& \stackrel{r\rightarrow\infty}{\simeq} &\,\,
-\frac{\pi}{6r}|h^1|^2+\frac{\pi}{6r^3}(4|h^1|^2-|h^1|^4)+{\cal
O}({1\over r^5}) \\2\pi r c_4^G(r)& \stackrel{r\rightarrow\infty}{\simeq} & \,\,
\frac{832\pi}{21r^9}|h^1|^4-\frac{43\pi}{30r^9}|h^1|^6+\frac{\pi}{48r^9}|h^1|^8+
\frac{\pi}{768r^9}|h^1|^{10}+{\cal O}({1\over r^{11}})
\end{eqnarray*}}
{\small\begin{eqnarray*} 2\pi r{\rm tr}c_5^I(r)& \stackrel{r\rightarrow\infty}{\simeq} &
\,\,\frac{\pi}{30r}|h^1|^2+\frac{\pi}{30r^3}(2|h^1|^2-|h^1|^4)+{\cal
O}({1\over r^5}) \\ 2\pi r{\rm tr}c_5^O(r)& \stackrel{r\rightarrow\infty}{\simeq} & \,\,
-\frac{\pi}{30r}|h^1|^2+\frac{\pi}{30r^3}(-4|h^1|^2+|h^1|^4)+{\cal
O}({1\over r^5})
\\2\pi r c_5^G(r)& \stackrel{r\rightarrow\infty}{\simeq} & \,\,-\frac{\pi}{r^9}
\left(\frac{2048}{105}|h^1|^2-\frac{416}{189}|h^1|^4+\frac{43}{540}|h^1|^6-{1\over
864}|h^1|^8+{1\over 138240}|h^1|^{10}\right)+{\cal O}({1\over
r^{11}})
\end{eqnarray*}}
{\small\begin{eqnarray*} 2\pi r{\rm tr}c_6^I(r)& \stackrel{r\rightarrow\infty}{\simeq} & \,\,\frac{\pi}{180 r}|h^1|^2+\frac{\pi}{15
r^3}({1\over 7}|h^1|^2-{1\over 12}|h^1|^4)+{\cal O}({1\over r^5})
\\ 2\pi r{\rm tr}c_6^O(r)& \stackrel{r\rightarrow\infty}{\simeq} &\,\,
\frac{\pi}{180r}|h^1|^2+\frac{\pi}{45r^3}(|h^1|^2-{1\over
4}|h^1|^4)+{\cal O}({1\over r^5})
\\
\end{eqnarray*}}
The key observation is the appearance of infrared logarithmic
divergences in the Seeley coefficients ${\rm tr}c^I_n(K)$ and ${\rm
tr}c^O_n(K)$ for all $n$. The ghost coefficients $c^G_n(K^G)$,
however, are infrared convergent. The combination of the signs that
we have seen in the previous sub-sections with the long $r$ behavior
shows that one-loop mass shifts of semi-local topological solitons
tend to $+\infty$ in the infinite area limit. Semi-local topological
defects grow infinitely massive due to the infrared effects of
one-loop fluctuations. This phenomenon seems to be amazingly close
to the non-existence of Goldstone bosons in (1+1)-dimensions.

There is a very important exception: for ANO vortices, $|h^1|=0$ and
only the first-order coefficients are infrared divergent. However,
the contribution of these coefficients is totally canceled by mass
renormalization counter-terms. Our results suggest that only the ANO
vortices between all the semi-local topological solitons survive
one-loop quantum fluctuations. It would be very interesting to try a
more analytic approach to this problem in order to fully elucidate
this delicate issue.

\subsubsection{Circle symmetric self-dual
Abrikosov-Nielsen-Olesen vortices: one-loop mass shifts up to $l=4$}

Finally, we consider the problem of computing the one-loop mass
shifts for superimposed ANO vortices at the $\kappa=1$ limit up to
four quanta -$8\pi$- of magnetic flux in the Abelian Higgs model (no
fluctuations in the $a=5,6$ directions and $2l$ zero modes).
\begin{figure}[h]
\centerline{\includegraphics[height=1.9cm]{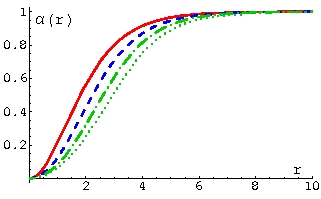}\hspace{0.8cm}
\includegraphics[height=1.9cm]{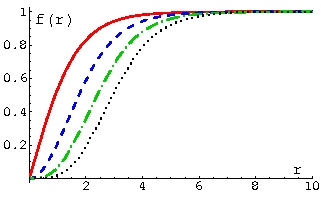}\hspace{0.8cm}
\includegraphics[height=1.9cm]{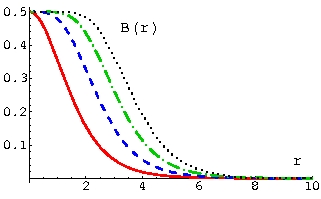}\hspace{0.8cm}
\includegraphics[height=1.9cm]{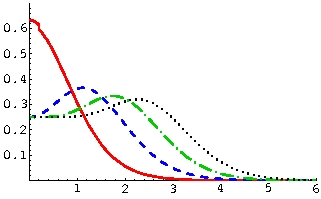}}
\caption{Plots of the field profiles $\alpha(r)$ (a) and $f(r)$ (b),
the magnetic field $B(r)$ (c), and the energy density
$\varepsilon(r)$ for self-dual vortices with $l=1$ (solid line),
$l=2$ (broken line), $l=3$ (broken-dotted line) and $l=4$ (dotted
line).}
\end{figure}
The Figure 13 shows the field profiles, the magnetic field, and the
energy density for the numerically generated solutions for $l=1$,
$l=2$, $l=3$, and $l=4$ in \cite{AMJW1}. The Figure 14, however,
encompass the 3D plots of these solutions.

\begin{figure}[h]
\centerline{\includegraphics[height=3.8cm]{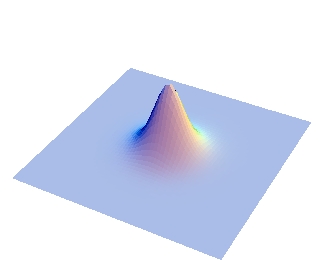}\hspace{0.8cm}
\includegraphics[height=3.8cm]{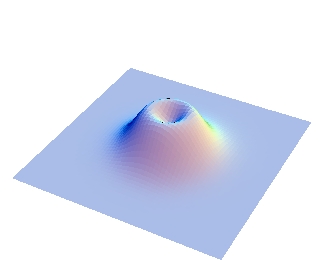}\hspace{0.8cm}}
\end{figure}
\begin{figure}[h]
\centerline{\includegraphics[height=3.8cm]{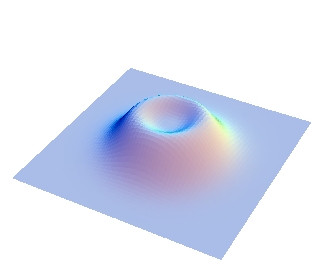}\hspace{0.8cm}
\includegraphics[height=3.8cm]{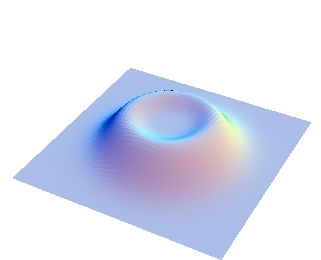}}
\caption{3D graphics of the energy density for $l=1$, $l=2$, $l=3$
and $l=4$ self-dual symmetric ANO vortices.}
\end{figure}
The Seeley coefficients for the different values of $l$ as well as
the one-loop mass shifts are displayed in the Tables below
{\footnote{ The Seeley densities corresponding to self-dual ANO
vortices superposed at the origin up to four quanta of magnetic flux
are given in \cite{AMJW1}.}}

\begin{table}[h]
{\footnotesize
\hspace{0.2cm} \begin{tabular}{|c|cc|cc|}
\hline & \multicolumn{2}{|c|}{$l=1$} & \multicolumn{2}{|c|}{$l=2$}
\\ \hline
$n$ & $ {\rm tr}{c}_n(K)$ & $ {c}_n(K^G)$ & $ {\rm tr}{c}_n(K)$ & $
{c}_n(K^G)$
\\ \hline
2 & 30.36316 & 2.60773 & 61.06679 & 6.81760    \\
3 & 12.94926 & 0.31851 & 25.61572 & 1.34209    \\
4 &  4.22814 &  0.022887 & 8.21053 & 0.20481   \\
5 & 1.05116 &  0.0011928 & 2.02107 & 0.023714  \\
6 &  0.20094 & 0.00008803 & 0.40233 & 0.002212  \\
\hline
\end{tabular}
\begin{tabular}{|c|cc|cc|}
\hline & \multicolumn{2}{|c|}{$l=3$} & \multicolumn{2}{|c|}{$l=4$}
\\ \hline
$n$ & $ {\rm tr}{c}_n(K)$ & $ {c}_n(K^G)$ & ${\rm tr}{c}_n(K)$ & $
{c}_n(K^G)$
\\ \hline
2 & 90.20440 & 11.51035 & 118.67540 & 16.46895 \\
3 & 36.68235 &  2.60898 &  46.01141 & 4.00762 \\
4 & 11.69979 &  0.46721 &  14.64761 & 0.77193 \\
5 & 2.86756  & 0.067279 &   3.58906 & 0.11747 \\
6 & 0.566227  & 0.0079269 & 0.667202 & 0.01620 \\
\hline
\end{tabular}}
\caption{Seeley Coefficients for $l=1$, $l=2$, $l=3$
and $l=4$ self-dual symmetric ANO vortices.}
\end{table}

\begin{table}[h]
\hspace{4cm} \begin{tabular}{|c|cccc|}  \hline
$N_0$ & $\Delta M_V (N_0)$ & $\Delta M_V (N_0)$& $\Delta M_V
(N_0)$ & $\Delta M_V (N_0)$ \\
 & $l=1$ & $l=2$ & $l=3$ & $l=4$
\\ \hline
2 & -1.02951 & -2.03787  &  -3.01187 & -3.97025 \\
3 & -1.08323 & -2.14111  &  -3.15680 & -4.14891 \\
4 & -1.09270 & -2.15913  &  -3.18208 & -4.18014 \\
5 & -1.09427 & -2.16212  &  -3.18628 & -4.18534 \\
6 & -1.09449 & -2.16257  &  -3.18690 & -4.18606 \\ \hline
\end{tabular}
\caption{One-loop mass shift for $l=1$, $l=2$, $l=3$
and $l=4$ self-dual symmetric ANO vortices.}
\end{table}
\noindent and the last Table provides the one-loop mass shifts of
circle symmetric self-dual ANO vortices up to four quanta of
magnetic flux taking into account $N_0=6$ Seeley coefficients.
\begin{center}
\begin{tabular}{|c|c|}
\hline $l$ & $\Delta M_V/\hbar m$ \\ \hline 1 & -1.09449  \\
2 & -2.16257 \\ 3 & -3.18690 \\  4 & -4.18606 \\
\hline
\end{tabular} \hspace{0.7cm}
\begin{tabular}{c}\includegraphics[height=2.2cm]{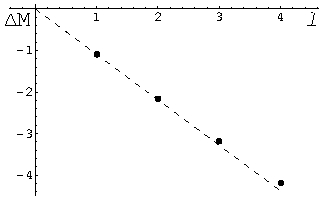}
\end{tabular}
\end{center}
It is remarkable that the one-loop mass shift seems to be linear in
$l$. We have found that the one-loop correction to the mass of a
circle symmetric vortex of magnetic flux $8\pi$ is extremely close
to four times the one-loop mass shift of a similar vortex of
magnetic flux $2\pi$. This result strongly supports our method:
after extremely sophisticated calculations we end with the natural
answer.

\section{Prospects on the future of the subject}

We finish by offering a summary of other possible approaches to this
subject as well as pointing to other playgrounds where similar
methods may work and should be applied.

\begin{enumerate}

\item In \cite{BC}, and \cite{BC1} Blas and Carrion explored a generalized sine-Gordon
model containing as many scalar fields as the rank of the ${\mathbb
S}{\mathbb L}(N,{\mathbb C})$ Lie group. The potential energy
density is determined from the simple roots in such a way that the
integrability of the sine-Gordon model is enjoyed by the generalized
model. Clearly, there is a correspondence between this model and the
ordinary sine-Gordon model that is analogous to the correspondence
between the deformed linear ${\mathbb O}(N)$-sigma model and the
$\lambda\phi^4$ model. There are several types of kinks in this
system, which have also been dealt with by the authors in a quantum
setting. Nevertheless, it seems to us that the generalized
sine-Gordon models offer another excellent arena for applying and
improving the quantization method described in this report.

\item In our Lectures we have addressed a massive non-linear ${\mathbb
S}^2$-sigma model having a rich kink manifold, see \cite{AMAJ2}. In
fact, we also computed the one-loop correction to the classical mass
of the topological kinks using the Cahill-Comtet-Glauber formula in
\cite{AMAJ3}. We look forward to attacking this problem by means of
the heat kernel/zeta function approach, not only in this system but
also in other similar non-linear sigma models where we vary the
potential energy density; e.g., adding quartic terms, and/or we
change the target space, e.g., to ${\mathbb S}^3$.

\item In a very interesting paper \cite{FMVW} Vafa et al. analyzed
${\cal N}=2$ supersymmetric Landau-Ginzburg models that are
integrable deformations of the ${\cal N}=2$ supersymmetric $A_k$
minimal series. The BPS states of these integrable
$(1+1)$-dimensional field theories were identified by these authors
as \lq\lq holomorphic" kinks with very noticeable properties. Almost
ten years later several of us went through the same set of kinks in
\cite{AMAJ4}. We focused only upon the bosonic sector of the model
and chose a real analytic point of view. We believe that the
techniques developed in these Lectures provide a procedure for
computing the one-loop kink mass shifts due only to bosonic
fluctuations. $k^2$ is the threshold of the wells and hence we
expect to need at least $N_0=20$ coefficients in order to reach a
good approximation. In the ${\cal N}=2$ supersymmetric version of
this model, the one-loop correction including bosonic and fermionic
fluctuations is zero. Therefore, there is no need to compute the
effect of fermionic fluctuations: bosons and fermions created in the
kink background cancel each other exactly.

\item Another strategy may be fruitful. The heat trace, or partition
function, which is basic in our approach, admits another conceptual
understanding as a path integral over closed world lines:
\begin{eqnarray}
S_E[z_1, \cdots , z_N]&=&\int_0^1 \, d\tau \,
\left[\frac{1}{2}\sum_{a=1}^N \,
\frac{dz_a}{d\tau}\cdot\frac{dz_a}{d\tau}+U[z_1(\tau), \cdots
z_N(\tau)]\right] \nonumber
\\ Z(\beta ,l)&=&{\rm Tr}_{L^2}e^{-\beta K}=\Pi_{a=1}^N\,
\int_{-\frac{l}{2}}^\frac{l}{2}\, dx_a \,
\int_{z_a(0)=x_a}^{z_a(1)=x_a}\, {\cal D}z(\tau)\, e^{-\beta
S_E[z_1, \cdots , z_N]} \label{pfwl} \qquad .
\end{eqnarray}
Here, $\tau=i t$ is imaginary time, $S_E$ is the Euclidean action
for a particle moving in a cube ${\mathbb I}^N$ with varying
positions $z_a(\tau)$. The path integral is over all the closed
paths, and ordinary integration over all the base points of the
loops is necessary. It is well known that the Feynman treatment of
these path integrals in the high-temperature regime \cite{Feyn}
reproduces the heat kernel expansion. What seems more promising is
the numerical computation of these world line path integrals using
Montecarlo methods, see e.g \cite{GL}. Applied to the ANO vortices
and semilocal strings this should provide reliable results to be
contrasted against our calculations.

\item The Lagrangian density
\begin{equation}
{\cal L}=\frac{1}{2}\left\{\partial_\mu\phi\partial^\mu\phi
-\left[\frac{\lambda_-}{l}\delta(z+\frac{l}{2})
+\frac{\lambda_+}{l}\delta(z-\frac{l}{2})+4\sigma^2-\frac{6\sigma^2}{{\rm
cosh}^2\sigma z}\right]\phi^2(x_\mu)\right\} \label{caki}
\end{equation}
describes the dynamics of a {\it rara avis} scalar quantum field
theory. The mesons do not move freely even though they do not
interact because they are constrained by a background. If
$\sigma=0$, the background in (\ref{caki}) is formed by two parallel
plates located at a distance $l$ from each other in two planes
orthogonal to the $z$-axis. The effect of the plates is mimicked by
two $\delta$-function potentials of strength $\lambda_\pm$. When
$l=\infty$ and $\sigma=1$, the background corresponds to a kink
living in the $z$-axis, or to a solitonic/thick domain wall
orthogonal to the $z$-axis. Other choices of the couplings provide
more complex backgrounds built from these two basic backgrounds.

In Reference \cite{KM} Milton wrote the Green's function for the
two-plate setup (using Dirichlet boundary conditions). This leads to
the energy momentum tensor encoding the Casimir energy -essentially
the $T_{00}$ component- and the Casimir force -essentially the
$T_{zz}$ component-. It is tempting to perform the same calculation
for the kink background. This should provide not only information
about the one-loop kink mass shift but should also shed light on the
qualitative nature of the forces exerted by the scalar fluctuations
on the kink profiles.

\item Over the last year, an interesting paper by Baacke and
Kevlishvili was published \cite{BK1} in which the one-loop shifts to
the classical masses of Nielsen-Olesen vortices were obtained by
using Green's function methods. The remarkable fact is that the
authors gave the quantum corrections with no restriction in the
ratio of Higgs and particle masses.

It seems that the time is ready to tackle the problem of the quantum
corrections of $Z$-electroweak strings. These topological defects
are embedded NO vortices in the neutral $Z_\mu$ massive vector field
of electroweak theory, and the calculation may will be of
experimental interest (even though some imaginary contribution to
the energy will arise, because electroweak strings are unstable). In
fact, there is an interesting paper on this subject, see
\cite{Weigel1}, at the non-physical value of the weak angle equal to
zero, and even a status report \cite{Weigel2}.

\item In a longer perspective, one might think of studying the
quantum fluctuations of BPS magnetic monopoles. The work in
\cite{RvNW4} suggests that the bosonic sector of ${\cal N}=2$
supersymmetric Yang-Mills is the right model to look into this
problem. Even though van Nieuwenhuizen et al. succeeded in computing
the one-loop mass shift to ${\cal N}=2$ SUSY monopoles, the
difficulties in a purely bosonic framework seem insurmountable.
First, the second-order operator governing the fluctuations is (for
the ${\mathbb SU}(2)$ group) a $21\times 21$ matrix-Schr$\ddot{\rm
o}$dinger operator in three dimensions. Second, the theory is
renormalizable, not super-renormalizable. The coupling constant also
receives one-loop divergent contributions that must be canceled by
the secod-order Seeley coefficient. So who is afraid of this big bad
wolf?

\end{enumerate}

\end{document}